%% file: main.tex
\documentclass[aps,prx,superscriptaddress,amsmath,amsfonts,amssymb,showpacs,floatfix,reprint,nofootinbib,longbibliography]{revtex4-2}

\usepackage{multirow}
\usepackage{graphicx}
\usepackage{bm}
\usepackage{dcolumn}
\usepackage[colorlinks=true,linkcolor=blue,urlcolor=blue,citecolor=blue]{hyperref}
\usepackage{natbib}

\usepackage{amsmath}
\usepackage{times}
\usepackage{xcolor}
\usepackage{soul}
\usepackage{comment}
\usepackage[normalem]{ulem}
\usepackage{textcomp,mathcomp}
\usepackage[varg]{txfonts}
\usepackage{physics}
\usepackage{siunitx}
\usepackage{gensymb} 
\usepackage{graphicx}
\usepackage{array}
\newcommand{\BTmZO}{\ch{Ba_3Tm_2Zn_5O_11}}
\newcommand{\BYZO}{\ch{Ba_3Yb_2Zn_5O_11}}
\newcommand{\BLuZO}{\ch{Ba_3Lu_2Zn_5O_11}}

\newcommand{\angstrom}{\textup{\AA}}

\usepackage{chemformula}

\usepackage{amssymb}
\usepackage{xcolor}

\begin{document}

\title{Synthesis and characterization of the novel breathing pyrochlore compound \BTmZO}

\author{Lalit Yadav}
\affiliation{Department of Physics, Duke University, Durham, NC, USA}

\author{Rabindranath Bag}
\affiliation{Department of Physics, Duke University, Durham, NC, USA}

\author{Ramesh Dhakal}
\affiliation{Department of Physics and Center for Functional Materials, Wake Forest University, Winston-Salem, NC 27109, USA}

\author{Stephen M. Winter}
\affiliation{Department of Physics and Center for Functional Materials, Wake Forest University, Winston-Salem, NC 27109, USA}

\author{Jeffrey G. Rau}
\affiliation{Department of Physics, University of Windsor, 401 Sunset Avenue, Windsor, Ontario, N9B 3P4, Canada}

\author{Sachith E. Dissanayake}
\affiliation{Department of Physics, Duke University, Durham, NC, USA}

\author{Alexander I. Kolesnikov}
\affiliation{Neutron Scattering Division, Oak Ridge National Laboratory, Oak Ridge, TN 37831, USA}

\author{Andrey A. Podlesnyak}
\affiliation{Neutron Scattering Division, Oak Ridge National Laboratory, Oak Ridge, TN 37831, USA}

\author{Craig M. Brown}
\affiliation{Center for Neutron Research, National Institute of Standards and Technology, MS 6100 Gaithersburg, MD 20899, USA}

\author{Nicholas P. Butch}
\affiliation{Center for Neutron Research, National Institute of Standards and Technology, MS 6100 Gaithersburg, MD 20899, USA}

\author{David Graf}
\affiliation{National High Magnetic Field Laboratory and Department of Physics, Florida State University, Tallahassee, FL 32310, USA.}

\author{Michel J.P. Gingras}
\affiliation{Department of Physics and Astronomy, University of Waterloo, Ontario, N2L 3G1, Canada}

\author{Sara Haravifard}
\email[email:]{sara.haravifard@duke.edu} 
\affiliation{Department of Physics, Duke University, Durham, NC, USA} 
\affiliation{Department of Mechanical Engineering and Materials Science, Duke University, Durham, NC, USA}


\begin{abstract}
In this study, a novel material from the rare-earth based breathing pyrochlore family, \BTmZO, was successfully synthesized. Powder X-ray diffraction and high-resolution powder neutron diffraction confirmed phase purity and crystal structure, while thermogravimetric analysis revealed incongruent melting behavior compared to its counterpart, \BYZO. High-quality single crystals of \BTmZO were grown using the traveling solvent floating zone technique and assessed using Laue X-ray diffraction and single crystal X-ray diffraction. Thermodynamic characterization indicated paramagnetic behavior down to 0.05 K, and inelastic neutron scattering measurements identified distinct crystal electric field bands, with the crystal electric field model predicting a single-ion singlet ground state and an energy gap of $\sim 9$ meV separating it from the first excited (singlet) state. Additional low-energy excitation studies on single crystals revealed dispersionless bands at 0.8 and 1 meV. Computed phonon dispersions from first principles calculations ruled out phonons as the source of these modes, further illustrating the puzzling and unique properties of \BTmZO.
\end{abstract}

\maketitle

\section{Introduction}

Frustrated quantum magnets have over the past thirty years become a fruitful playground for the exploration of exotic magnetic ground states with suppressed tendency towards conventional long-range order and displaying unconventional excitations~\cite{Balents2010SpinMagnets,Knolle2019ALiquids,Springer_frust}. The competing interactions in these systems can originate from a frustrated lattice geometry, spin-spin interactions beyond nearest neighbors or anisotropic spin-spin interactions. Geometrically frustrated lattices are commonly achieved using spins residing on either elementary triangular or tetrahedral units and forming regular two and three dimensional lattices. In three dimensions, the pyrochlore lattice of corner-sharing tetrahedra has become the paradigmatic example of high magnetic frustration~\cite{Balents2010SpinMagnets,Springer_frust}.

The pyrochlore architecture is prominently featured in the $A_2B_2$O$_7$ magnetic pyrochlores~\cite{Gardner2010MagneticOxides,
Hallas-AnnRevCMP,Rau-AnnRevCMP}. In these materials, the trivalent $A^{3+}$ and tetravalent $B^{4+}$ ions respectively span two independent inter-penetrating pyrochlore lattices, with either or both sites capable of hosting a magnetic moment depending on the cation. Magnetic materials with non-magnetic $A^{3+}$ (e.g. Y$^{3+}$ or Lu$^{3+}$) and magnetic transition metal $B^{4+}$ ions have been studied~\cite{Gardner2010MagneticOxides} (e.g. Y$_2$Mn$_2$O$_7$
~\cite{PhysRevB.54.7189}, Y$_2$Mo$_2$O$_7$\cite{Gingras-Y2Mo2O7,Gardner-Y2Mo2O7,Silverstein-Y2Mo2O7}, Lu$_2$Mo$_2$O$_7$~\cite{Molybdate_2014},  Lu$_2$Mo$_2$O$_5$N$_2$~\cite{Molybdate_2014} and Lu$_2$V$_2$O$_7$~\cite{Onose-Lu2V2O7}), and found to display a range of interesting thermodynamic, magnetic as well as thermal transport~\cite{Onose-Lu2V2O7} properties. 
However, a wider variety of materials displaying a gamut of magnetic behaviors has been afforded by the rare-insulating rare-earth compounds in which $A$ is a magnetic trivalent 4f lanthanide rare-earth element (e.g. $A=$ Ce, Pr, Nd, Sm, Gd, Tb, Dy, Ho, Er, Yb) and $B$ is a non-magnetic transition metal element (e.g. $B=$ Ti, Sn, Ge, Zr, Hf)~\cite{Gardner2010MagneticOxides,Sm2Ti2O7_2018, Nd2Zr2O7_2015,Gd2SnTi2O7_2003,Hallas-AnnRevCMP,Rau-AnnRevCMP}. In these systems, the strong single-ion crystal field anisotropy typically results in a magnetic crystal field ground state doublet described by an effective spin-$1/2$ degree of freedom. The interactions (e.g. magnetic, quadrupolar, etc) between the ions are then expressed in terms of anisotropic bilinear ``exchange-like'' couplings between those pseudospins $1/2$ ~\cite{Rau-AnnRevCMP,Ross-PRX,Lee-nonKramers,Huang-DO}.

\begin{figure}[tp]
\centering
  \includegraphics[scale=0.29]{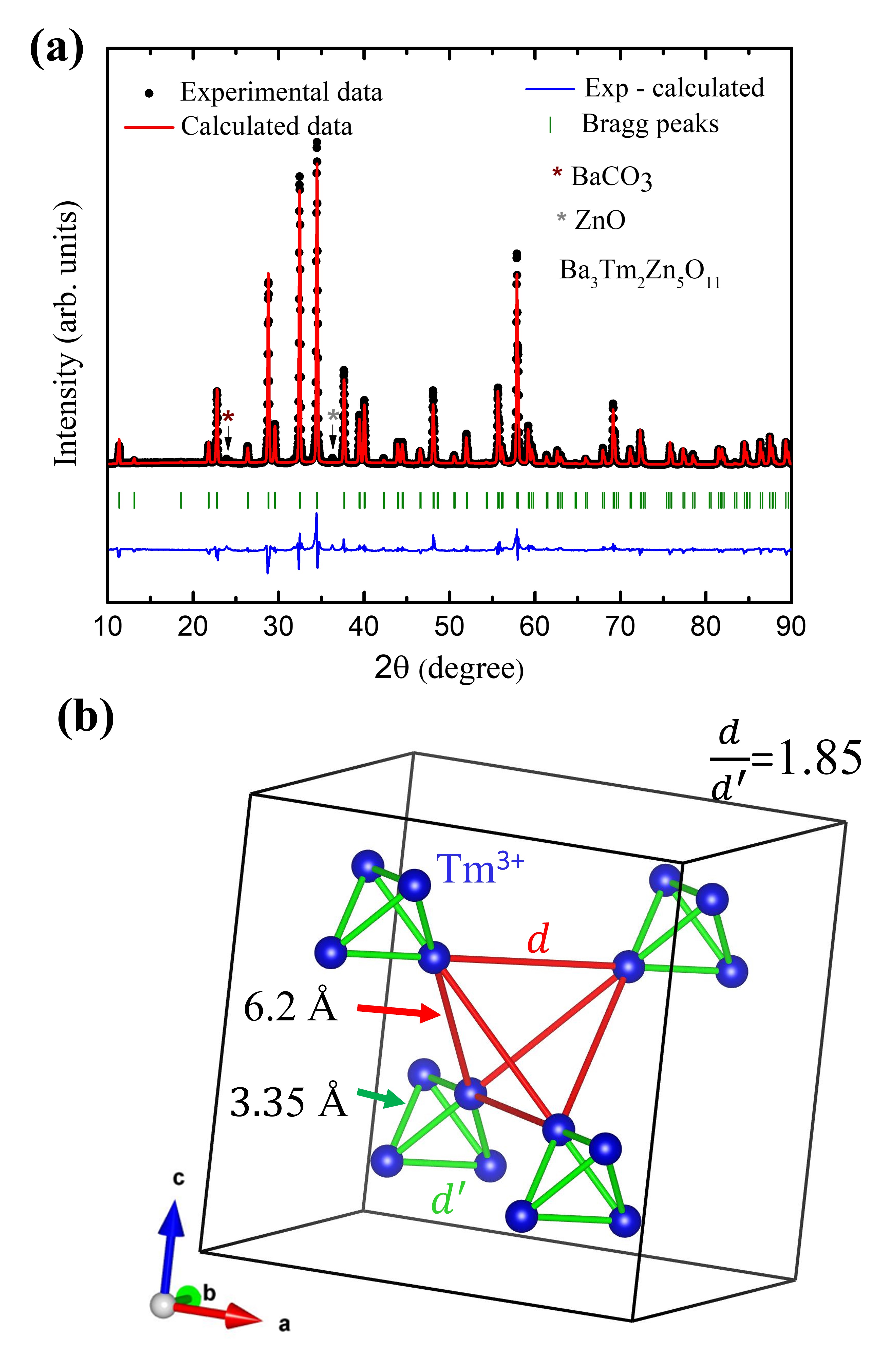}
  \caption{(a) Powder X-ray diffraction pattern of \BTmZO{} at room temperature is shown as black circle. The red, blue and green lines represent the calculated intensity, the difference between observed and calculated intensity, and the expected Bragg's positions, respectively. The additional non-magnetic \(\ch{ZnO}\) and \(\ch{BaCO3}\) impurity peaks are identified as $\ast$, specifically at the strongest peaks of these phases.  (b) Crystal structure of \ch{Ba3Tm2Zn5O11}, highlighting only the Tm$^{3+}$ ions, is shown with the breathing ratio ($d/d'$) and the alternating large and small corner sharing tetrahedra.}\label{pxrd2}
\end{figure}

In addition to the $A_2B_2$O$_7$ pyrochlore oxides, the spinels $AB_2X_4$ (e.g. $A$=Zn, Mg; $B$=Cr, Fe, $X$=O, S), in which the magnetic transition metal $B^{3+}$ ions reside on a pyrochlore lattice, have also been extensively studied
~\cite{spinels-review}.
The latter have provided great opportunities to study the physics of magnetic frustration on the pyrochlore lattice with transition metal ions (Cr$^{3+}$, Fe$^{3+}$) beyond that explored with the $A_2B_2$O$_7$ compounds with non-magnetic $A^{3+}$ and magnetic $B^{4+}$ transition metal ions
~\cite{PhysRevB.54.7189,Gingras-Y2Mo2O7,Gardner-Y2Mo2O7,Silverstein-Y2Mo2O7,Molybdate_2014,Onose-Lu2V2O7}.
About ten years ago~\cite{Okamoto-BP}, a new variety of magnetic spinels, $AA'B_4X_8$ (e.g. $A$=Li, Cu; $A'$=Ga, In; $B$=Cr; $X$=O, S, Se),  with the non-magnetic cations on the $A$ and $A'$ sites being crystallographically ordered, began attracting attention \cite{Okamoto-BP,Ghosh2019,Sharma2022}.
In these materials, the magnetic Cr$^{3+}$  ions reside on a so-called  \emph{breathing pyrochlore} (BP) lattice.
This architecture consists of an alternating arrangement of expanded (large) and contracted (small) corner-shared tetrahedra, with the nearest-neighbor bond distances 
$d$ and $d'$ for the large and small tetrahedra, respectively
(see. Fig.~\ref{pxrd2}(b) for an example of a breathing pyrochlore lattice with \emph{breathing ratio} $d/d'=1.85$).

The availability of BPs offers opportunities to explore interesting new and  enhanced frustration-driven phenomena. For example, theoretical studies have predicted that the parameter phase space of BP systems, which comprises different exchange couplings ($J_A$ and $J_B$) and  Dzyaloshinskii-Moriya (DM) interactions ($D_A$ and $D_B$) between large ($A$) and small ($B$) tetrahedra may contribute to stabilize exotic ground states such as quantum spin ice~\cite{Savary2016QuantumLattice}, spin liquid~\cite{Yan2020Rank-2Lattice}, and magnetic hedgehog-lattice~\cite{aoyama2023zero}  and Weyl magnons~\cite{Li2016WeylAntiferromagnets}.
In terms of  experimental realizations, Cr-based BP systems such as \ch{Li(Ga,In)Cr_4(O,S)_8}~\cite{Okamoto-BP,Pokharel2020ClusterS8}, CuInCr$_4$S$_8$~\cite{GenMasakiandOkamotoYoshihikoandMoriMasakiandTakenakaKoshiandKohama2020MagnetizationT} and CuAlCr$_4$S$_8$~\cite{Sharma2022} have been synthesized and investigated experimentally, as well as modelled theoretically~\cite{Ghosh2019}.  To a large extent, these Cr-based BP systems have a similar structure compared to a regular (i.e. non-breathing) pyrochlore since breathing ratio is nearly unity
($d/d'\approx 1.05 \pm 0.02$). Moreover, the Cr$^{3+}$ has effective $S=3/2$ spin for which quantum fluctuations are suppressed and are expected to behave nearly classically~\cite{Ghosh2019}.

A potential route to explore a  different regime outside the $d/'d\approx 1$ breathing ratio limit and the large $S$ value  characterizing the \ch{Li(Ga,In)Cr_4(O,S,Se)_8} materials is the rare (RE)  earth breathing pyrochlore (BP) family, known as \ch{Ba_3RE_2Zn_5O_{11}}~\cite{Haku2016CrystalBa3Yb2Zn5O11}. So far, only one member has been systematically explored: the effective spin-$\frac{1}{2}$ system (\BYZO{}). Unlike other breathing pyrochlores, the breathing ratio ($d/d'$) of the \BYZO{} BP system is approximately $1.8$ and thus much larger than the ratio for Cr-based BPs. At the very least, in this limit, one may hope to develop a better understanding of the microscopic aspects of magnetism  associated to frustration as well as the nature of the effective spin-$1/2$ interactions in rare-earth systems~\cite{Rau2018FrustrationOctahedra,Chibotaru} without the need to address the complexity of collective behavior.  In further contrast to the Cr-based BPs, \BYZO{} exhibits no signs of long-range ordering down to 100 mK~\cite{Haku2016CrystalBa3Yb2Zn5O11, Rau2018BehaviorField}. The limit of nearly isolated tetrahedra describes the physics of this compound quite well~\cite{Kimura2014ExperimentalAntiferromagnet, rau2016anisotropic, Haku2016CrystalBa3Yb2Zn5O11, haku2016low, haku2017neutron,Rau2018BehaviorField, Dissanayake2021TowardsStudy}, implying that the exchange interactions between the tetrahedra in \BYZO{} are small~\cite{Haku2016CrystalBa3Yb2Zn5O11}. On the other hand, \BYZO{} has been suggested as a candidate for rank-$2$ $U(1)$ QSL due to the strong DM interaction~\cite{Yan2020Rank-2Lattice} and has been found to exhibit low-temperature properties that hint at the significance of interactions between tetrahedra~\cite{Rau2018BehaviorField,bag2023beyond}.

Theoretical calculations aimed at assessing the 
stability of breathing pyrochlore compounds conclude that the size of small and large tetrahedra could be either slightly different ($d/d' \approx 1$) or significantly different ($d/d' \approx 2$)~\cite{Talanov2020FormationAspects}. In the first case, materials such as \ch{Li(Ga,In)Cr_4(O,S,Se)_8} and \ch{Pb_2Ir_2O_7} are closer to a standard $d/d'=1$ pyrochlore lattice with strongly coupled tetrahedra, while in the second case, the rare-earth BP compounds \ch{Ba_3RE_2Zn_5O_11} behave, to a large extent~\cite{Rau2018BehaviorField,bag2023beyond}, as a system of quasi-isolated tetrahedra. Nonetheless, in the second scenario, where a significant distance between small tetrahedra hinders inter-tetrahedron interactions, application of external pressure or the substitution with a magnetic ion possessing a larger magnetic moment may enhance the inter-tetrahedron interactions. Moreover, the ability of these inter-tetrahedron interactions to alter the single-tetrahedron physics is highly dependent on the specific single-ion manifold, which varies between different RE ions. Additionally, the intra-tetrahedron interactions, which, in principle and generally speaking, can also be strongly influenced by the choice of RE ion, may play a significant role. To explore the role of inter- and intra-tetrahedra interactions in providing a clearer picture of magnetism in BPs, it is thus important to search for new materials in the \ch{Ba_3RE_2Zn_5O_11} family that may harbor stronger inter-tetrahedra interactions or display interesting physics qualitatively distinct from the one found in \ch{Ba_3Yb_2Zn_5O_11}.

Building upon our previous synthesis and single-crystal growth achievements~\cite{Dissanayake2021TowardsStudy} and drawing from our accumulated expertise, we report in the present work a groundbreaking synthesis, single crystal growth, and comprehensive characterization of the novel rare-earth BP compound \BTmZO{}. Neutron diffraction data confirms the phase purity and structure of \BTmZO{} down to a temperature of 4 K. Unlike the relatively straightforward growth of single crystals of \BYZO{}~\cite{Dissanayake2021TowardsStudy}, growing single crystals of this new \BTmZO{} compound proved challenging due to the compound's incongruent melting behavior. To overcome this difficulty, we employed the traveling-solvent-floating zone (TSFZ) technique, utilizing a solvent disk composed of \ch{BaZnO_2}, which ultimately yielded large high-quality single crystals of \BTmZO{}. Magnetic and heat capacity studies on these single crystals conclusively indicate the absence of any long-range magnetic ordering down to 0.1 K. Additionally, neutron scattering was employed to investigate the crystal electric field (CEF) levels and low-energy excitations of this material.

\section {Experimental methods}
\label{sec:exp-methods}

Polycrystalline \BTmZO{} powder was prepared by a solid-state reaction route using the starting precursors of \ch{Tm_2O_3} (99.9\%, Alfa Aesar) with \ch{ZnO} (99.99\%, Alfa Aesar) and \ch{BaCO_3} (99.99\%, Alfa Aesar).The starting precursors were weighed in the (1:5.05:3) molar ratio, respectively, and the well-ground mixture was sintered at 800\degree ~C for 8 hours, 1000\degree ~C for 12 hours, and 1110\degree ~C for 48 hours with intermediate grinding. For the non-magnetic analogue, \ch{Ba3Lu2Zn5O11} was synthesized from the precursors \ch{Lu2O3} (99.9\%, Alfa Aesar) with \ch{ZnO} (99.99\%, Alfa Aesar) and \ch{BaCO3} (99.99\%, Alfa Aesar) weighed in the (1:5:3) molar ratio, respectively, and sintered at 1100°C for 48 hours. Additionally, attempts to synthesize \ch{Ba3Er2Zn5O11} revealed that it is highly unstable and likely not synthesizable. The phase purity and crystal structure were examined using a powder X-ray diffraction (PXRD) technique with a MiniFlex diffractometer from RIGAKU equipped with Cu K-$\alpha$
radiation source at room temperature. The temperature-dependent structural stability and long-range ordering were assessed through high-resolution powder neutron diffraction using the BT-1 spectrometer at NIST, Gaithersburg on $\approx$ 7 g of \BTmZO{} powder sample. The powder sample was packed in a vanadium can filled with helium-4 exchange  gas so that the sample could be cooled down to the base temperature. Measurements were systematically performed at 4 K,  45 K, and 100 K using a helium-4 type closed-cycle refrigerator (CCR).  The Cu (311) monochromator was used to produce incident neutron of wavelength $\SI{1.54}{\angstrom}$ along with a 15' collimator installed upstream of the sample. The diffraction pattern was analyzed using the FullProf suite \cite{Fulprof}. 

Once the phase purity of the powder sample was confirmed, the thermogravimetric analysis (TGA/DSC 3+, METTLER TOLEDO) was used to understand the intricacies of the melting behavior of this compound.
Subsequently, single crystals of \ch{Ba3Tm2Zn5O11} were grown using a four semi-ellipsoidal mirrors optical floating zone furnace (Model: FZ-T-12000-X-VII-VPO-PC, Crystal System Corporation, Japan) equipped with Xenon lamps with a maximum power of 12 kW. These mirrors focus the radiation from the Xenon lamps to achieve a large vertical temperature gradient. Additionally, alumina shutters control the feed rod's exposure to the optical rays, adjusting the temperature gradient and aiding in tuning the liquid zone size. The quality of the grown crystals was examined using a Laue diffractometer (MULTIWIRE LABS MWL120). 
Single crystal X-ray diffraction of \BTmZO{} was performed at the UNC Chapel Hill Department of Chemistry using a Bruker KAPPA Apex II diffractometer and measured at 230 K. 
The structure was solved using Superflip3 \cite{superflip} and refined using full-matrix least squares with the Crystals software from the  University of Oxford \cite{crystals}. Magnetic characterization on the grown crystal was done using a vibrating sample magnetometer (VSM) and heat capacity data was collected using helium-4 (1.8 K $\leq$ T $\leq$ 300 K) and dilution refrigeration (0.06 K $\leq$ T $\leq$ 2 K) set-up attached to the Physical Properties Measurement Systems, Quantum design (PPMS Dynacool, QD, USA). 
To investigate the magnetic properties of \BTmZO{}  down to \(0.02 \, \text{K}\) and in high magnetic fields, we performed tunnel diode oscillator (TDO) measurements at the DC Field Facility of the National High Magnetic Field Laboratory in Tallahassee. A bar-shaped single-crystal sample, with a length of approximately \(2 \, \text{mm}\), a width of \(1 \, \text{mm}\), and a thickness of \(0.5 \, \text{mm}\), was wound inside a detection coil, aligning the \([111]\) \BTmZO{} crystallographic direction along the coil axis.

The sample and coil together constituted the inductive component of an LC circuit connected to a tunnel diode operating in a negative resistance region which was fine-tuned to achieve resonance within a frequency range of \(10 \, \text{to} \, 50 \, \text{MHz}\). The shift in the resonance frequency \(f\), which relates to the change in the sample magnetization \(M\) (\(df/dH \propto d^2 M/dH^2\)) \cite{Clover1970,Shi2019}, was recorded as the field was swept to \(18 \, \text{T}\) using superconducting magnets.

Inelastic neutron scattering (INS) measurements were conducted using the SEQUOIA spectrometer \cite{granroth2010sequoia} at the spallation neutron source (SNS), Oak Ridge National Laboratory. Approximately 6 grams each of pure \BTmZO{} and \BLuZO{} (for the determination of the phonon contributions) powder samples were loaded into aluminium cans and sealed under helium-4 atmosphere. Measurements were collected at incident neutron energy ($E_i$) \(E_i=150\) meV and \(100\) meV in high flux mode and at \(E_i=50\) meV and \(25\) meV in high-resolution mode over the temperature range \(5 \, \text{K} \leq T \leq 150 \, \text{K}\). The data were analyzed using the DAVE MSlice \cite{azuah2009dave} while the PyCrystalField~\cite{scheie2021pycrystalfield} and MCphase~\cite{rotter2004using} packages were utilized to determine the parameters of the crystal electric field (CEF) Hamiltonian for Tm$^{3+}$ in 
\BTmZO{}.
The low-energy inelastic scattering experiments were performed at the Cold Neutron Chopper Spectrometer (CNCS) \cite{ehlers2011new} at the Spallation Neutron Source (SNS) at Oak Ridge National Lab on six pieces of co-aligned high-quality single crystal samples of \BTmZO{} with a total mass of \(\approx\) 0.9 gm. The crystals were mounted on a copper holder using superglue (see Fig.~\ref{sc_laue}), aligning in the $[h+k, -h+k, -2k]$ scattering plane, which is orthogonal to the $[111]$ crystallographic direction, with the magnetic field ${\bf H}$ along the $[111]$ direction. The measurements were conducted in a dilution refrigerator with a base temperature of 0.1 K as well as an 8 T magnet insert, using \(E_i=3.32\) meV.

\section{Results}
\label{sec:results}

\subsection{Crystal structure and phase purity}
\BTmZO{} is isostructural to \BYZO{} and crystallizes into the cubic structure with a space group of $F\overline{4}3m$ (no. 216). We show in Fig.~\ref{pxrd2}, the powder X-ray diffraction pattern collected on \BTmZO{} powder sample and the crystal structure obtained from the refined diffraction pattern.
Figure~\ref{pxrd2}(a) shows the experimental powder X-ray diffraction data and the calculated data obtained using Rietveld refinement along with the expected Bragg positions. The phase analysis confirms the phase purity of the synthesized \BTmZO{} powder sample. More than 99\% of the sample mass is in the BP phase with space group $F\overline{4}3m$. A negligible percentage of non-magnetic  impurity (BaCO$_3$ and ZnO) are identified with the associated peaks are shown with $\ast$ symbols in Fig.~\ref{pxrd2}(a). The crystal structure of \BTmZO{} thus corresponds to a breathing pyrochlore network where the Tm$^{3+}$ magnetic ions form large and small tetrahedra, as shown in \ref{pxrd2}(b). 

\begin{figure}[tp]
\centering
  \includegraphics[scale=0.4]{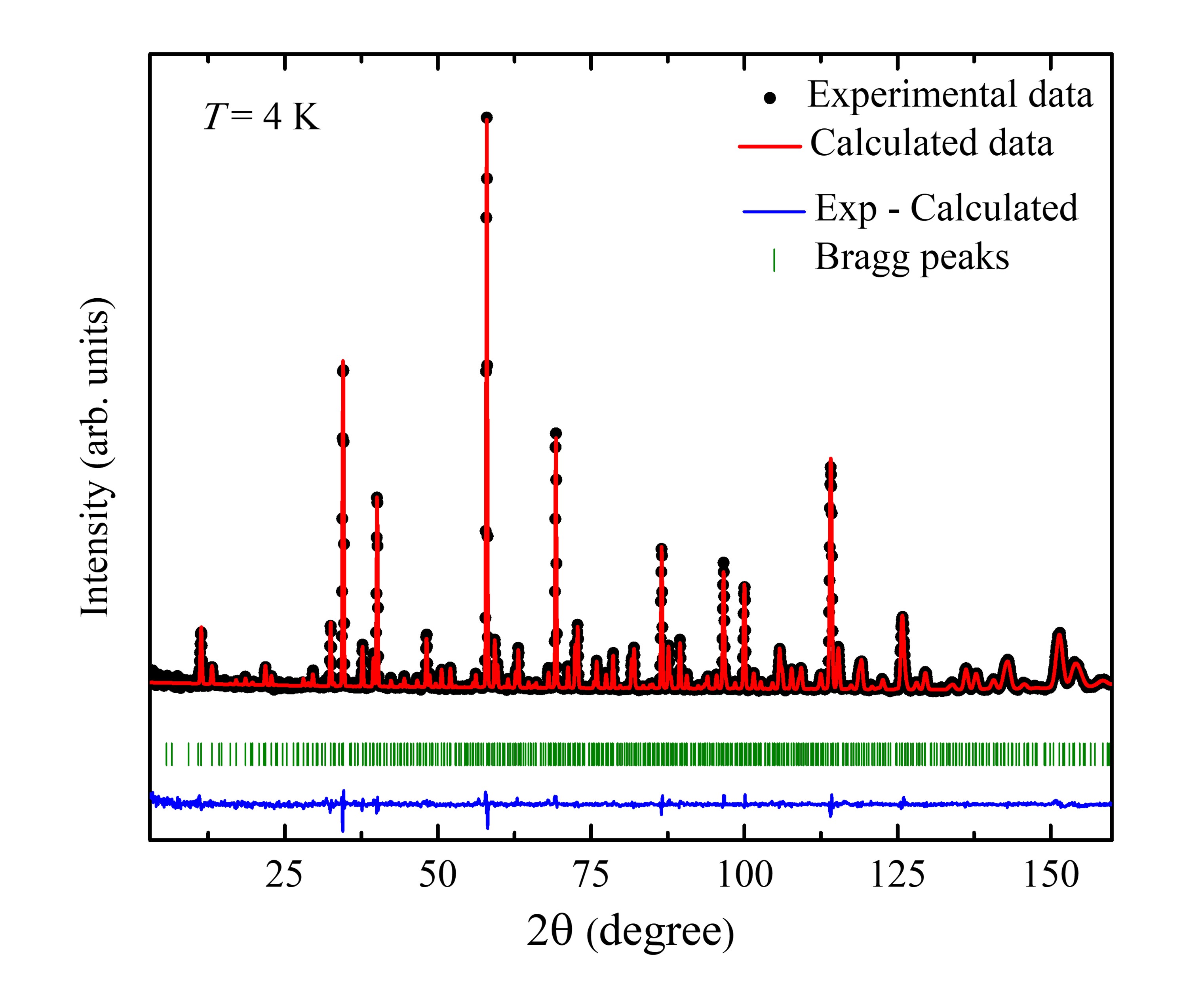}
  \caption{Neutron powder diffraction data of \BTmZO{} collected at 4 K using incident neutron of wavelength $\SI{1.54}{\angstrom}$. The black circles, red and blue lines represent the observed intensity, calculated intensity and difference between observed and calculated intensity, respectively. The short vertical green lines show the expected locations of Bragg peaks associated with the breathing pyrochlore structure.}\label{bt1}
\end{figure}

To examine the structural stability of the breathing pyrochlore structure, high-resolution powder neutron diffraction was carried out on \BTmZO{} powder sample at the selected temperatures $T =$ 4 K, 45 K, and 100 K. The measured diffraction data was structurally refined using the FullProf software, with no structural transition being observed down to 4 K. Apart from the expected lattice contraction with decreasing temperature, no significant changes in the structure were detected. In Fig.~\ref{bt1}, one representative powder neutron diffraction data collected at 4 K is shown along with the calculated intensity. It is found that all the observed Bragg peaks can be indexed based on the space group $F\overline{4}3m$ at 4 K. The  small value of chi-square (1.59) and Bragg R-factor (4.5) obtained from Rietveld refinement confirm a good agreement with the 
BP structure.
{\subsection{PDF analysis}

\label{pdf_analysis}

To investigate the local structure of \BTmZO{}, total neutron scattering measurements were conducted on a powder sample using the Nanoscale-Ordered Materials Diffractometer (NOMAD) \cite{NEUEFEIND201268} at Oak Ridge National Laboratory. The data were collected across selected temperatures ranging from 2 K to 300 K. After normalization, the NOMAD data were Fourier transformed to obtain the pair distribution function (PDF), \(G(r)\), a real-space representation of atomic correlations that quantifies the likelihood of finding a pair of atoms separated by a distance \(r\), averaged over time:

\begin{equation}
G(r) = A \times \left[ \int_{Q_{\text{min}}}^{Q_{\text{max}}} Q(S(Q) - 1) \sin(Qr) \, dQ \right] . \quad 
\end{equation}

Here, \(Q_{\text{max}}\) was set to 30 \AA\(^{-1}\) and \(A\) is an arbitrary scale parameter determined from the analysis. The momentum transfer \(Q\) is defined as \(Q = 4\pi\sin(\theta)/\lambda\), where \(\lambda\) is the neutron wavelength and \(\theta\) is the scattering angle. The PDF data \(G(r)\) for the selected temperature range, displayed in Fig.~\ref{pdf} from 1.8 \AA\ to 10 \AA, have been shifted vertically for clarity. 
Small-box refinements of the PDF data were performed using the PDFgui \cite{Farrow_2007} software. The average structure, modeled by the space group \(F\bar{4}3m\), fits the data well with no peak splitting or significant mismatches observable. The inset of Fig.~\ref{pdf} shows the anisotropic atomic displacement parameter \(U_{11}\) for various atoms as a function of temperature, derived from the PDF fits. \(U_{11}\) decreases with temperature, which is expected as thermal fluctuations decrease. All \(U_{11}\) values are on the order of 10\(^{-3}\), indicating no disorder or atomic tunneling between configurations.


\begin{figure}[]

\includegraphics[width=0.47\textwidth]{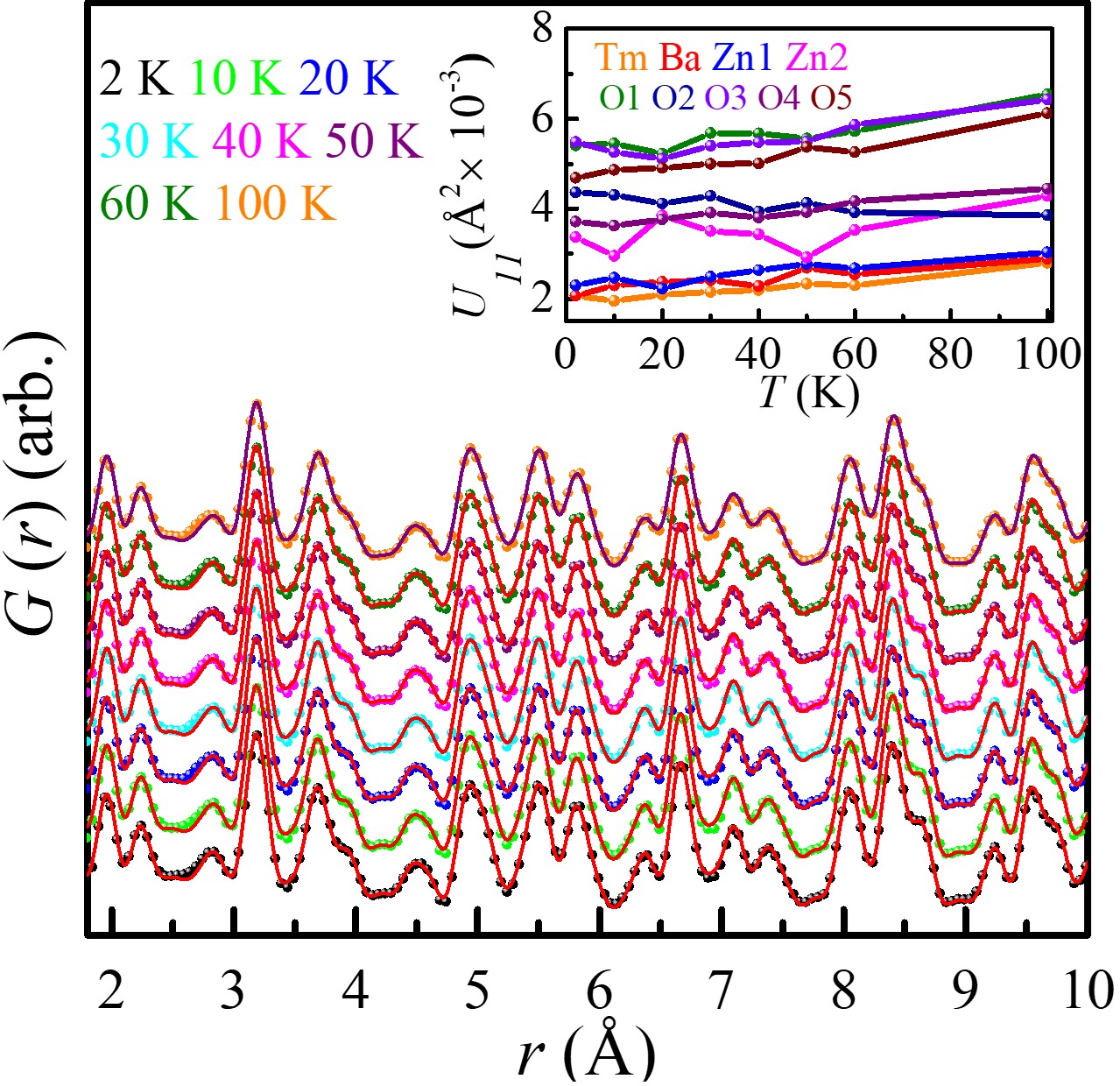}

\caption{The pair distribution function, $G(r)$, of the local structure of \BTmZO{} collected at selected temperatures between 2 K and 100 K (scatter markers), compared with the average model (solid red lines) refined using PDFGUI software with the assumed space group $F\bar{4}3m$ . The data has been shifted vertically for clarity. The inset shows the thermal displacement factor $U_{11}$ for different atoms. As can be seen, no anomalous temperature dependence or anomalously large value of any thermal factor is observed.}

\label{pdf}
     \hfill
\end{figure}

\subsection{Single crystal growth}

In order to grow single crystals of \BTmZO{}, the melting behavior of \BTmZO{} was studied by thermogravimetric analysis, as detailed in Appendix \ref{melting}. Upon heating, \BTmZO{} was found to melt incongruently, decomposing into products according to the reaction~\eqref{chmeq}. 
Among the decomposed products, \ch{BaZnO_2} is the only component in the liquid phase which can be used as a flux solvent for subsequent crystal growth. 
Consequently, a traveling-solvent-floating-zone (TSFZ) technique was employed to grow large and high-quality single crystals of the \BTmZO{} compound using \ch{BaZnO_2} as flux. The preparation details of the feed rod and flux are elaborated upon in Appendix~\ref{growth}.

About 250 mg-300 mg of \ch{BaZnO_2} solvent disk is used depending on the \BTmZO{} feed rods' dimensions. The growth process is performed in two steps: firstly, the feed rod is placed at the seed rod holder (lower shaft of the FZ furnace), and the solvent disk is melted partially and is stuck to the top end of the feed rod under a 0.1 \% of the maximum lamp power. Secondly, the feed rod with the attached\ch{BaZnO_2} solvent disk was suspended from the upper shaft of the furnace, and the growth process was started. 
The crystal growth is performed while bringing down the feed rod into the heating zone under  0.1-1.0 \% of the maximum lamp power and shutter length $\approx$ 70-65 mm. The melted \ch{BaZnO_2} flux is partially absorbed by the feed rod and thereafter dissolves the \BTmZO{}. This eventually percolates down and forms a liquid blob with solid particles at the end of the feed rod.
At this stage, the liquid blob contains mostly the \ch{BaZnO_2} liquid from the provided external flux and other phases \ch{Ba_5Tm_8Zn_4O_{21}} (S), \ch{Ba_2Tm_2Zn_8O_{13}} (S), \ch{ZnO} (S) and \ch{BaZnO_2} (L) from the decomposition of \BTmZO{} feed rod. Once the amount of the \ch{BaZnO_2} liquid appeared to be sufficient on the surface of the feed rod, the liquid zone was formed, and both feed and seed rods were moved downwards with a growth speed of 1-3 mm/h. The lamp power was gradually reduced to its lowest value of 0.1\% of its maximum power.
The feed and seed were made to rotate (8-20 rpm) in opposite directions to achieve a homogeneous mixture in the liquid zone. During the growth process, the \BTmZO{} decomposes at the liquid zone's upper region, while at the lower region, the decomposed phases recombine, crystallizing into the breathing pyrochlore phase.

To obtain centimeter-scale \BTmZO{} single crystals, the best growth conditions were identified as follows: the oxygen pressure was set to a maximum of 0.95 MPa to minimize bubbling. A `two-scanning' method was used, where the feed rod was initially passed through the heating zone at a high speed ($V_s$ $>$ 1 mm/hr) to pre-melt the crystal. This pre-melted grown crystal was then used as the feed rod for the slow growth process ($V_s$ = 0.4 mm/hr), which improved molten zone stability by preventing liquid penetration into the highly dense pre-melted crystal feed rod.
The optimized growth parameters are tabulated on Table \ref{growthtable} and representative grown single crystals are shown in Fig.~\ref{sc_laue}.

\begin{figure}[tp]
\includegraphics[width=0.45\textwidth]{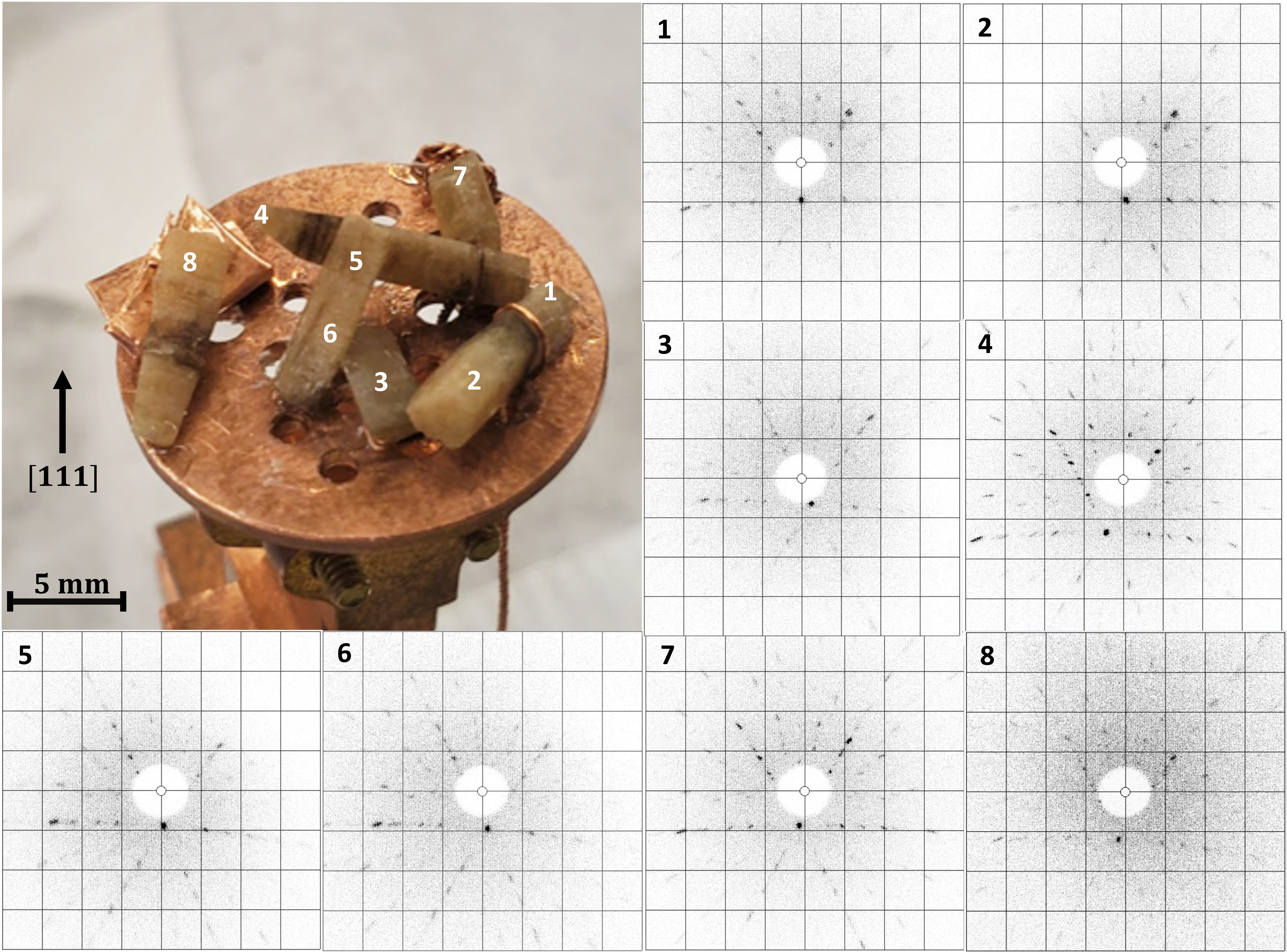}
 \caption{\BTmZO{} single crystals extracted from different growths mutually coaligned along the $[111]$ direction and mounted on the sample holder for the  neutron experiments. Representative Laue patterns from different regions of single crystals are shown. The central bright spot corresponds to the $(111)$ reflections. }
 \label{sc_laue}
\end{figure}

\subsection{Single crystal characterization}

Several high-quality single crystals with a few millimeter sizes were extracted from the different growths and oriented using a Laue diffractometer. 
The crystals were oriented along the crystallographic $[111]$ direction, and the magnetic and thermal characterizations were carried out on these oriented crystals. Figure~\ref{sc_laue} shows the co-aligned \BTmZO{} single crystals (around 1 g) mounted on a copper-based sample holder for neutron study with some representative images of collected Laue patterns  also shown.
Each crystal specimen was aligned along the $[111]$ direction and placed on the sample holder with Cu-wire and super-glue as shown in Fig.~\ref{sc_laue}. The similar Laue patterns obtained from the different mounted crystal specimens confirm the orientation of all \BTmZO{} crystals to be  along the crystallographic $[111]$ direction.  

Single crystal X-ray diffraction (XRD) was conducted on the \BTmZO{} crystal to confirm the crystal structures,  with refined atomic positions presented in Table~\ref{atomicposition}. The \ch{Tm^{3+}}--\ch{Tm^{3+}} and Tm--O bond distances, bond angles,  and the breathing ratio ($d/d'$) obtained from these positions  are separately tabulated in Table~\ref{bond_distances}.  The breathing ratio ($d/d'$) determined  for \BTmZO{} is 1.8822(9), larger than the 1.82(2) value found for \BYZO{}~ \cite{Dissanayake2021TowardsStudy}.

\begin{table}[tb]
\centering
\footnotesize
\caption{Single crystal x-ray diffraction refinement results at 230 K using the space group $F\overline{4}3m$. Fractional atomic coordinates and equivalent isotropic displacement parameters ($\AA^2$) for \ch{Ba_3Tm_2Zn_5O_11}. $U_{\rm iso}$ is defined as 1/3 of the trace of the orthogonalized $U_{ij}$ tensor. Values in parentheses indicate $\pm 1~ \sigma$.}\label{atomicposition}
\begin{tabular}{| m{2.5cm} | m{1.3cm}| m{1.3cm} | m{1.3cm}| m{1.3cm}|}
\hline \hline
\multicolumn{5}{l}{\begin{tabular}[c]{@{}l@{}}$T = 230\;$ K  \;\;\;\;\;\;\;\; $a=b=c =13.4971(9)$ \AA$ \;\;\;\;\;\;\;\;$  $\rm{wR_2}$ = 0.0629\end{tabular}} \\
\hline
Atom (Wyckoff site) & x & y  & z  & $U_{\rm iso}$\\
\hline
Tm1 (16e) & 0.6633(1) & 0.6633(1) & 0.6633(1) & 0.014(1) \\
\hline
Ba1 (24f) & 0.5000 & 0.5000 & 0.7950(1) & 0.018(1) \\
\hline
Zn1 (16e) & 0.4167(1) & 0.5833(1) & 0.5833(1) & 0.015(1) \\
\hline
Zn2 (24g) & 0.4167(2) & 0.7500 & 0.7500 & 0.016(1) \\
\hline
O1 (4a) & 0.5000 & 0.5000 & 1.0000 & 0.014(3) \\
\hline
O2 (4b) & 0.5000 & 0.5000 & 0.5000 & 0.019(4) \\
\hline
O3 (16e) & 0.3448(7) & 0.6552(7) & 0.8448(7) & 0.009(2) \\
\hline
O4 (16e) & 0.8314(9) & 0.6686(9) & 0.6686(9) & 0.022(2) \\
\hline
O5 (48h) & 0.4985(9) & 0.6668(5) & 0.6668(5) & 0.017(2) \\
\hline

\end{tabular}
\end{table}


\begin{table}[tb]

\caption{Selected bond distances (\si{\angstrom}) and bond angles (\si{\degree}) of \ch{Ba3Tm2Zn5O11} obtained from the refined single crystal X-ray diffraction data collected at 230 K. Values in parentheses indicate $\pm 1~ \sigma$.}\label{bond_distances}
\begin{tabular}{| >{\centering\arraybackslash}m{12em} | >{\centering\arraybackslash}m{2cm}|}
\hline
Tm-Tm (small tetrahedron) & 3.3113(14) \\ \hline
Tm-Tm (large tetrahedron)& 6.2325(15) \\ \hline
Breathing ratio d/d'& 1.8822(9) \\ \hline
Tm-O4 Bond length& 2.272(13) \\ \hline
Tm-O5 Bond length& 2.225(13) \\ \hline
O5-Tm-O5 Bond angles& 92.3(3) \\ \hline
O5-Tm-O4 Bond angles& 86.3(5) \\ \hline
\end{tabular}
\end{table}

\begin{figure*}
\centering
\includegraphics[width=\textwidth]{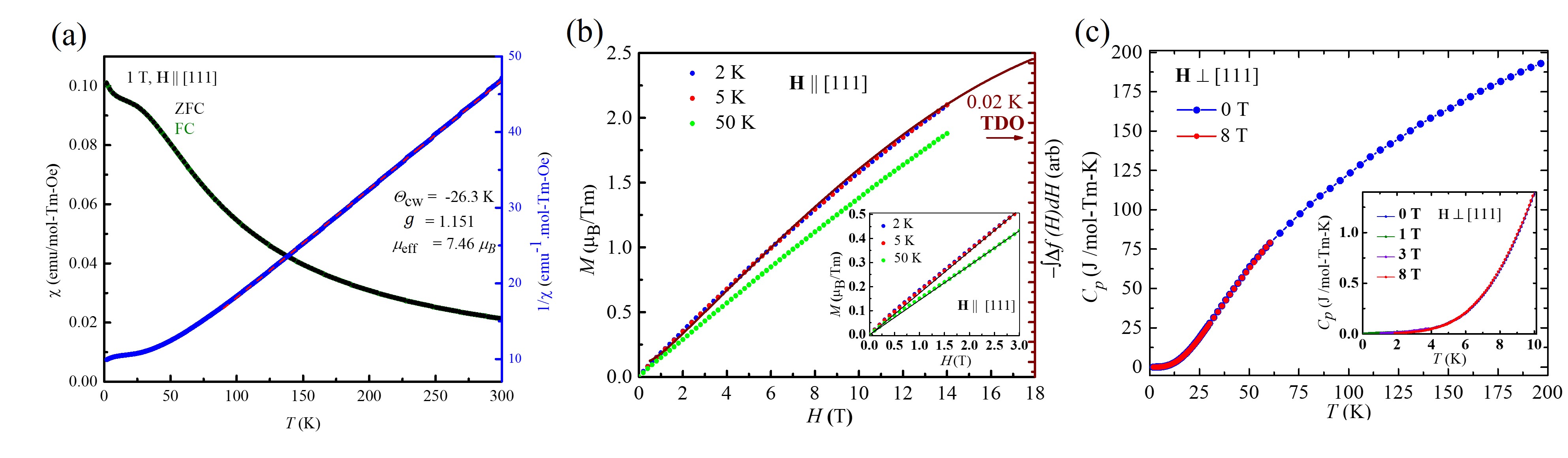}
  \caption{(a) Temperature-dependence of susceptibility $\chi(T)$ and inverse susceptibility 1/$\chi(T)$, measured at 1 T, revealing no significant difference between ZFC and FC down to 2 K. The dashed red line is the Curie-Weiss fit. Note that $1 \, \text{emu}/(\text{mol Oe}) = 4\pi \times 10^{-6} \, \text{m}^3/\text{mol}$ (b) Isothermal magnetization $M(H)$ curves at 2 K, 5 K, and 50 K. The inset illustrates the zoomed-in low-field magnetization, exhibiting no hysteresis during field ramping. The black linear guidelines are included to highlight the linear response. The right axis shows the TDO measurement taken at 0.02 K, where $\Delta f$ is integrated with respect to the field. This integrated curve is proportional to the magnetization and has been scaled to match the $M(H)$ data collected at 2 K. The similarity in trends between the TDO measurement and the $M(H)$ data at 2 K indicates no anomalous behavior or phase transitions down to 0.02 K. (c) Temperature-dependent total heat capacity of \BTmZO{} single crystal with applied magnetic field, ${\bf H}$, perpendicular to $[111]$ direction, down to 100 mK. Minimal alterations are observed in the heat capacity trace at 8 T, and the inset focuses on intermediate fields in the low-temperature regimes. The lack of any sharp anomaly in heat capacity indicates the absence of phase transitions. }
 \label{chi_hc}
     \hfill
\end{figure*}

\subsection{Magnetic and thermodynamic characterization}

Magnetic measurements were carried out on the oriented \BTmZO{} single crystal specimen. 
Magnetic susceptibility was measured in the presence of an external magnetic field of $H$ = 1 T along  ${\bf H}\parallel[111]$. The measurement was performed from 300 K to 2 K under ZFC (zero-field cooling) and FC (field cooling) conditions. 
Figure~\ref{chi_hc} (a) shows the temperature-dependent magnetic susceptibility $\chi$ with ${\bf H}\parallel[111]$ under ZFC protocol. No splitting was observed between the ZFC and FC data, and the results are consistent with measurements on the powder sample, indicating no directional anisotropy. Low magnetic field (e.g. 0.01 T) $\chi$ measurements were also carried out. Although these measurements showed fluctuations in $\chi$ on the order of less than 1\%, no FC-ZFC splitting was observed. The inverse magnetic susceptibility was fitted using the Curie-Weiss (CW) law: $1/\chi = (T-\theta_{\rm CW})/C$; where $C$ is the Curie constant and $\theta_{\rm CW}$ is the Curie-Weiss temperature. The effective moment $\mu_{\rm{eff}}$ can be calculated from the Curie constant by $\mu_{\rm eff}=\sqrt{3k_{\rm B}C/N_A}$, where $N_A$ is Avogadro's number. The Land\'e $g$-factor can be determined from $C=ng^2\mu_{\rm B}^2 J(J+1)/3k_{\rm B}$, where $n$ is the number of free spins per formula unit. The CW fit was performed in the high-temperature range (90 K-300 K) regime where $1/\chi$ follows a linear temperature dependence and yielded $C$ = 6.96 emu-K-mol$^{-1}$Oe$^{-1}$ and 
$\theta_{\rm CW}$ = $-26.3$ K. 

This (negative) $\theta_{\rm CW}$ parameter should not be strictly viewed as a sole measure of ``antiferromagnetic'' interactions in this compound since the $1/\chi=0$ temperature intercept reflects some combination of the crystal electric field (CEF) levels and Tm$^{3+}$-Tm$^{3+}$ exchange and magnetic dipolar interactions~\cite{Gingras-Tb2Ti2O7}.
The calculated effective moment $\mu_{\rm eff}$ of 7.46 $\mu_{\rm B}$ and $g=1.151$ agrees well with the theoretical value of $\mu_{\rm eff}=7.55$ $\mu_{\rm B}$ and $g=\frac{7}{6}$ for free Tm$^{3+}$ ($L$=5, $S$=1, $J$=6) ion at ambient temperature. 
As the temperature decreases, the higher energy CEF energy levels depopulate, reducing the effective moment and hence changing the slope of $1/\chi$. Near $T \approx$ 45 K, a shoulder-like feature emerges in $\chi(T)$ as the ion enters its CEF ground state. As we will see below from the inelastic neutron scattering analysis, the CEF ground state is a singlet and this plateauing of the susceptibility at low-temperature is the van Vleck component of $\chi$. 
However, a slight upturn in $\chi$ at temperatures below 20 K cannot be explained by the van Vleck component, and its origin is unknown. We have investigated whether this up-turn could be due to parasitic paramagnetic phases or dilute impurities, but we found no evidence that would confirm such scenario.
Finally, the lack of a sharp feature in the susceptibility data of \BTmZO{} indicates an absence of long-range ordering down to 2 K while the indistinguishable FC and ZFC susceptibility further suggests a lack of spin glass freezing, also above 2 K.

\begin{figure*}[t]
\centering
\includegraphics[width=\textwidth]{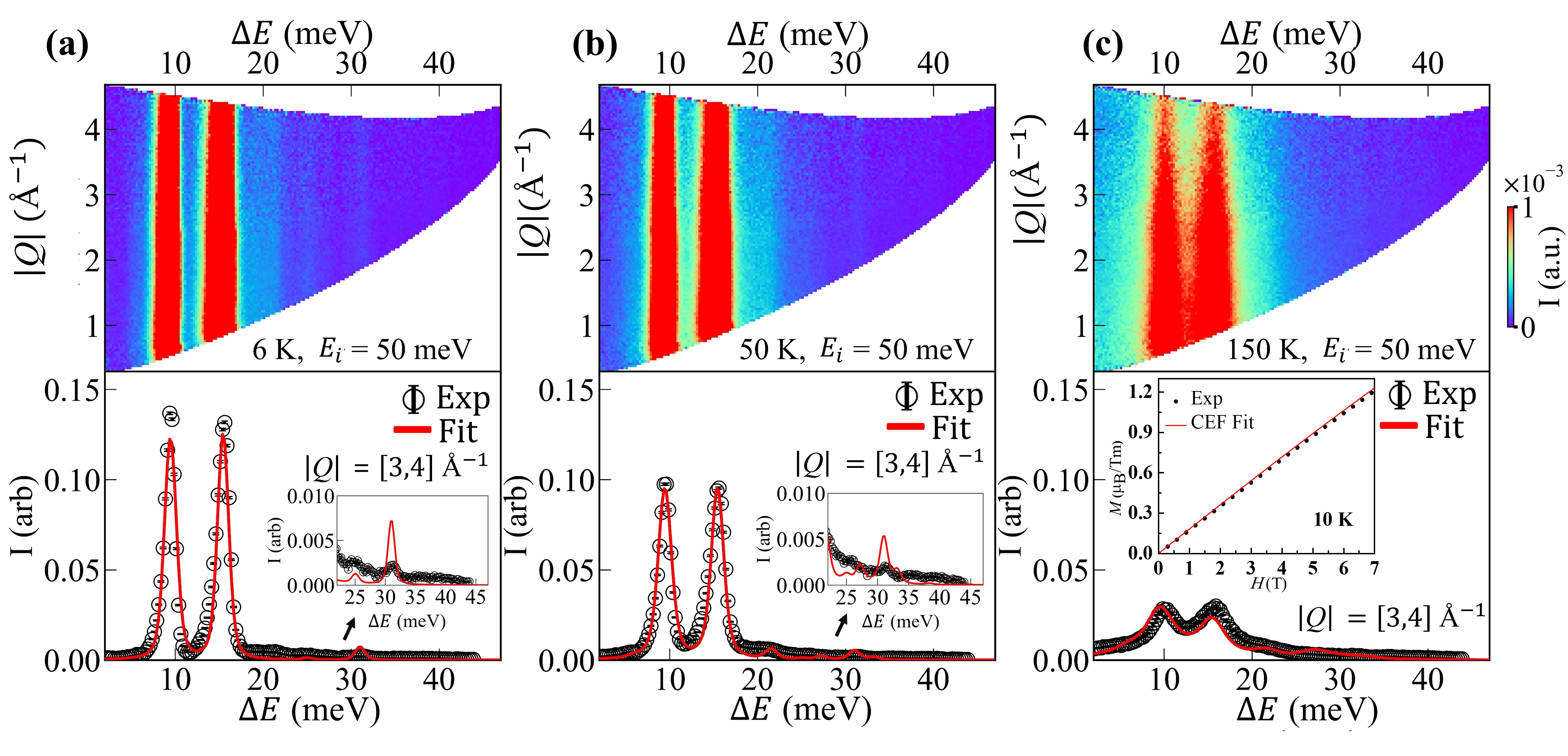}                           
\caption{Inelastic neutron scattering measurements and CEF model fits for polycrystalline \BTmZO{}. (a-c, top) $\Delta E$ vs. $|Q|$ slices at temperatures of 6 K, 50 K, and 150 K, respectively, with incident energy $E_i = 50$ meV. The spectrum from \ch{Ba3Lu2Zn5O11} has been subtracted to remove the non-magnetic background. Two prominent CEF bands located at 9.47 and 15.45 meV, alongside lower-intensity bands centered at 25 and 31.4 meV are visible. (a-c, bottom) Constant-$Q$ cuts with an integration window of 3-4 $\AA^{-1}$  with overlayed CEF fits derived from the most optimal solution discussed in the main text. The inset in the bottom panels of (a) and (b) shows the zoomed-in line cuts at higher energy, while the inset in the bottom panel of (c) shows the fit to the isothermal powder magnetization at 10 K.}
 \label{CEF}
\hfill 
\end{figure*}

The isothermal magnetization $M(H)$ measurements were performed at selected temperatures ($T =$ 2 K, 5 K, and 50 K) as shown in Fig.~\ref{chi_hc} (b). The magnetic field was ramped up and down to 14 Tesla with no hysteresis observed in the field sweeps. At both low and high fields, $M(H)$ shows a linear trend, consistent with the Van Vleck paramagnetism of a CEF ground state singlet. The linear trend becomes more stepper at 5 K and 2 K, indicating a paramagnetic behavior.

The right axis of Fig.~\ref{chi_hc} (b) shows the integrated frequency response of TDO measured at 0.02 K up to 18 Tesla. The integrated frequency response is proportional to magnetization and shows a linear behavior, similar to 2 K data, suggesting no anomalous behavior or phase transition down to 0.02 K. 

Figure~\ref{chi_hc} (c) shows the temperature-dependent total heat capacity of \BTmZO{} single crystal a magnetic field  ${\bf H}\perp[111]$. The heat capacity shows no sharp anomalies down to 100 mK, again indicating the absence of any phase transition.  Upon applying a field up to 8 T, minimal changes are observed in the heat capacity as function of temperature (see inset of Fig.~\ref{chi_hc} (c) which focuses on the low temperature limit). Additionally, no differences were observed between the powder and single crystal measurements. Interestingly, the low-temperature heat capacity does not follow the expected Debye like phonon contribution $\propto T^3$ as fits to such a form proved unsuccessful with no temperature windows yielding a good fit (See Appendix~\ref{App_hc} for details). Intriguingly, the low-temperature heat capacity of \BLuZO{} can be modeled well by a Debye like form, suggesting the presence of other additional non-phonon contributions to the heat capacity of \BTmZO{}.

\begin{figure*}[]
\centering
\includegraphics[width=\textwidth]{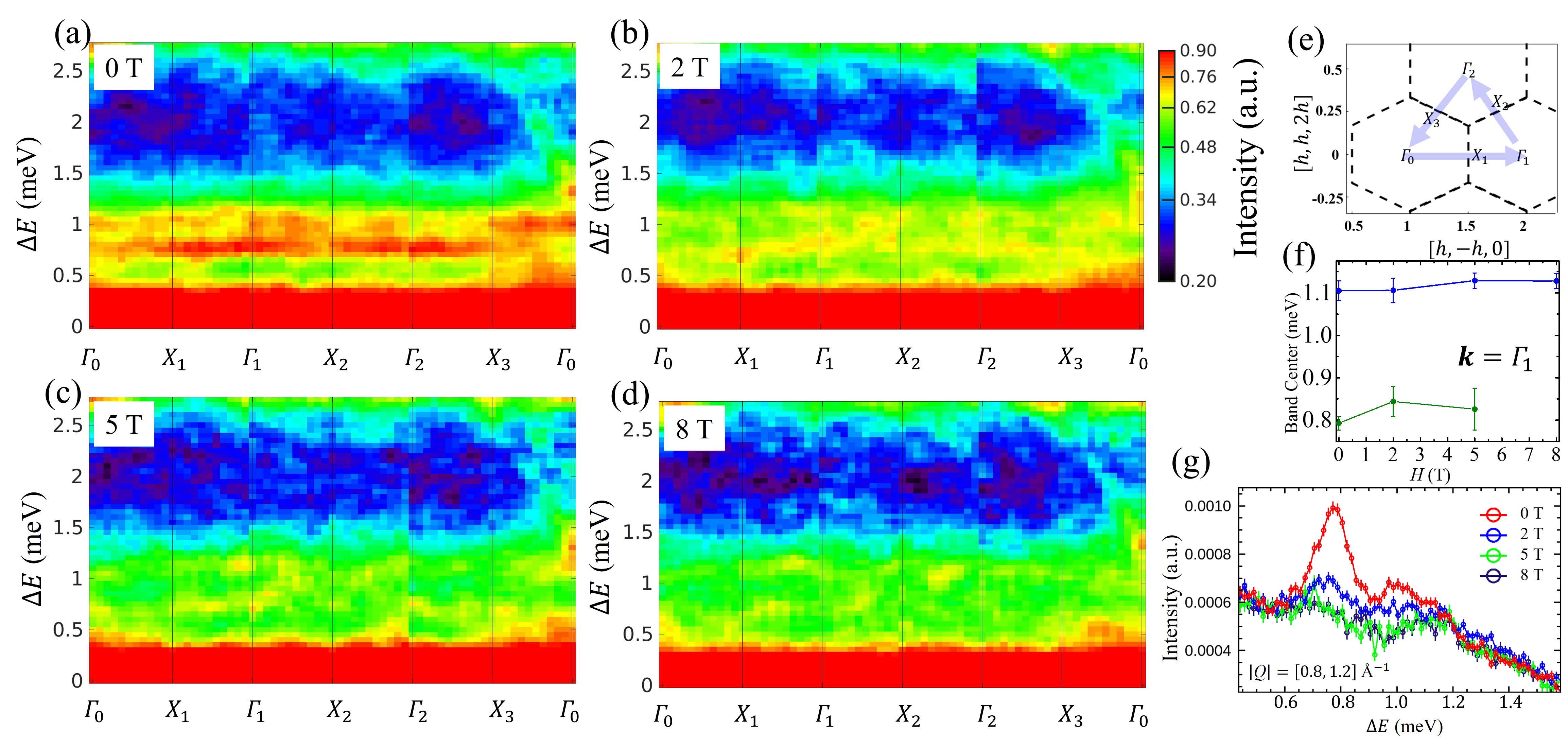}                           
\caption{Scattering intensity as a function of energy transfer for \BTmZO{}. Panels (a-d) display data from INS experiments conducted on high-quality, co-aligned single crystals with an applied field ${\bf H} \parallel [111]$, measured at 100 mK, and traversing through the high-symmetry points $\mathit{\Gamma}_0-X_1-\mathit{\Gamma}_1-X_2-\mathit{\Gamma}_2-X_3-\mathit{\Gamma}_0$
, as illustrated in cartoon (e). Notably, two dispersionless excitations  identified at \(0.8 \, \text{meV}\) and \(1.1 \, \text{meV}\) in a \(0 \, \text{T}\) field exhibit decreased intensities upon application of the field. (f) The band centers for the upper and lower bands at ${\bm k} = \Gamma_1$ show no significant energy shifts with the applied field.
(g) However, constant-$Q$ cuts, after applying powder averaging, reveal significant field-induced changes in the intensities of these excitation bands, suggesting a magnetic origin. Error bars indicate $\pm 1~ \sigma$.
}
 \label{cncs}
\hfill 
\end{figure*}
 
\subsection{Inelastic Neutron Scattering}

Inelastic neutron scattering (INS) was used to probe the crystal electric field (CEF) levels of \BTmZO{}. The CEF levels originate from the splitting of Hund's ground state due to the crystal environment around the magnetic ion. The \ch{Tm^{3+}} ion in \ch{Ba3Tm2Zn5O11} reside at a site with $C_{3v}$ point group symmetry whose local crystal environment splits the \ch{Tm^{3+}} ion's ${}^3H_6$ 13 degenerate states into 5 singlets and 4 doublets. The effect of this crystalline electric field can be modelled by the single-ion CEF Hamiltonian, $H_{\rm CEF}$, acting within the ${}^3H_6$ manifold 
\begin{align}
\label{eq:cef-ham}
H_{\rm CEF} &= B_{20}O_{20}+B_{40}O_{40}+B_{43}O_{43} \nonumber\\
& + B_{60}O_{60}+B_{63}O_{63}+B_{66}O_{66},
\end{align}
where the $O_{lm}$ are the Stevens operators~\cite{hutchings1964point,stevens1952matrix,scheie2021pycrystalfield} and the $B_{lm}$ are parameters characterizing the  crystal field potential within the set of $J=6$ states.

In Fig.~\ref{CEF}, the top three panels show a representative INS spectrum sets collected using an incident neutron energy of $E_i=50$ meV at temperatures of 6 K, 50 K, and 150 K. As shown in the bottom panels, we observed two distinct transitions at 9.4 meV and 15.4 meV, as well as low-intensity transitions centered at 25 meV and 31.4 meV, whose intensities are less than 1\% of the most intense transition.
While \ch{Ba3Lu2Zn5O11} was used to subtract the non-magnetic background, discerning very weak CEF levels atop the large contributions of nearby phonon excitations proved challenging due to limited momentum coverage.
Therefore, experimentally, we can only reliably observe four of the eight possible CEF transitions out of the CEF ground state. To fit the six CEF parameters in Eq.~(\ref{eq:cef-ham}) using this data, we employed the PyCrystalField software\cite{scheie2021pycrystalfield}. As an additional constraint, we included isothermal magnetization data in the fitting procedure, as elaborated in further detail in Appendix~\ref{App_cef}. The best fit parameters were ultimately determined to be (in meV)  \( B_{20} = -7.83 \times 10^{-2} \), \( B_{40} = -1.01 \times 10^{-3} \), \( B_{43} = 1.84 \times 10^{-2} \), \( B_{60} = -1.59 \times 10^{-5} \), \( B_{63} = 5.24 \times 10^{-4} \), and \( B_{66} = -5.42 \times 10^{-4} \). Table \ref{CEF_Eigenvectors} lists the energy levels and the corresponding eigenvectors determined from the fit. The bottom three panels in Fig.~\ref{CEF} show the calculated constant-$Q$ cuts overlaid on the experimental data. The fit reproduces reasonably well the positions and relative intensities of the four observed transitions. 
It also predicts additional transitions at 42.6 meV, 53.8 meV, and 57.3 meV with intensity less than 1\% of the 9.4 meV transition, consistent with our inability to discern the higher energy transitions against the experimental background. The ground eigenstate of $H_{\rm CEF}$ is identified as a singlet, with no other CEF level (of either low or high intensity) until the first excited state at 9.4 meV. The proposed CEF scheme also fits the isothermal magnetization data very well as shown in the inset of Fig.~\ref{CEF}(c). Furthermore, the fitted CEF parameters reproduce the weak magnetic field dependence of the 9.4 meV and 15.4 meV transitions as shown in  Fig.~\ref{CEF_field}. 
Given the singlet ground state and the substantial $\approx$ 10 meV gap to the first excited state, we should expect to observe no significant features in the INS spectrum below 10 meV. Moreover, estimating the typical scale of the Tm$^{3+}$$-$Tm$^{3+}$ interactions, $H_{\rm Tm-Tm}$, to be in the range of 0.01 meV $-$1 meV, which is typical for insulating rare-earth oxides, we would not expect the development of long-range order since the first excited crystal field level (at 9.4 meV) is more than a factor 10 compared to $H_{\rm Tm-Tm}$, and thus causing only a small admixing between the ground singlet and the lowest excited crystal field states.
This is similar to what is seen in the Tm$_2$Ti$_2$O$_7$ pyrochlore compound~\cite{Zinkin-Tm2Ti2O7}.

\begin{table*}
{\caption{Energy levels and their associated eigenvectors for the single-ion CEF Hamiltonian, obtained by fitting the \BTmZO{} INS data. The wavefunctions are represented in the \(\ket{m_J}\) basis. The energies are labeled as singlets (S) or doublets (D) in brackets next to the corresponding energy level.} 

\begin{ruledtabular}
\begin{tabular}{c|ccccccccccccc}
E (meV) &\(|{-6\rangle}\) & \(|{-5\rangle}\) & \(|{-4\rangle}\) & \(|{-3\rangle}\) & \(|{-2\rangle}\) & \(|{-1\rangle}\) & \(|{0\rangle}\) & \(|{1\rangle}\) & \(|{2\rangle}\) & \(|{3\rangle}\) & \(|{4\rangle}\) & \(|{5\rangle}\) & \(|{6\rangle}\) \tabularnewline
 \hline 
0.000 (S) & -0.4726 & 0.0 & 0.0 & -0.526 & 0.0 & 0.0 & 0.0 & 0.0 & 0.0 & -0.526 & 0.0 & 0.0 & 0.4726 \tabularnewline
9.467 (S) & 0.6472 & 0.0 & 0.0 & 0.2399 & 0.0 & 0.0 & 0.2169 & 0.0 & 0.0 & -0.2399 & 0.0 & 0.0 & 0.6472 \tabularnewline
15.450 (D) & 0.0 & 0.0 & 0.5661 & 0.0 & 0.0 & -0.1319 & 0.0 & 0.0 & 0.7455 & 0.0 & 0.0 & -0.3262 & 0.0 \tabularnewline
15.450 (D) & 0.0 & -0.3262 & 0.0 & 0.0 & -0.7455 & 0.0 & 0.0 & -0.1319 & 0.0 & 0.0 & -0.5661 & 0.0 & 0.0 \tabularnewline
25.052 (D) & 0.0 & 0.5168 & 0.0 & 0.0 & -0.2014 & 0.0 & 0.0 & 0.8026 & 0.0 & 0.0 & -0.2196 & 0.0 & 0.0 \tabularnewline
25.052 (D) & 0.0 & 0.0 & 0.2196 & 0.0 & 0.0 & 0.8026 & 0.0 & 0.0 & 0.2014 & 0.0 & 0.0 & 0.5168 & 0.0 \tabularnewline
31.012 (S) & 0.0336 & 0.0 & 0.0 & 0.314 & 0.0 & 0.0 & -0.8947 & 0.0 & 0.0 & -0.314 & 0.0 & 0.0 & 0.0336 \tabularnewline
31.037 (S) & -0.526 & 0.0 & 0.0 & 0.4726 & 0.0 & 0.0 & 0.0 & 0.0 & 0.0 & 0.4726 & 0.0 & 0.0 & 0.526 \tabularnewline
42.643 (D)& 0.0 & 0.5213 & 0.0 & 0.0 & 0.3437 & 0.0 & 0.0 & -0.4281 & 0.0 & 0.0 & -0.6532 & 0.0 & 0.0 \tabularnewline
42.643 (D) & 0.0 & 0.0 & 0.6532 & 0.0 & 0.0 & -0.4281 & 0.0 & 0.0 & -0.3437 & 0.0 & 0.0 & 0.5213 & 0.0 \tabularnewline
53.874 (D)& 0.0 & 0.0 & -0.4523 & 0.0 & 0.0 & -0.3939 & 0.0 & 0.0 & 0.5344 & 0.0 & 0.0 & 0.5956 & 0.0 \tabularnewline
53.874 (D) & 0.0 & -0.5956 & 0.0 & 0.0 & 0.5344 & 0.0 & 0.0 & 0.3939 & 0.0 & 0.0 & -0.4523 & 0.0 & 0.0 \tabularnewline
57.365 (S)& 0.2828 & 0.0 & 0.0 & -0.5864 & 0.0 & 0.0 & -0.3903 & 0.0 & 0.0 & 0.5864 & 0.0 & 0.0 & 0.2828 \tabularnewline
\end{tabular}\end{ruledtabular}
\label{CEF_Eigenvectors}}
\end{table*}

With the CEF scheme established, we conducted INS experiments on high-quality coaligned single crystals of \BTmZO{} to probe the low-energy excitations in the compound with applied magentic field ${\bf H}\parallel[111]$. We show in Fig.~\ref{cncs} the scattering intensity as a function of energy transfer, with a path taken through the high symmetry points of $\mathit{\Gamma}_0-X_1-\mathit{\Gamma}_1-X_2-\mathit{\Gamma}_2-X_3-\mathit{\Gamma}_0$
 for selected fields dispersionless bands are observed at $\approx$ 0.8 meV and $\approx$ 1 meV at 0 T. Applying a magnetic field reveals a systematic shift in the band intensities and centers. The band intensity gradually decreases from 0 T to 8 T but remains visible even at 8 T. Within the resolution of the present data, no significant field-dependent shifts in the band energies were observed, however, the intensity of this excitation band clearly changes as a function of the applied field, as shown in the line-cuts of Figs.~\ref{cncs} (g). The origin of this signal is rather mysterious and we have not been able to find a compelling explanation for it.

\section{Discussion} 

From the experimental results presented in Sec.~\ref{sec:results},  the following picture of \BTmZO{} emerges: the Tm$^{3+}$ ion in \BTmZO{} has a singlet ground state with a gap of $9.4\ {\rm meV}$ to the lowest crystal field level. This gap is significantly larger than the expected intra-tetrahedron exchange interactions  (e.g. in \BYZO{} they are roughly $J \approx 1\ {\rm meV}$ or so~\cite{rau2016anisotropic}) and thus also the inter-tetrahedron exchange interactions. The singlet ground state should thus lead to a ``trivial'' quantum paramagnetic ground state in \BTmZO{}, a product of CEF singlets on each Tm site, with the first excitation -- a kind of CEF ``exciton'' appearing at $\approx 10\ {\rm meV}$. This picture, similar to what is found in the Tm$_2$Ti$_2$O$_7$ pyrochlore~\cite{Zinkin-Tm2Ti2O7}, appears consistent with the thermodynamic data, susceptibility and magnetization data, which do not show any evidence of low-lying excitations below the CEF gap (e.g. via a significant entropy release).

This description of \BTmZO{}, however, presents an interesting puzzle: what is the excitation observed in inelastic neutron scattering data at $\approx 1\ {\rm meV}$? From the above discussion, there should be nothing magnetic from the Tm$^{3+}$ ions near $1\ {\rm meV}$, as it lies well below the CEF gap. The CEF analysis rules out a low-lying CEF mode, as detailed in Appendix \ref{App_cef}. This additional mode likely points to other interactions or physical phenomena that manifest not only in the INS data but also in the heat capacity. We briefly mention here a number of possible natural origins for this excitation:

\begin{enumerate}
    \item \emph{Low-lying phonon excitation:} While optical phonons typically lie higher in energy than $1\ {\rm meV}$, one might ask whether proximity to a structural instability could result in an anomalously low energy phonon excitation. 
     Indeed, we have experimentally found that the BP family becomes unstable for lanthanides lighter than Tm, suggesting that \BTmZO{}  may sit close to an instability.
    \item \emph{Magnetic Disorder:} One possibility is that the low energy excitations at $E\approx 1$ meV are due to a small fraction of Tm$^{3+}$ sites that have a different local environment than the bulk and thus harbor a low-lying crystal field excitation in the $\approx 1$ ${\rm meV}$ range. Alternatively, oxygen vacancies, if present, can create a charge imbalance around the Tm$^{3+}$ ion, potentially reducing its oxidation state to Tm$^{2+}$ and thereby creating low-lying crystal field excitations. 
    \item \emph{Atomic Tunnelling:} If the crystal structure of \BTmZO{} has two nearby local minima in the location of one of the atoms, quantum tunnelling could result in a superposition of states and an associated tunnel splitting~\cite{RevModPhys.42.201,tunnel1,tunnel2,tunnel3}. This excitation would be beyond the phonon picture of lattice dynamics and could potentially be observable in inelastic neutron scattering~\cite{sun2020high}.
    \item \emph{Spin-Lattice Coupling:} Significant coupling of the Tm$^{3+}$ CEF excitations to a lattice vibration could result in a change of character of those excitations and lead to new (and potentially low-lying) excitations.
\end{enumerate}

We now briefly discuss each of these possibilities in order. \\

\emph{Low-lying Phonon Excitation:} In order to rule out phonons as the identity of the low-energy modes in the range of 0.8 meV to 1 meV, we computed the expected phonon dispersion from first principles calculations. Full details of these calculations are presented in Appendix~\ref{App_phononcalc}. 
In Fig.~\ref{phonons}, we show the computed phonon dispersions. We find that the $F\bar{4}3m$ structure is energetically stable and that there are no anomalous low-energy phonons that could explain the nearly dispersionless modes appearing in the inelastic neutron scattering data. 
We estimate that the lowest optical phonon at the $\Gamma$-points occurs at 2.5 meV, but that it is sufficiently dispersive to reach 5.7 meV at the $X$-points. Similarly, the lowest acoustic branch reaches the same energy of 5.7 meV at the $X$-points. Considering these calculations, we conclude that the dispersionless excitations detected in the 0.8 meV to 1 meV range are unlikely to be of phononic origin.

 \emph{Magnetic Disorder:}
To rule out any magnetic disorder, such as potential site mixing where Tm$^{3+}$ might occupy a different atomic site or oxygen vacancies that could change the oxidation state to Tm$^{2+}$ and thus experience a different crystal field around it, which could result in a 1 meV band, we analyzed the powder neutron diffraction data shown in Fig.~\ref{bt1}. We specifically investigated scenarios where the Tm$^{3+}$ ion could be situated at other sites. We found that substituting the Tm$^{3+}$ ion at another site results in an occupancy close to 0\% with no improvement in fitting quality. Additionally, we did not find any evidence of oxygen vacancy as the occupancy stayed close to 1. It, therefore, seems unlikely that the 1 meV band originates from a fraction of Tm$^{3+}$ ions occupying a different local CEF environment or from the presence of Tm$^{2+}$ ions. Another possibility could be the presence of a magnetic impurity phase in the single crystal sample that produces these excitations. However, in a powder inelastic scattering experiment conducted at the SEQUOIA spectrometer (see Appendix \ref{seq_lowE}), we observed the same $\sim 1$ meV excitation band. The powder sample contains less than 1\% magnetic impurity phase which would make it impossible to detect a signal from such a small quantity. Thus, this magnetic disorder scenario also appears unlikely.

 \emph{Atomic Tunnelling:}
In scenarios typically associated with atomic tunneling, atoms exhibit transitions between two nearby position configurations~\cite{RevModPhys.42.201,tunnel1,tunnel2,tunnel3}, which can manifest as anomalously large atomic displacement parameters~\cite{sun2020high}.
 In our study of \BTmZO{}, such anomalies were not observed. Single-crystal X-ray diffraction data collected at 230 K indicates that the isotropic displacement parameters ($U_{\text{iso}}$) for all atoms are of the order of $10^{-2}$ \AA$^2$, as documented in Table~\ref{atomicposition}.
 This smallness in the displacement parameters suggests no significant deviations from the expected atomic positions.
 
To further investigate the possibility of atomic tunneling, we utilized neutron pair distribution function (PDF) analysis on powder samples, as detailed in section \ref{pdf_analysis}. Neutron scattering is particularly sensitive to oxygen, thus providing complementary insights to the X-ray analysis. The PDF analysis, conducted across temperatures from 100 K to 2 K, affirmed that the average structure, corresponding to the space group $F\bar{4}3m$, accurately reflects the experimental data (see Figure \ref{pdf}). This agreement between the local and average structures indicates that atomic tunneling is unlikely. Details concerning the behavior of atomic displacement parameters across the temperature range are further discussed in section \ref{pdf_analysis}. The parameters observed align with typical thermal behaviors, showing no evidence for the disorder or anomalies typically indicative of tunneling phenomena.

 \emph{Spin-Lattice Coupling:}
Although phonons and crystal electric field (CEF) excitations are often considered decoupled \cite{Adroja_2012}, strong magnetoelastic coupling between the lattice and the orbital degrees of freedom that can in principle bridge these excitations~\cite{Loewenhaupt_1979, Thalmeier_1982, Thalmeier_1984, Loong_1999}. Two primary conditions facilitate this coupling \cite{Thalmeier_1982, Thalmeier_1984}: 1) the presence of a phonon mode in the proximity of a CEF excitation, and 2) coupling between the phonon and CEF modes being allowed by symmetry. If these conditions are satisfied, magnetoelastic interactions can result in the mixing of phonon and CEF modes, leading to the formation of a new type of excitation known as a vibronic mode. 
In the case of \BTmZO{}, the lowest CEF excited state is located at approximately 10 meV. Given that the observed low energy mode is around 1 meV, a very large magnetoelastic coupling would be required to reduce the gap by 9 meV and bring either excitation within the observed 1 meV range. Therefore, the possibility of such extensive coupling in \BTmZO{} appears unlikely.

\begin{figure}[!htb]
    \centering
    \includegraphics[width=1.0\linewidth]{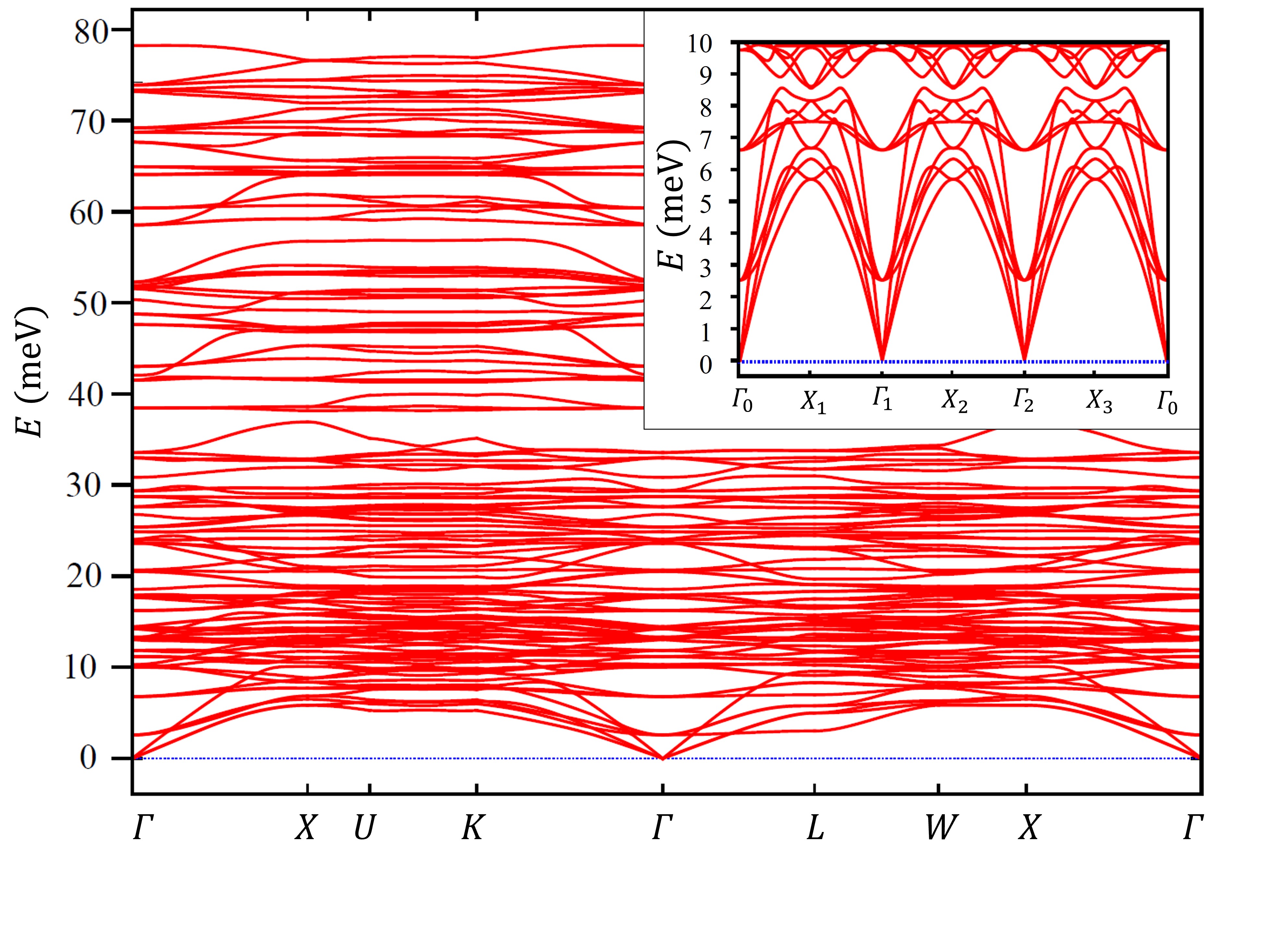}
    \caption{Computed phonon dispersions for \BTmZO, showing absence of low-energy dispersionless modes (kpath as suggested in SeeK-path \cite{hinuma2017band,togo2018texttt}). Inset: Phonon dispersion along the $k$-path plotted in Fig.~\ref{cncs}.}
    \label{phonons}
\end{figure}

\section{Conclusion}
In conclusion, we have successfully synthesized \BTmZO{}, a novel magnetic material within the rare-earth based breathing pyrochlore family, and grown high-quality single crystals of this compound using the traveling solvent floating zone technique. Investigation into its magnetic and heat capacity properties revealed the absence of long-range magnetic ordering down to 0.05 K. Neutron scattering techniques were employed to characterize the crystal electric field levels and unveiling two additional magnetic bands at 0.8 and 1 meV despite the presence of singlet CEF ground state and 9.4 meV gap to the first excited CEF state. The absence of low-energy dispersionless modes in computed phonon dispersions excludes phonons as the identity of the observed excitations. These findings provide valuable insights into the structural, magnetic, and vibrational properties of \BTmZO{}, contributing to understanding its puzzling behavior at the atomic and electronic levels and paving the way for further studies.

\begin{acknowledgments}
Work performed at Duke University is supported by the U.S. Department of Energy, Office of Science, Office of Basic Energy Sciences, under Award Number DE-SC0023405. R.B. acknowledges the support provided by Fritz London Endowed Post-doctoral Research Fellowship. The research at the University of Windsor (J.G.R) and the University of Waterloo (M.J.P.G.) was funded by the NSERC of Canada~(M.J.P.G, J.G.R) and the Canada Research Chair Program~(M.J.P.G, Tier I). A portion of this research used resources at the Spallation Neutron Source, a DOE Office of Science User Facility operated by the Oak Ridge National Laboratory. A portion of this work was performed at the National High Magnetic Field Laboratory, which is supported by National Science Foundation Cooperative Agreement No. DMR-2128556 and the State of Florida. We acknowledge the support of the National Institute of Standards and Technology, U.S. Department of Commerce, in providing the neutron research facilities used in this work. The identification of any commercial product or trade name does not imply endorsement or recommendation by the National Institute of Standards and Technology.
\end{acknowledgments}
\appendix

\section{Melting behaviour}
\label{melting}
To understand the melting behavior of the  \ch{Ba3Tm2Zn5O11} compound, simultaneous differential scanning calorimetry and thermogravimetric analysis were performed on a  \BTmZO{} powder sample. The sample was contained in an alumina crucible, and another alumina crucible was used as the reference while \ch{N_2} gas was used as the method gas. The samples were heated up to 1380 \degree C with a heating rate 10 \degree C/min. In Fig.~\ref{tga}, the percentage heat flow change and the change of \BTmZO{} sample weight are shown in a comparison with the \BYZO{} sample. Note that the synthesis of the \BYZO{} sample has been previously reported in \cite{Dissanayake2021TowardsStudy}. It is found that \BYZO{} shows a sharp endothermic dip in heat flow at $T \approx 1220^\circ$C with no significant weight loss, which is a sign of melting as the phase transition from solid to liquid happens over a narrow temperature range. In contrast, \BTmZO{} shows a broader endothermic dip in heat flow at $T \approx 1200^\circ$C with significant weight loss, indicating decomposition over a wider temperature range. Thus the behavior of heat flow and weight loss suggests that the melting behavior of \BTmZO{} is different (i.e. incongruent in nature) compared to \BYZO{}, which melts congruently.

\begin{figure}[!htb]
    \centering
    \includegraphics[width=0.42\textwidth]{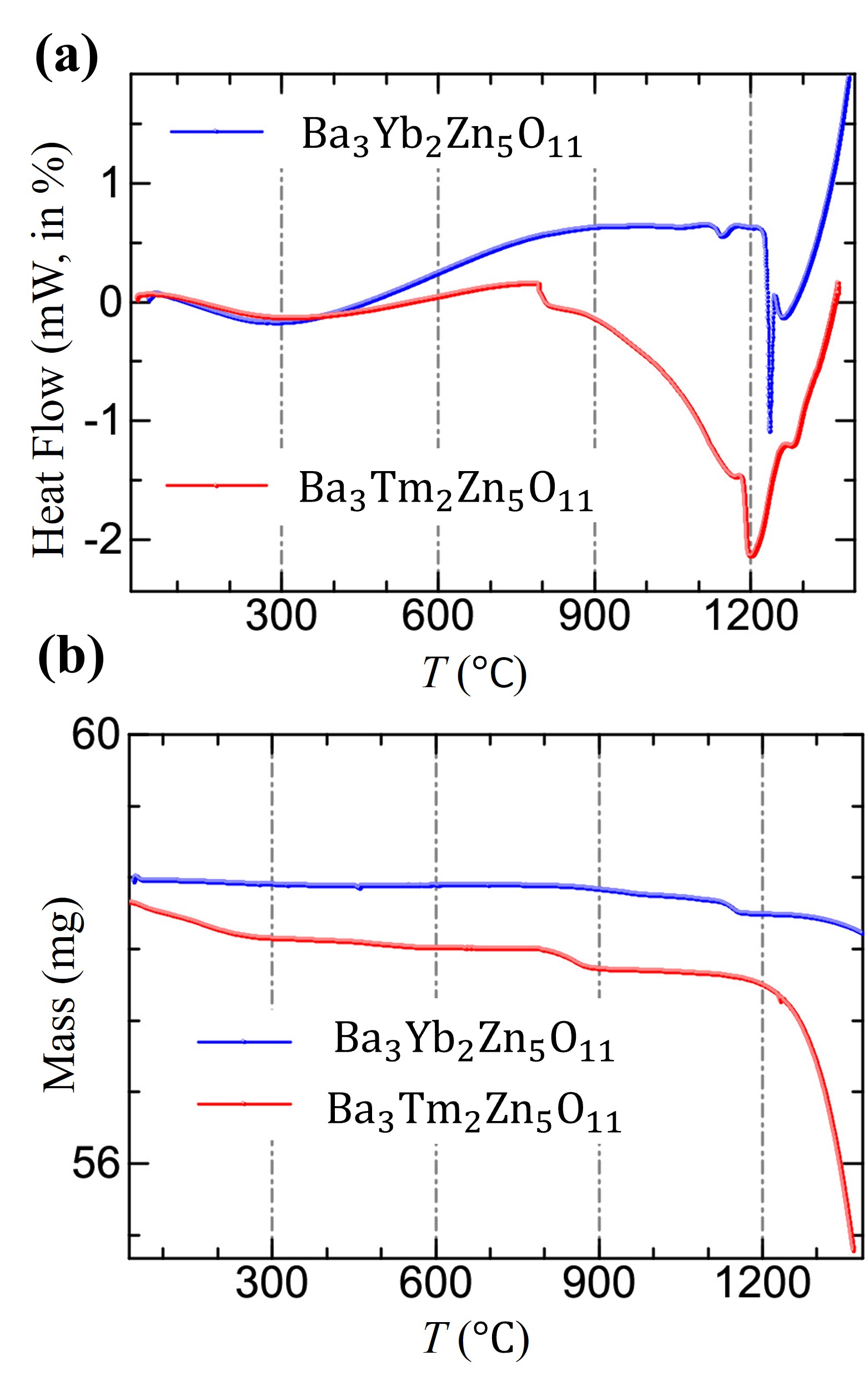}
       \caption{The percentage of heat flow and weight loss of \BYZO{} (blue) and \BTmZO{} (red) with respect to temperature while heating the sample up to 1380 \degree C in TGA. The sharp endothermic dip around 1210 \degree C is the region where melting event occurs. Melting in \BYZO{} is not accompanied by significant weight loss whereas in \BTmZO{}, weight reduces drastically suggesting \BTmZO{} melts in-congruently.}
    \label{tga}
\end{figure}

To further validate our conclusion, PXRD data were collected on both \BYZO{} and \BTmZO{} melted samples left after the TGA/DSC experiment. 
The PXRD analysis confirms that the \BYZO{} is chemically stable after melting and this agrees with our previously crystal growth report \cite{Dissanayake2021TowardsStudy}.
In contrast, the \BTmZO{} decomposes into three different solid (S) and one liquid (L) phases: 

 \begin{equation}
  \begin{split}
13~ \BTmZO{} ({\text S})  \longrightarrow 3~  \ch{Ba_5Tm_8Zn_4O_{21}} ({\text S}) +  \\
\ch{Ba_2Tm_2Zn_8O_{13}} ({\text S}) + 22~\ch{BaZnO_2}
({\text L}) + 23~\ch{ZnO} ({\text S}) 
 \end{split}
\label{chmeq}
\end{equation}

This incongruent melting behavior of \BTmZO{} poses a significant hurdle in growing single crystals as it cannot be achieved by the simple floating-zone (FZ) technique similar that used to grow single crystals of  the \BYZO{} compound~\cite{Dissanayake2021TowardsStudy}. A traveling-solvent-floating-zone (TSFZ) technique was used to grow the large and high-quality single crystal of \BTmZO{} compound, with the detailed growth process discussed in Section~\ref{sec:exp-methods} the main text.  

\section{Growth preparation}
\label{growth}
As described in Eq.~\eqref{chmeq}, \ch{BaZnO_2} is one of the products of the decomposition, and it is the only liquid in the decomposition above 1200 \degree C, which is used as a solvent/flux in TSFZ growth process of \BTmZO{}. 
The \ch{BaZnO_2} powder was separately synthesized via a solid-state reaction route using BaCO$_3$ and ZnO as starting precursors.
The well-ground mixture was first sintered at 970 \degree C for 16 hours under an \ch{Ar} atmosphere with the ground mixture then resintered at 1000 \degree C for 48 hours under \ch{N_2} flow. 
It was found that \ch{BaZnO_2} is not stable below a temperature of 525 \degree C as it decomposes into \ch{BaO_2} and \ch{ZnO}. 
To avoid decomposition, the sintered pellet of the \ch{BaZnO_2} sample was rapidly cooled down to room temperature from 1000 \degree C, resulting in a pure phase of \ch{BaZnO_2}. 
In order to prepare a ceramic rod of \BTmZO{} for growth, the pure \BTmZO{} powder was compressed into cylindrical-shaped feed and seed rods using a hydrostatic pressure of approximately 700 bar. These cylindrical feed rods typically weighed between 9 to 12 grams and measured about 8 to 11 cm in length.  These rods were sintered at 1130 \degree C for 24 hours under an oxygen environment in a vertical tube furnace (Crystal System Corporation) to achieve higher density.

\input{growth_table}
The details of the crystal growth parameters are tabulated in Table~\ref{growthtable}. The initial attempt (Growth no.~1), conducted without a solvent disk, resulted in a crystal with impurity phases from the decomposition of \BTmZO{}, although traces of the \BTmZO{} BP phase were detected on the crystal's surface. 
Furthermore, without using any solvent disk to sustain the liquid zone, the lamp power had to be increased to 5-6 \% of its maximum power, which likely contributed to decomposition. We attempted  to use  \ch{BaCO3} or a combination of (\ch{BaCO3} + ZnO) as a flux, but the resulting grown crystal was found to exhibit the presence of the impurity phases identified in Eq.~\eqref{chmeq}.
However, subsequent incorporation of a \ch{BaZnO2} solvent disk and optimizing the shutter length to approximately 70 mm not only mitigated this issue, but also allowed for the lamp power to be substantially reduced to 0.1\% of its maximum operating power.
The crystal growth process was hampered by a number of challenges, such as maintaining a stable liquid zone, with growth speeds above 5 mm/hr leading to incomplete BP phase formation and speeds below 1 mm/hr causing a liquid zone instability. 
Furthermore, the release of trapped oxygen disrupted the growth, resulting in multiple crystalline grains instead of single crystals. To overcome these difficulties, the growth parameters were iteratively optimized in each growth, progressively enhancing crystal quality. The oxygen pressure was increased to a maximum  of 0.95 MPa value in order to minimize bubbling and ultimately centimeter-scale \BTmZO{} single crystals were obtained using a `two-scanning' method.

\begin{figure}[]
\includegraphics[width=0.47\textwidth]{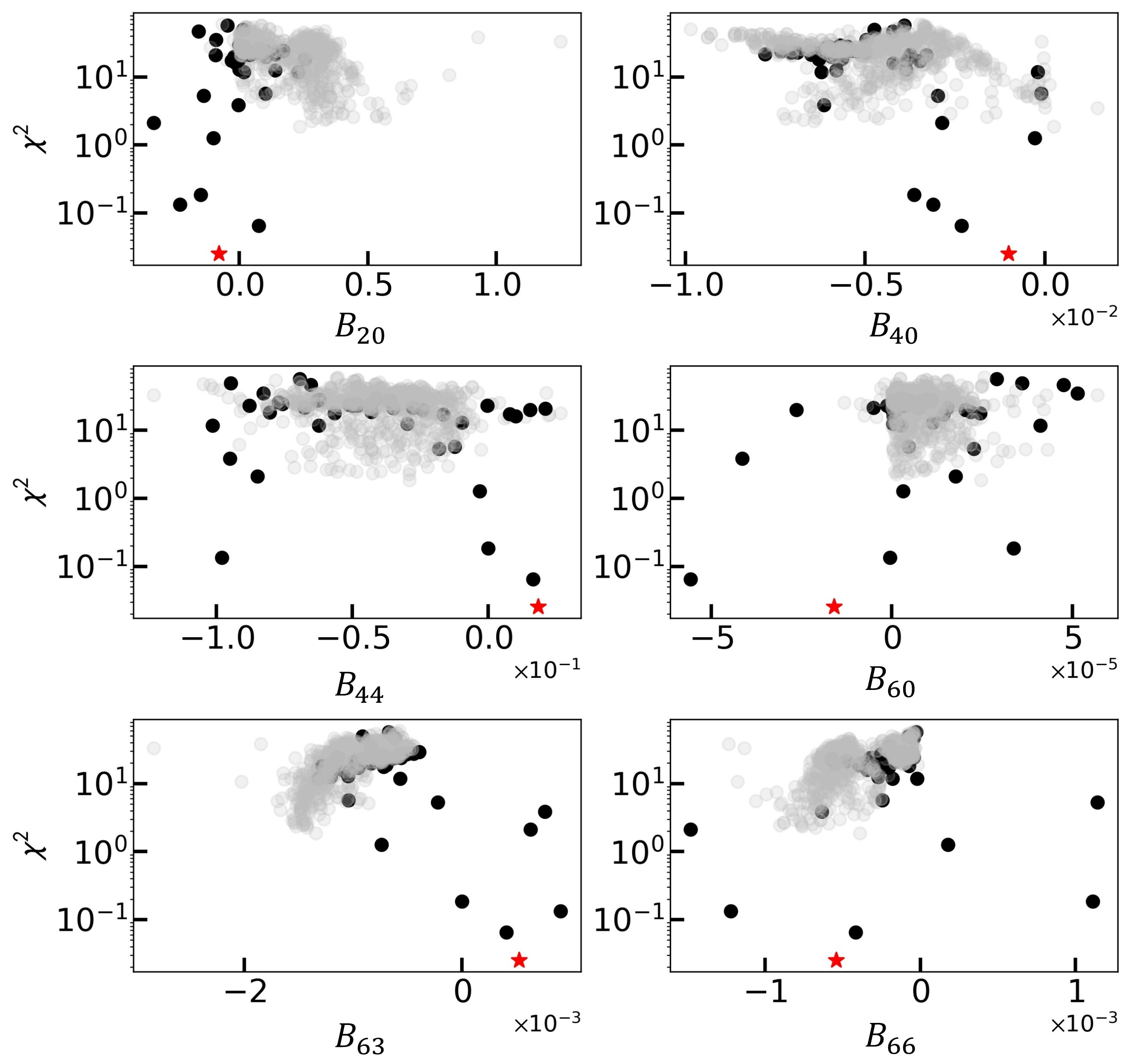}

\caption{The fitted crystal field parameters versus $\chi^2$ error obtained through various strategies are depicted. The red star data point represents the best parameter, while all other circular markers denote other converged solutions. Note that solutions predicting an intermediate CEF level between 0 and 8 meV, shown by gray color, exhibited significantly larger $\chi^2$ errors.}
 \label{chisq_Allblms}
     \hfill
\end{figure}

\begin{figure}[]
\includegraphics[width=0.47\textwidth]{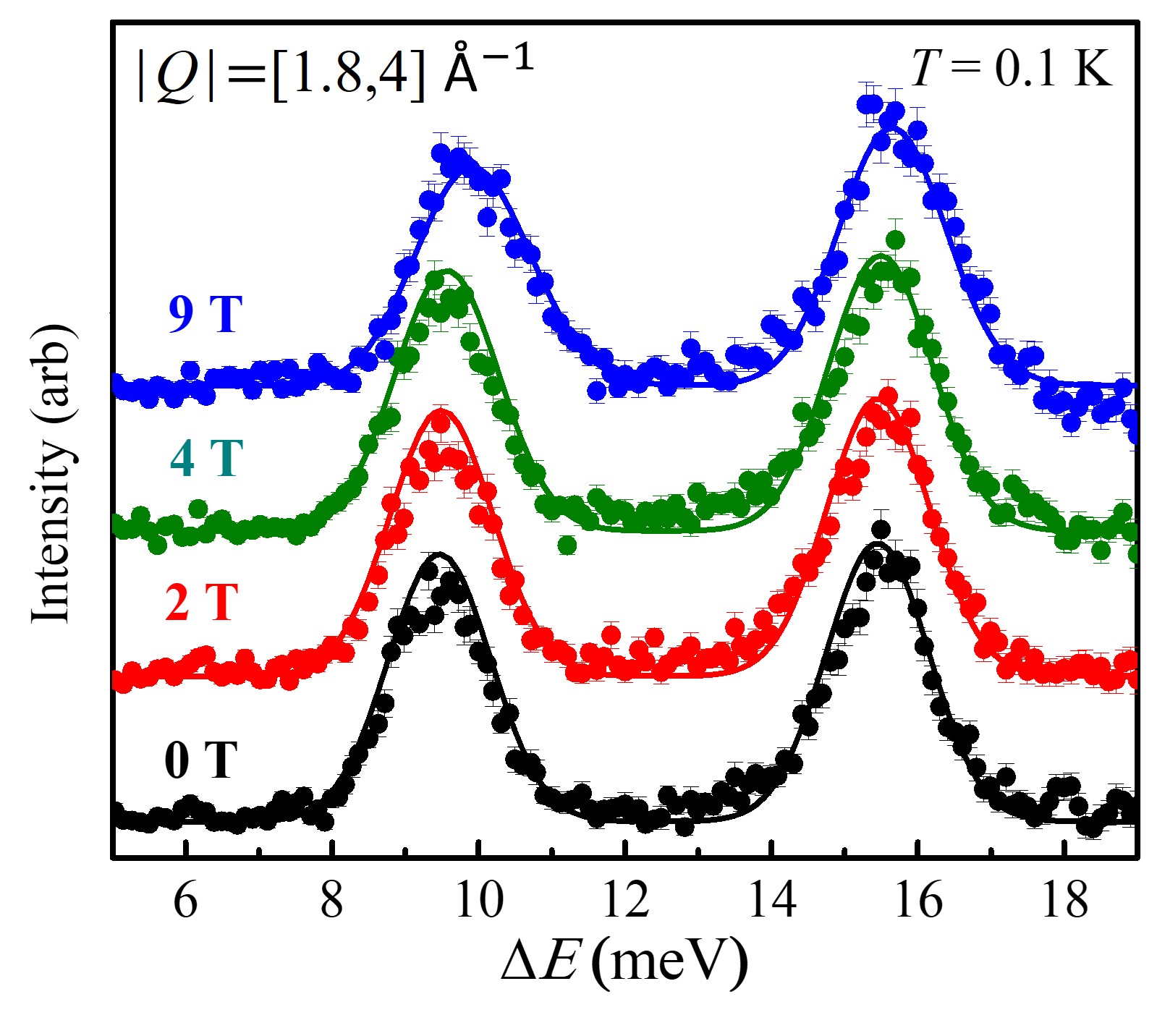}

\caption{The magnetic field dependence of CEF levels at 0.1 K. Constant-$Q$ cuts have been derived from $\Delta E$ vs. $|Q|$ slices by integrating within the window of 1.8-4 $\AA^{-1}$. The scatter points represent the experimental data, and the solid line represents the fit. The curves have been shifted along the y-axis for better clarity. Error bars indicate $\pm 1~ \sigma$.}
 \label{CEF_field}
     \hfill
\end{figure}

\section{CEF analysis}
\label{App_cef}

 In our study, we used the PyCrystalField \cite{scheie2021pycrystalfield} package to analyze the inelastic neutron scattering data at zero field (0 T) and determine the crystal electric field (CEF) $B_{lm}$ parameters of Eq.~\eqref{eq:cef-ham}. For simulating the field dependence of the CEF levels, we employed the MCPhase \cite{rotter2004using} package.
We designed a custom cost function to account for the unique challenges in the experimental data, including the presence of weak bands and the uncertainty of unobserved bands. Our cost function included multiple contributions, such as the $2D$ $Q$- and $\delta E$-dependent spectrum, background penalties for any excitation with intensity more than the background in areas where we did not see any energy band, integrated peak intensities to account for multiple energy levels within the observed peak width, powder magnetization and eigenvalue penalties to ensure the presence of eigenvalues at the positions of clearly observed bands. 

Considering the multiple contributions to the cost function, we used adaptive weighing of the cost contributions to ensure that all sources of cost receive appropriate weight. The CEF fitting is heavily influenced by the choices of initial parameters and tends to get trapped in local minima. To solve this issue, we used various methods for the initial parameter estimation. For example, we used a point charge model to obtain the starting values for the $B_{lm}$ parameters. We also tried rescaling the published CEF parameters from the sister compound \ch{Ba3Yb2Zn5O11} \cite{Haku2016CrystalBa3Yb2Zn5O11} using
\begin{equation}
B_{l m}=\frac{\theta^{(l)}\left\langle r^l\right\rangle}{\theta_0^{(l)}\left\langle r^l\right\rangle_0}\left(\frac{a}{a_0}\right)^{-l+1} B_{l m}^0
\end{equation}
as discussed for other systems~\cite{bertin2012crystal,Rau2018FrustrationOctahedra}. In addition,  we also used logarithmic random sampling to sample values for initial parameters as the $B_{lm}$ parameters can span a wide range of magnitudes. This strategy helped us explore a wide range of parameter space and reduced the risk of being trapped in local minima during the optimization process. In all, we explored about 100,000 random starting parameters using a computer cluster to run the program in parallel.

In Fig.~\ref{chisq_Allblms}, we present all the converged solutions obtained through various strategies. The y-axis displays the error, $\chi^2$, on a logarithmic scale, while the x-axis shows the values of the $B_{lm}$ parameters. The optimal solution is highlighted by a red star while other converged solutions are shown as circular markers. We observed that solutions predicting low-lying energy levels between 0 and 8 meV exhibited significantly higher experimental disagreement, as indicated by the large $\chi^2$ error, which is represented by the gray circular markers. Thus, our CEF analysis indicates that the crystal field ground state is a singlet with no intermediate CEF level between the ground state and the first excited state at 9.47 meV. The complete list of the 13 CEF energy levels calculated using the best fit $B_{lm}$ parameters (all in meV) are as follows: 9.47, 15.45, 15.45, 25.05, 25.05, 31.01, 31.04, 42.64, 42.64, 53.87, 53.87, 57.36. Finally, we present the field dependence of the CEF excitations measured at the disk chopper spectrometer(DCS, NIST). Figure \ref{CEF_field} shows the evolution of the 9.47 meV and 15.45 meV band as the field is applied up to 9 T. The CEF model fit is shown by the line and agrees with the observed trend.     
\section{Low Energy Powder INS Data}
\label{seq_lowE}
\begin{figure}[]
\includegraphics[width=0.45\textwidth]{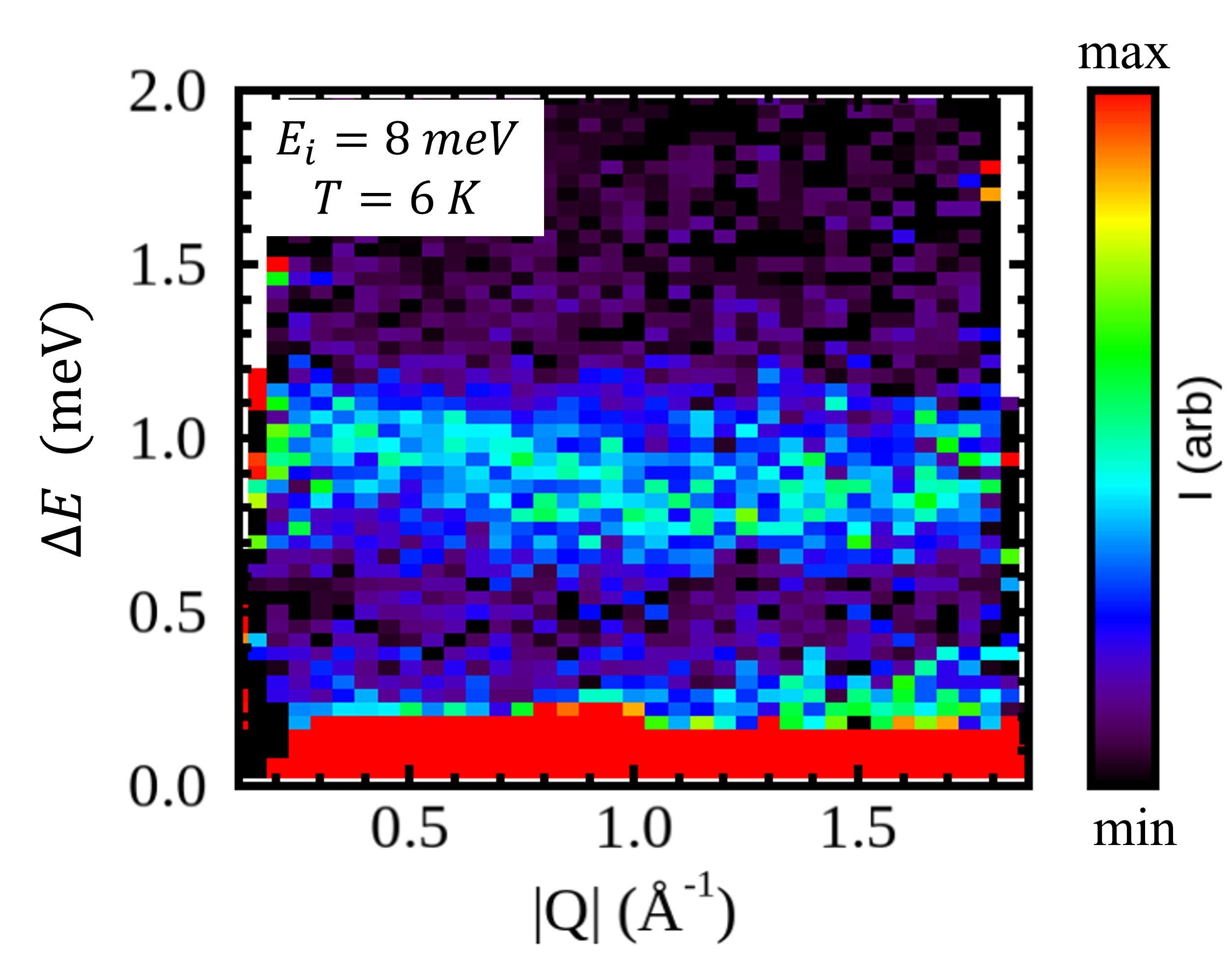}
  \caption{Powder INS data collected at the SEQUOIA spectrometer with \(E_i = 8 \text{ meV}\) at 6 K, highlighting the anomalous 1 meV energy mode.
    }
 \label{fig_seqlowe}
     \hfill
\end{figure}
In the main results sections, we reported an anomalous energy mode at approximately 1 meV, observed in the single crystal INS experiment conducted at the CNCS spectrometer. This mode is also present in the powder INS data collected at the SEQUOIA spectrometer during our investigation of the CEF levels. In addition to the \(E_i\) mentioned in the methods section, we used a lower \(E_i\) of 8 meV, which provided sufficient resolution to resolve the 1 meV band. As shown in Figure \ref{fig_seqlowe}, the 1 meV mode is clearly visible in the powder data as well. This consistency rules out the possibility that the observed band arises from scattering due to superglue, magnets, or other non-sample-related sources.

\section{Heat capacity phononic contribution}
\label{App_hc}

\begin{figure}[]
\includegraphics[width=0.45\textwidth]{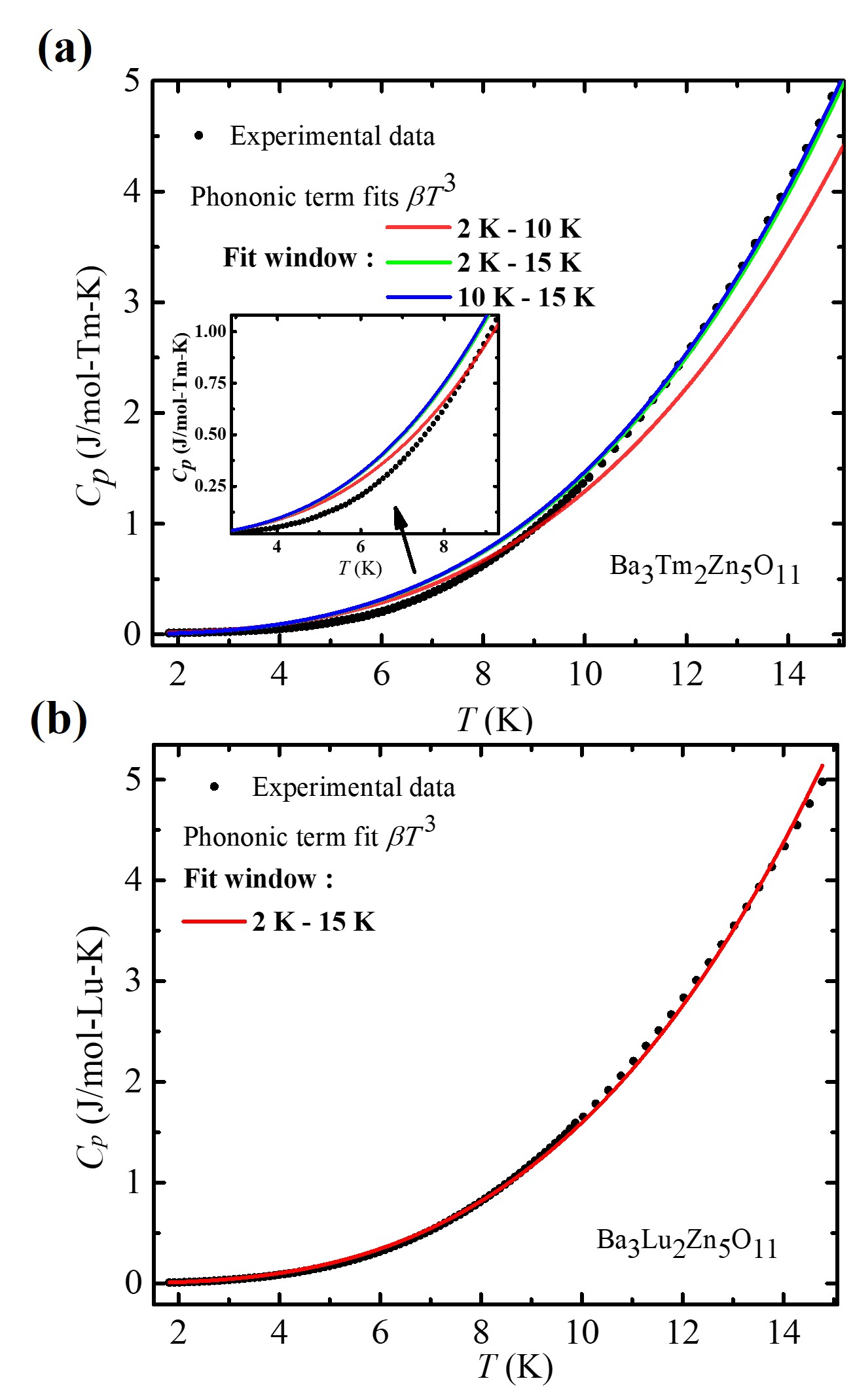}
  \caption{Comparison of the total heat capacity and phononic term fitting for \BTmZO{} and non-magnetic \BLuZO{} compounds. Black scatter points represent the measured heat capacity, while the lines indicate the phononic term $\beta T^3$ fitting across various temperature windows. (a) For \BTmZO, all such fits fail to fit the total heat capacity at the lower temperature end, irrespective of the temperature window chosen. (b) In contrast, the heat capacity of \BLuZO{} agrees well an $\beta  T^3$ term across the entire range of 2 K to 15 K.
    }
 \label{HC_Tm_Lu}
     \hfill
\end{figure}

We attempted to extract the phononic contribution in the total heat capacity  $C_p$ of \BTmZO{} by subtracting the heat capacity of the non-magnetic analogue \BLuZO{}. However, due to the larger heat capacity of \BLuZO{} below $\approx$ 18 K, the subtraction was unsuccessful. As a result, we fitted the low-temperature heat capacity data using the $\beta T^3$ term, where $\beta$ is a proportionality constant. We explored various fitting windows between 2 K to 15 K, as illustrated in Fig.~\ref{HC_Tm_Lu}. However, we could not find any temperature window between 2 K to 15 K where the fit does not overestimate the total heat capacity at the lower temperature. Additionally, incorporating a linear $\alpha T$ term did not improve the fit. In contrast, nonmagnetic \BLuZO{} total heat capacity can easily be fitted by the $\beta T^3$ term for the full temperature range between 2 K to 15 K. The inability to fit a $\beta T^3$ term to \BTmZO{} suggests that there are additional contributions to its heat capacity, possibly from magnetic excitations, which are not present in \BLuZO{}. Further, the $C_p$ of \BTmZO{} lacks a Schottky anomaly peak centered at 4 K that would correspond to the 1 meV excitation observed in the INS spectra.

\section{\textit{Ab initio} phonon calculations}
\label{App_phononcalc}
In order to rule out phonons as the origin of the low-energy modes in the inelastic neutron scattering (INS) data in Fig.~\ref{cncs},  we performed \textit{Ab initio} calculations using the plane wave pseudopotential density functional theory method as implemented in the Vienna \textit{Ab Initio} Simulation Package 
(VASP)~\cite{kresse1996efficient,kresse1996efficiency,kresse1993ab}. 
For the purpose of this calculation, we used the projector augmented wave (PAW) 
pseudopotential~\cite{kresse1999ultrasoft,blochl1994projector} for all the elements. The exchange-correlation potential employed was the modified Perdew-Burke-Ernzerhof generalized gradient approximation for solids (PBEsol GGA)~\cite{perdew2008restoring}. 
In order to perform phonon calculations, we started by relaxing imposing no final target symmetry, but starting from coordinates taken from the $F\bar{4}3m$ structure.
We also performed relaxations starting from lower symmetry (perturbed) structures, but found that the $F\bar{4}3m$ symmetry was restored under relaxation, confirming the stability of the $F\bar{4}3m$ structure. 
The energy cutoff for the plane-wave basis set and energy convergence threshold for electronic self-consistent cycle was set to 600 eV and $10^{-8}$ eV, respectively. 
The Blocked-Davidson iteration scheme was used for the electronic minimization algorithm. 
We used a $k$-point grid of $8\times 8\times 8$ for structure relaxation. For the phonon calculations, we employed the finite displacement method~\cite{kresse1995ab,parlinski1997first} in conjunction with Phonopy~\cite{phonopy-phono3py-JPCM,phonopy-phono3py-JPSJ}, with a displacement magnitude of 0.01\AA.
We used a real-space supercell of $2\times 2\times 2$ of the primitive unit cell, and a $k$-point grid of $2\times 2\times 2$.

\bibliography{references}

\end{document}

%% file: growth_table.tex
\begin{table*}[!htb]
\small

\caption{Experimental details of \ch{Ba_3Tm_2Zn_5O_{11}} crystal growth using optical floating zone technique.  }\label{growthtable}

 	 \begin{tabular}{ |>{\centering\arraybackslash}m{1cm} >{\centering\arraybackslash}m{1.5cm} >{\centering\arraybackslash}m{1.45cm} >{\centering\arraybackslash}m{1.45cm} >{\centering\arraybackslash}m{1.45cm} >{\centering\arraybackslash}m{1.35cm} >{\centering\arraybackslash}m{1.35cm} >{\centering\arraybackslash}m{1.25cm} >{\centering\arraybackslash}m{1.95cm}| } 
 \hline
Growth no. &	\ch{O_2} Pressure (MPa) 	&Rotation speed (rpm)&	Feed speed (mm/h) & Growth speed (mm/h)  &	Shutter length (mm)	& Solvent&	Lamp power & Results\\
 \hline	
1&	0.3-0.6 & 9-10&1-2.5 & 1.2-2 &65-70  &No flux& 2-5\%	& Decomposed products\\
 \hline 
2&	0.3-0.6 & 9-10&5-15 & 5-15 &65-70  &\ch{BaZnO_2}& 2-5\%	& \ch{Ba_3Tm_2Zn_5O_{11}} in the crystal's shell\\
 \hline
3&	0.6 & 9-10&1-2.5 & 1.2-2 &65-70  &BaZnO2&	0.10\% & \ch{Ba_3Tm_2Zn_5O_{11}} prominent but no crystal\\
 \hline
4&	0.6 & 9-10&1.8-1.9&	1.0&70&	BaZnO2&	0.10\% &Largest grain size: 0.13 gm\\
 \hline

5&	0.95 &14-15&	1.15-2.19&1.15&	70 &	BaZnO2&	0.10\% &Largest grain size: 0.2 gm\\
\hline
6 a&	 0.95 &	20&	4.5-3.3	&1.5&70 &	BaZnO2&	0.10\%&      Crystallized feed rod for 6 b\\
 \hline
6 b&	0.95 &15-18	&0.55-0.75&	0.4&70 &	BaZnO2&	0.10\% & Largest grain size: 0.33 gm\\
 \hline
\end{tabular}
\end{table*}

%% file: main.bbl
\begin{thebibliography}{78}%
\makeatletter
\providecommand \@ifxundefined [1]{%
 \@ifx{#1\undefined}
}%
\providecommand \@ifnum [1]{%
 \ifnum #1\expandafter \@firstoftwo
 \else \expandafter \@secondoftwo
 \fi
}%
\providecommand \@ifx [1]{%
 \ifx #1\expandafter \@firstoftwo
 \else \expandafter \@secondoftwo
 \fi
}%
\providecommand \natexlab [1]{#1}%
\providecommand \enquote  [1]{``#1''}%
\providecommand \bibnamefont  [1]{#1}%
\providecommand \bibfnamefont [1]{#1}%
\providecommand \citenamefont [1]{#1}%
\providecommand \href@noop [0]{\@secondoftwo}%
\providecommand \href [0]{\begingroup \@sanitize@url \@href}%
\providecommand \@href[1]{\@@startlink{#1}\@@href}%
\providecommand \@@href[1]{\endgroup#1\@@endlink}%
\providecommand \@sanitize@url [0]{\catcode `\\12\catcode `\$12\catcode
  `\&12\catcode `\#12\catcode `\^12\catcode `\_12\catcode `\%12\relax}%
\providecommand \@@startlink[1]{}%
\providecommand \@@endlink[0]{}%
\providecommand \url  [0]{\begingroup\@sanitize@url \@url }%
\providecommand \@url [1]{\endgroup\@href {#1}{\urlprefix }}%
\providecommand \urlprefix  [0]{URL }%
\providecommand \Eprint [0]{\href }%
\providecommand \doibase [0]{https://doi.org/}%
\providecommand \selectlanguage [0]{\@gobble}%
\providecommand \bibinfo  [0]{\@secondoftwo}%
\providecommand \bibfield  [0]{\@secondoftwo}%
\providecommand \translation [1]{[#1]}%
\providecommand \BibitemOpen [0]{}%
\providecommand \bibitemStop [0]{}%
\providecommand \bibitemNoStop [0]{.\EOS\space}%
\providecommand \EOS [0]{\spacefactor3000\relax}%
\providecommand \BibitemShut  [1]{\csname bibitem#1\endcsname}%
\let\auto@bib@innerbib\@empty
\bibitem [{\citenamefont {Balents}(2010)}]{Balents2010SpinMagnets}%
  \BibitemOpen
  \bibfield  {author} {\bibinfo {author} {\bibfnamefont {L.}~\bibnamefont
  {Balents}},\ }\bibfield  {title} {\bibinfo {title} {{Spin liquids in
  frustrated magnets}},\ }\href {https://doi.org/10.1038/nature08917}
  {\bibfield  {journal} {\bibinfo  {journal} {Nature}\ }\textbf {\bibinfo
  {volume} {464}},\ \bibinfo {pages} {199} (\bibinfo {year}
  {2010})}\BibitemShut {NoStop}%
\bibitem [{\citenamefont {Knolle}\ and\ \citenamefont
  {Moessner}(2019)}]{Knolle2019ALiquids}%
  \BibitemOpen
  \bibfield  {author} {\bibinfo {author} {\bibfnamefont {J.}~\bibnamefont
  {Knolle}}\ and\ \bibinfo {author} {\bibfnamefont {R.}~\bibnamefont
  {Moessner}},\ }\bibfield  {title} {\bibinfo {title} {{A field guide to spin
  liquids}},\ }\href {https://doi.org/10.1146/annurev-conmatphys-031218-013401}
  {\bibfield  {journal} {\bibinfo  {journal} {Annual Review of Condensed Matter
  Physics}\ }\textbf {\bibinfo {volume} {10}},\ \bibinfo {pages} {451}
  (\bibinfo {year} {2019})}\BibitemShut {NoStop}%
\bibitem [{\citenamefont {Lacroix}\ \emph {et~al.}(2011)\citenamefont
  {Lacroix}, \citenamefont {Mendels},\ and\ \citenamefont
  {Mila}}]{Springer_frust}%
  \BibitemOpen
  \bibfield  {author} {\bibinfo {author} {\bibfnamefont {C.}~\bibnamefont
  {Lacroix}}, \bibinfo {author} {\bibfnamefont {P.}~\bibnamefont {Mendels}},\
  and\ \bibinfo {author} {\bibfnamefont {F.}~\bibnamefont {Mila}},\ }\href@noop
  {} {\emph {\bibinfo {title} {Introduction to Frustrated Magnetism}}}\
  (\bibinfo  {publisher} {Springer},\ \bibinfo {year} {2011})\BibitemShut
  {NoStop}%
\bibitem [{\citenamefont {Gardner}\ \emph {et~al.}(2010)\citenamefont
  {Gardner}, \citenamefont {Gingras},\ and\ \citenamefont
  {Greedan}}]{Gardner2010MagneticOxides}%
  \BibitemOpen
  \bibfield  {author} {\bibinfo {author} {\bibfnamefont {J.~S.}\ \bibnamefont
  {Gardner}}, \bibinfo {author} {\bibfnamefont {M.~J.}\ \bibnamefont
  {Gingras}},\ and\ \bibinfo {author} {\bibfnamefont {J.~E.}\ \bibnamefont
  {Greedan}},\ }\bibfield  {title} {\bibinfo {title} {{Magnetic pyrochlore
  oxides}},\ }\href {https://doi.org/10.1103/RevModPhys.82.53} {\bibfield
  {journal} {\bibinfo  {journal} {Reviews of Modern Physics}\ }\textbf
  {\bibinfo {volume} {82}},\ \bibinfo {pages} {53} (\bibinfo {year}
  {2010})}\BibitemShut {NoStop}%
\bibitem [{\citenamefont {Hallas}\ \emph {et~al.}(2018)\citenamefont {Hallas},
  \citenamefont {Gaudet},\ and\ \citenamefont {Gaulin}}]{Hallas-AnnRevCMP}%
  \BibitemOpen
  \bibfield  {author} {\bibinfo {author} {\bibfnamefont {A.~M.}\ \bibnamefont
  {Hallas}}, \bibinfo {author} {\bibfnamefont {J.}~\bibnamefont {Gaudet}},\
  and\ \bibinfo {author} {\bibfnamefont {B.~D.}\ \bibnamefont {Gaulin}},\
  }\bibfield  {title} {\bibinfo {title} {{Experimental Insights into
  Ground-State Selection of Quantum XY Pyrochlores}},\ }\href
  {https://doi.org/10.1146/annurev-conmatphys-031016-025218} {\bibfield
  {journal} {\bibinfo  {journal} {{Annu. Rev. Condens. Matter Phys.}}\ }\textbf
  {\bibinfo {volume} {9}},\ \bibinfo {pages} {105} (\bibinfo {year}
  {2018})}\BibitemShut {NoStop}%
\bibitem [{\citenamefont {Rau}\ and\ \citenamefont
  {Gingras}(2019)}]{Rau-AnnRevCMP}%
  \BibitemOpen
  \bibfield  {author} {\bibinfo {author} {\bibfnamefont {J.~G.}\ \bibnamefont
  {Rau}}\ and\ \bibinfo {author} {\bibfnamefont {M.~J.~P.}\ \bibnamefont
  {Gingras}},\ }\bibfield  {title} {\bibinfo {title} {{Frustrated Quantum
  Rare-Earth Pyrochlores}},\ }\href
  {https://doi.org/10.1146/annurev-conmatphys-022317-110520} {\bibfield
  {journal} {\bibinfo  {journal} {{Annu. Rev. Condens. Matter Phys.}}\ }\textbf
  {\bibinfo {volume} {10}},\ \bibinfo {pages} {357} (\bibinfo {year}
  {2019})}\BibitemShut {NoStop}%
\bibitem [{\citenamefont {Greedan}\ \emph {et~al.}(1996)\citenamefont
  {Greedan}, \citenamefont {Raju}, \citenamefont {Maignan}, \citenamefont
  {Simon}, \citenamefont {Pedersen}, \citenamefont {Niraimathi}, \citenamefont
  {Gmelin},\ and\ \citenamefont {Subramanian}}]{PhysRevB.54.7189}%
  \BibitemOpen
  \bibfield  {author} {\bibinfo {author} {\bibfnamefont {J.~E.}\ \bibnamefont
  {Greedan}}, \bibinfo {author} {\bibfnamefont {N.~P.}\ \bibnamefont {Raju}},
  \bibinfo {author} {\bibfnamefont {A.}~\bibnamefont {Maignan}}, \bibinfo
  {author} {\bibfnamefont {C.}~\bibnamefont {Simon}}, \bibinfo {author}
  {\bibfnamefont {J.~S.}\ \bibnamefont {Pedersen}}, \bibinfo {author}
  {\bibfnamefont {A.~M.}\ \bibnamefont {Niraimathi}}, \bibinfo {author}
  {\bibfnamefont {E.}~\bibnamefont {Gmelin}},\ and\ \bibinfo {author}
  {\bibfnamefont {M.~A.}\ \bibnamefont {Subramanian}},\ }\bibfield  {title}
  {\bibinfo {title} {Frustrated pyrochlore oxides, \ch{Y2Mn2O7}, \ch{Ho2Mn2O7}
  and \ch{Yb2Mn2O7}: Bulk magnetism and magnetic microstructure},\ }\href
  {https://doi.org/10.1103/PhysRevB.54.7189} {\bibfield  {journal} {\bibinfo
  {journal} {Phys. Rev. B}\ }\textbf {\bibinfo {volume} {54}},\ \bibinfo
  {pages} {7189} (\bibinfo {year} {1996})}\BibitemShut {NoStop}%
\bibitem [{\citenamefont {Gingras}\ \emph {et~al.}(1997)\citenamefont
  {Gingras}, \citenamefont {Stager}, \citenamefont {Raju}, \citenamefont
  {Gaulin},\ and\ \citenamefont {Greedan}}]{Gingras-Y2Mo2O7}%
  \BibitemOpen
  \bibfield  {author} {\bibinfo {author} {\bibfnamefont {M.~J.~P.}\
  \bibnamefont {Gingras}}, \bibinfo {author} {\bibfnamefont {C.~V.}\
  \bibnamefont {Stager}}, \bibinfo {author} {\bibfnamefont {N.~P.}\
  \bibnamefont {Raju}}, \bibinfo {author} {\bibfnamefont {B.~D.}\ \bibnamefont
  {Gaulin}},\ and\ \bibinfo {author} {\bibfnamefont {J.~E.}\ \bibnamefont
  {Greedan}},\ }\bibfield  {title} {\bibinfo {title} {{Static Critical Behavior
  of the Spin-Freezing Transition in the Geometrically Frustrated Pyrochlore
  Antiferromagnet {${\mathrm{Y}}_{2}{\mathrm{Mo}}_{2}{\mathrm{O}}_{7}$}}},\
  }\href {https://doi.org/10.1103/PhysRevLett.78.947} {\bibfield  {journal}
  {\bibinfo  {journal} {Phys. Rev. Lett.}\ }\textbf {\bibinfo {volume} {78}},\
  \bibinfo {pages} {947} (\bibinfo {year} {1997})}\BibitemShut {NoStop}%
\bibitem [{\citenamefont {Gardner}\ \emph {et~al.}(1999)\citenamefont
  {Gardner}, \citenamefont {Gaulin}, \citenamefont {Lee}, \citenamefont
  {Broholm}, \citenamefont {Raju},\ and\ \citenamefont
  {Greedan}}]{Gardner-Y2Mo2O7}%
  \BibitemOpen
  \bibfield  {author} {\bibinfo {author} {\bibfnamefont {J.~S.}\ \bibnamefont
  {Gardner}}, \bibinfo {author} {\bibfnamefont {B.~D.}\ \bibnamefont {Gaulin}},
  \bibinfo {author} {\bibfnamefont {S.-H.}\ \bibnamefont {Lee}}, \bibinfo
  {author} {\bibfnamefont {C.}~\bibnamefont {Broholm}}, \bibinfo {author}
  {\bibfnamefont {N.~P.}\ \bibnamefont {Raju}},\ and\ \bibinfo {author}
  {\bibfnamefont {J.~E.}\ \bibnamefont {Greedan}},\ }\bibfield  {title}
  {\bibinfo {title} {{Glassy Statics and Dynamics in the Chemically Ordered
  Pyrochlore Antiferromagnet
  {${\mathrm{Y}}_{2}{\mathrm{Mo}}_{2}{\mathrm{O}}_{7}$} }},\ }\href
  {https://doi.org/10.1103/PhysRevLett.83.211} {\bibfield  {journal} {\bibinfo
  {journal} {Phys. Rev. Lett.}\ }\textbf {\bibinfo {volume} {83}},\ \bibinfo
  {pages} {211} (\bibinfo {year} {1999})}\BibitemShut {NoStop}%
\bibitem [{\citenamefont {Silverstein}\ \emph {et~al.}(2014)\citenamefont
  {Silverstein}, \citenamefont {Fritsch}, \citenamefont {Flicker},
  \citenamefont {Hallas}, \citenamefont {Gardner}, \citenamefont {Qiu},
  \citenamefont {Ehlers}, \citenamefont {Savici}, \citenamefont {Yamani},
  \citenamefont {Ross}, \citenamefont {Gaulin}, \citenamefont {Gingras},
  \citenamefont {Paddison}, \citenamefont {Foyevtsova}, \citenamefont
  {Valenti}, \citenamefont {Hawthorne}, \citenamefont {Wiebe},\ and\
  \citenamefont {Zhou}}]{Silverstein-Y2Mo2O7}%
  \BibitemOpen
  \bibfield  {author} {\bibinfo {author} {\bibfnamefont {H.~J.}\ \bibnamefont
  {Silverstein}}, \bibinfo {author} {\bibfnamefont {K.}~\bibnamefont
  {Fritsch}}, \bibinfo {author} {\bibfnamefont {F.}~\bibnamefont {Flicker}},
  \bibinfo {author} {\bibfnamefont {A.~M.}\ \bibnamefont {Hallas}}, \bibinfo
  {author} {\bibfnamefont {J.~S.}\ \bibnamefont {Gardner}}, \bibinfo {author}
  {\bibfnamefont {Y.}~\bibnamefont {Qiu}}, \bibinfo {author} {\bibfnamefont
  {G.}~\bibnamefont {Ehlers}}, \bibinfo {author} {\bibfnamefont {A.~T.}\
  \bibnamefont {Savici}}, \bibinfo {author} {\bibfnamefont {Z.}~\bibnamefont
  {Yamani}}, \bibinfo {author} {\bibfnamefont {K.~A.}\ \bibnamefont {Ross}},
  \bibinfo {author} {\bibfnamefont {B.~D.}\ \bibnamefont {Gaulin}}, \bibinfo
  {author} {\bibfnamefont {M.~J.~P.}\ \bibnamefont {Gingras}}, \bibinfo
  {author} {\bibfnamefont {J.~A.~M.}\ \bibnamefont {Paddison}}, \bibinfo
  {author} {\bibfnamefont {K.}~\bibnamefont {Foyevtsova}}, \bibinfo {author}
  {\bibfnamefont {R.}~\bibnamefont {Valenti}}, \bibinfo {author} {\bibfnamefont
  {F.}~\bibnamefont {Hawthorne}}, \bibinfo {author} {\bibfnamefont {C.~R.}\
  \bibnamefont {Wiebe}},\ and\ \bibinfo {author} {\bibfnamefont {H.~D.}\
  \bibnamefont {Zhou}},\ }\bibfield  {title} {\bibinfo {title} {{Liquidlike
  correlations in single-crystalline
  {${\mathrm{Y}}_{2}{\mathrm{Mo}}_{2}{\mathrm{O}}_{7}$}: An unconventional spin
  glass}},\ }\href {https://doi.org/10.1103/PhysRevB.89.054433} {\bibfield
  {journal} {\bibinfo  {journal} {Phys. Rev. B}\ }\textbf {\bibinfo {volume}
  {89}},\ \bibinfo {pages} {054433} (\bibinfo {year} {2014})}\BibitemShut
  {NoStop}%
\bibitem [{\citenamefont {Clark}\ \emph {et~al.}(2014)\citenamefont {Clark},
  \citenamefont {Nilsen}, \citenamefont {Kermarrec}, \citenamefont {Ehlers},
  \citenamefont {Knight}, \citenamefont {Harrison}, \citenamefont {Attfield},\
  and\ \citenamefont {Gaulin}}]{Molybdate_2014}%
  \BibitemOpen
  \bibfield  {author} {\bibinfo {author} {\bibfnamefont {L.}~\bibnamefont
  {Clark}}, \bibinfo {author} {\bibfnamefont {G.~J.}\ \bibnamefont {Nilsen}},
  \bibinfo {author} {\bibfnamefont {E.}~\bibnamefont {Kermarrec}}, \bibinfo
  {author} {\bibfnamefont {G.}~\bibnamefont {Ehlers}}, \bibinfo {author}
  {\bibfnamefont {K.~S.}\ \bibnamefont {Knight}}, \bibinfo {author}
  {\bibfnamefont {A.}~\bibnamefont {Harrison}}, \bibinfo {author}
  {\bibfnamefont {J.~P.}\ \bibnamefont {Attfield}},\ and\ \bibinfo {author}
  {\bibfnamefont {B.~D.}\ \bibnamefont {Gaulin}},\ }\bibfield  {title}
  {\bibinfo {title} {{From Spin Glass to Quantum Spin Liquid Ground States in
  Molybdate Pyrochlores}},\ }\href
  {https://doi.org/10.1103/PhysRevLett.113.117201} {\bibfield  {journal}
  {\bibinfo  {journal} {Phys. Rev. Lett.}\ }\textbf {\bibinfo {volume} {113}},\
  \bibinfo {pages} {117201} (\bibinfo {year} {2014})}\BibitemShut {NoStop}%
\bibitem [{\citenamefont {Onose}\ \emph {et~al.}(2010)\citenamefont {Onose},
  \citenamefont {Ideue}, \citenamefont {Katsura}, \citenamefont {Shiomi},
  \citenamefont {Nagaosa},\ and\ \citenamefont {Tokura}}]{Onose-Lu2V2O7}%
  \BibitemOpen
  \bibfield  {author} {\bibinfo {author} {\bibfnamefont {Y.}~\bibnamefont
  {Onose}}, \bibinfo {author} {\bibfnamefont {T.}~\bibnamefont {Ideue}},
  \bibinfo {author} {\bibfnamefont {H.}~\bibnamefont {Katsura}}, \bibinfo
  {author} {\bibfnamefont {Y.}~\bibnamefont {Shiomi}}, \bibinfo {author}
  {\bibfnamefont {N.}~\bibnamefont {Nagaosa}},\ and\ \bibinfo {author}
  {\bibfnamefont {Y.}~\bibnamefont {Tokura}},\ }\bibfield  {title} {\bibinfo
  {title} {Observation of the {{Magnon Hall Effect}}},\ }\href
  {https://doi.org/10.1126/science.1188260} {\bibfield  {journal} {\bibinfo
  {journal} {Science}\ }\textbf {\bibinfo {volume} {329}},\ \bibinfo {pages}
  {297} (\bibinfo {year} {2010})}\BibitemShut {NoStop}%
\bibitem [{\citenamefont {Mauws}\ \emph {et~al.}(2018)\citenamefont {Mauws},
  \citenamefont {Hallas}, \citenamefont {Sala}, \citenamefont {Aczel},
  \citenamefont {Sarte}, \citenamefont {Gaudet}, \citenamefont {Ziat},
  \citenamefont {Quilliam}, \citenamefont {Lussier}, \citenamefont {Bieringer},
  \citenamefont {Zhou}, \citenamefont {Wildes}, \citenamefont {Stone},
  \citenamefont {Abernathy}, \citenamefont {Luke}, \citenamefont {Gaulin},\
  and\ \citenamefont {Wiebe}}]{Sm2Ti2O7_2018}%
  \BibitemOpen
  \bibfield  {author} {\bibinfo {author} {\bibfnamefont {C.}~\bibnamefont
  {Mauws}}, \bibinfo {author} {\bibfnamefont {A.~M.}\ \bibnamefont {Hallas}},
  \bibinfo {author} {\bibfnamefont {G.}~\bibnamefont {Sala}}, \bibinfo {author}
  {\bibfnamefont {A.~A.}\ \bibnamefont {Aczel}}, \bibinfo {author}
  {\bibfnamefont {P.~M.}\ \bibnamefont {Sarte}}, \bibinfo {author}
  {\bibfnamefont {J.}~\bibnamefont {Gaudet}}, \bibinfo {author} {\bibfnamefont
  {D.}~\bibnamefont {Ziat}}, \bibinfo {author} {\bibfnamefont {J.~A.}\
  \bibnamefont {Quilliam}}, \bibinfo {author} {\bibfnamefont {J.~A.}\
  \bibnamefont {Lussier}}, \bibinfo {author} {\bibfnamefont {M.}~\bibnamefont
  {Bieringer}}, \bibinfo {author} {\bibfnamefont {H.~D.}\ \bibnamefont {Zhou}},
  \bibinfo {author} {\bibfnamefont {A.}~\bibnamefont {Wildes}}, \bibinfo
  {author} {\bibfnamefont {M.~B.}\ \bibnamefont {Stone}}, \bibinfo {author}
  {\bibfnamefont {D.}~\bibnamefont {Abernathy}}, \bibinfo {author}
  {\bibfnamefont {G.~M.}\ \bibnamefont {Luke}}, \bibinfo {author}
  {\bibfnamefont {B.~D.}\ \bibnamefont {Gaulin}},\ and\ \bibinfo {author}
  {\bibfnamefont {C.~R.}\ \bibnamefont {Wiebe}},\ }\bibfield  {title} {\bibinfo
  {title} {{Dipolar-octupolar Ising antiferromagnetism in
  ${\text{Sm}}_{2}{\text{Ti}}_{2}{\text{O}}_{7}$: A moment fragmentation
  candidate}},\ }\href {https://doi.org/10.1103/PhysRevB.98.100401} {\bibfield
  {journal} {\bibinfo  {journal} {Phys. Rev. B}\ }\textbf {\bibinfo {volume}
  {98}},\ \bibinfo {pages} {100401} (\bibinfo {year} {2018})}\BibitemShut
  {NoStop}%
\bibitem [{\citenamefont {Xu}\ \emph {et~al.}(2015)\citenamefont {Xu},
  \citenamefont {Anand}, \citenamefont {Bera}, \citenamefont {Frontzek},
  \citenamefont {Abernathy}, \citenamefont {Casati}, \citenamefont
  {Siemensmeyer},\ and\ \citenamefont {Lake}}]{Nd2Zr2O7_2015}%
  \BibitemOpen
  \bibfield  {author} {\bibinfo {author} {\bibfnamefont {J.}~\bibnamefont
  {Xu}}, \bibinfo {author} {\bibfnamefont {V.~K.}\ \bibnamefont {Anand}},
  \bibinfo {author} {\bibfnamefont {A.~K.}\ \bibnamefont {Bera}}, \bibinfo
  {author} {\bibfnamefont {M.}~\bibnamefont {Frontzek}}, \bibinfo {author}
  {\bibfnamefont {D.~L.}\ \bibnamefont {Abernathy}}, \bibinfo {author}
  {\bibfnamefont {N.}~\bibnamefont {Casati}}, \bibinfo {author} {\bibfnamefont
  {K.}~\bibnamefont {Siemensmeyer}},\ and\ \bibinfo {author} {\bibfnamefont
  {B.}~\bibnamefont {Lake}},\ }\bibfield  {title} {\bibinfo {title} {{Magnetic
  structure and crystal-field states of the pyrochlore antiferromagnet
  ${\mathrm{Nd}}_{2}{\mathrm{Zr}}_{2}{\mathrm{O}}_{7}$}},\ }\href
  {https://doi.org/10.1103/PhysRevB.92.224430} {\bibfield  {journal} {\bibinfo
  {journal} {Phys. Rev. B}\ }\textbf {\bibinfo {volume} {92}},\ \bibinfo
  {pages} {224430} (\bibinfo {year} {2015})}\BibitemShut {NoStop}%
\bibitem [{\citenamefont {Bonville}\ \emph {et~al.}(2003)\citenamefont
  {Bonville}, \citenamefont {Hodges}, \citenamefont {Ocio}, \citenamefont
  {Sanchez}, \citenamefont {Vulliet}, \citenamefont {Sosin},\ and\
  \citenamefont {Braithwaite}}]{Gd2SnTi2O7_2003}%
  \BibitemOpen
  \bibfield  {author} {\bibinfo {author} {\bibfnamefont {P.}~\bibnamefont
  {Bonville}}, \bibinfo {author} {\bibfnamefont {J.~A.}\ \bibnamefont
  {Hodges}}, \bibinfo {author} {\bibfnamefont {M.}~\bibnamefont {Ocio}},
  \bibinfo {author} {\bibfnamefont {J.~P.}\ \bibnamefont {Sanchez}}, \bibinfo
  {author} {\bibfnamefont {P.}~\bibnamefont {Vulliet}}, \bibinfo {author}
  {\bibfnamefont {S.}~\bibnamefont {Sosin}},\ and\ \bibinfo {author}
  {\bibfnamefont {D.}~\bibnamefont {Braithwaite}},\ }\bibfield  {title}
  {\bibinfo {title} {{Low temperature magnetic properties of geometrically
  frustrated \ch{Gd2Sn2O7} and \ch{Gd2Ti2O7}}},\ }\href
  {https://doi.org/10.1088/0953-8984/15/45/016} {\bibfield  {journal} {\bibinfo
   {journal} {Journal of Physics: Condensed Matter}\ }\textbf {\bibinfo
  {volume} {15}},\ \bibinfo {pages} {7777} (\bibinfo {year}
  {2003})}\BibitemShut {NoStop}%
\bibitem [{\citenamefont {Ross}\ \emph {et~al.}(2011)\citenamefont {Ross},
  \citenamefont {Savary}, \citenamefont {Gaulin},\ and\ \citenamefont
  {Balents}}]{Ross-PRX}%
  \BibitemOpen
  \bibfield  {author} {\bibinfo {author} {\bibfnamefont {K.~A.}\ \bibnamefont
  {Ross}}, \bibinfo {author} {\bibfnamefont {L.}~\bibnamefont {Savary}},
  \bibinfo {author} {\bibfnamefont {B.~D.}\ \bibnamefont {Gaulin}},\ and\
  \bibinfo {author} {\bibfnamefont {L.}~\bibnamefont {Balents}},\ }\bibfield
  {title} {\bibinfo {title} {{Quantum Excitations in Quantum Spin Ice}},\
  }\href {https://doi.org/10.1103/PhysRevX.1.021002} {\bibfield  {journal}
  {\bibinfo  {journal} {Phys. Rev. X}\ }\textbf {\bibinfo {volume} {1}},\
  \bibinfo {pages} {021002} (\bibinfo {year} {2011})}\BibitemShut {NoStop}%
\bibitem [{\citenamefont {Lee}\ \emph {et~al.}(2012)\citenamefont {Lee},
  \citenamefont {Onoda},\ and\ \citenamefont {Balents}}]{Lee-nonKramers}%
  \BibitemOpen
  \bibfield  {author} {\bibinfo {author} {\bibfnamefont {S.}~\bibnamefont
  {Lee}}, \bibinfo {author} {\bibfnamefont {S.}~\bibnamefont {Onoda}},\ and\
  \bibinfo {author} {\bibfnamefont {L.}~\bibnamefont {Balents}},\ }\bibfield
  {title} {\bibinfo {title} {Generic quantum spin ice},\ }\href
  {https://doi.org/10.1103/PhysRevB.86.104412} {\bibfield  {journal} {\bibinfo
  {journal} {Phys. Rev. B}\ }\textbf {\bibinfo {volume} {86}},\ \bibinfo
  {pages} {104412} (\bibinfo {year} {2012})}\BibitemShut {NoStop}%
\bibitem [{\citenamefont {Huang}\ \emph {et~al.}(2014)\citenamefont {Huang},
  \citenamefont {Chen},\ and\ \citenamefont {Hermele}}]{Huang-DO}%
  \BibitemOpen
  \bibfield  {author} {\bibinfo {author} {\bibfnamefont {Y.-P.}\ \bibnamefont
  {Huang}}, \bibinfo {author} {\bibfnamefont {G.}~\bibnamefont {Chen}},\ and\
  \bibinfo {author} {\bibfnamefont {M.}~\bibnamefont {Hermele}},\ }\bibfield
  {title} {\bibinfo {title} {{Quantum Spin Ices and Topological Phases from
  Dipolar-Octupolar Doublets on the Pyrochlore Lattice}},\ }\href
  {https://doi.org/10.1103/PhysRevLett.112.167203} {\bibfield  {journal}
  {\bibinfo  {journal} {Phys. Rev. Lett.}\ }\textbf {\bibinfo {volume} {112}},\
  \bibinfo {pages} {167203} (\bibinfo {year} {2014})}\BibitemShut {NoStop}%
\bibitem [{\citenamefont {Tsurkan}\ \emph {et~al.}(2021)\citenamefont
  {Tsurkan}, \citenamefont {Krug~von Nidda}, \citenamefont {Deisenhofer},
  \citenamefont {Lunkenheimer},\ and\ \citenamefont {Loidl}}]{spinels-review}%
  \BibitemOpen
  \bibfield  {author} {\bibinfo {author} {\bibfnamefont {V.}~\bibnamefont
  {Tsurkan}}, \bibinfo {author} {\bibfnamefont {H.-A.}\ \bibnamefont {Krug~von
  Nidda}}, \bibinfo {author} {\bibfnamefont {J.}~\bibnamefont {Deisenhofer}},
  \bibinfo {author} {\bibfnamefont {P.}~\bibnamefont {Lunkenheimer}},\ and\
  \bibinfo {author} {\bibfnamefont {A.}~\bibnamefont {Loidl}},\ }\bibfield
  {title} {\bibinfo {title} {{On the complexity of spinels: Magnetic,
  electronic, and polar ground states}},\ }\href
  {https://doi.org/10.1016/j.physrep.2021.04.002} {\bibfield  {journal}
  {\bibinfo  {journal} {Physics Reports}\ }\textbf {\bibinfo {volume} {926}},\
  \bibinfo {pages} {1} (\bibinfo {year} {2021})}\BibitemShut {NoStop}%
\bibitem [{\citenamefont {Okamoto}\ \emph {et~al.}(2013)\citenamefont
  {Okamoto}, \citenamefont {Nilsen}, \citenamefont {Attfield},\ and\
  \citenamefont {Hiroi}}]{Okamoto-BP}%
  \BibitemOpen
  \bibfield  {author} {\bibinfo {author} {\bibfnamefont {Y.}~\bibnamefont
  {Okamoto}}, \bibinfo {author} {\bibfnamefont {G.~J.}\ \bibnamefont {Nilsen}},
  \bibinfo {author} {\bibfnamefont {J.~P.}\ \bibnamefont {Attfield}},\ and\
  \bibinfo {author} {\bibfnamefont {Z.}~\bibnamefont {Hiroi}},\ }\bibfield
  {title} {\bibinfo {title} {{Breathing Pyrochlore Lattice Realized in $A$-Site
  Ordered Spinel Oxides {${\mathrm{LiGaCr}}_{4}{\mathrm{O}}_{8}$} and
  {${\mathrm{LiInCr}}_{4}{\mathrm{O}}_{8}$}}},\ }\href
  {https://doi.org/10.1103/PhysRevLett.110.097203} {\bibfield  {journal}
  {\bibinfo  {journal} {Phys. Rev. Lett.}\ }\textbf {\bibinfo {volume} {110}},\
  \bibinfo {pages} {097203} (\bibinfo {year} {2013})}\BibitemShut {NoStop}%
\bibitem [{\citenamefont {Ghosh}\ \emph {et~al.}(2019)\citenamefont {Ghosh},
  \citenamefont {Iqbal}, \citenamefont {M{\"u}ller}, \citenamefont
  {Ponnaganti}, \citenamefont {Thomale}, \citenamefont {Narayanan},
  \citenamefont {Reuther}, \citenamefont {Gingras},\ and\ \citenamefont
  {Jeschke}}]{Ghosh2019}%
  \BibitemOpen
  \bibfield  {author} {\bibinfo {author} {\bibfnamefont {P.}~\bibnamefont
  {Ghosh}}, \bibinfo {author} {\bibfnamefont {Y.}~\bibnamefont {Iqbal}},
  \bibinfo {author} {\bibfnamefont {T.}~\bibnamefont {M{\"u}ller}}, \bibinfo
  {author} {\bibfnamefont {R.~T.}\ \bibnamefont {Ponnaganti}}, \bibinfo
  {author} {\bibfnamefont {R.}~\bibnamefont {Thomale}}, \bibinfo {author}
  {\bibfnamefont {R.}~\bibnamefont {Narayanan}}, \bibinfo {author}
  {\bibfnamefont {J.}~\bibnamefont {Reuther}}, \bibinfo {author} {\bibfnamefont
  {M.~J.~P.}\ \bibnamefont {Gingras}},\ and\ \bibinfo {author} {\bibfnamefont
  {H.~O.}\ \bibnamefont {Jeschke}},\ }\bibfield  {title} {\bibinfo {title}
  {Breathing chromium spinels: a showcase for a variety of pyrochlore
  {H}eisenberg hamiltonians},\ }\href
  {https://doi.org/10.1038/s41535-019-0202-z} {\bibfield  {journal} {\bibinfo
  {journal} {npj Quantum Materials}\ }\textbf {\bibinfo {volume} {4}},\
  \bibinfo {pages} {63} (\bibinfo {year} {2019})}\BibitemShut {NoStop}%
\bibitem [{\citenamefont {Sharma}\ \emph {et~al.}(2022)\citenamefont {Sharma},
  \citenamefont {Pocrnic}, \citenamefont {Richtik}, \citenamefont {Wiebe},
  \citenamefont {Beare}, \citenamefont {Gautreau}, \citenamefont {Clancy},
  \citenamefont {Ruff}, \citenamefont {Pula}, \citenamefont {Chen},
  \citenamefont {Yoon}, \citenamefont {Cai},\ and\ \citenamefont
  {Luke}}]{Sharma2022}%
  \BibitemOpen
  \bibfield  {author} {\bibinfo {author} {\bibfnamefont {S.}~\bibnamefont
  {Sharma}}, \bibinfo {author} {\bibfnamefont {M.}~\bibnamefont {Pocrnic}},
  \bibinfo {author} {\bibfnamefont {B.~N.}\ \bibnamefont {Richtik}}, \bibinfo
  {author} {\bibfnamefont {C.~R.}\ \bibnamefont {Wiebe}}, \bibinfo {author}
  {\bibfnamefont {J.}~\bibnamefont {Beare}}, \bibinfo {author} {\bibfnamefont
  {J.}~\bibnamefont {Gautreau}}, \bibinfo {author} {\bibfnamefont {J.~P.}\
  \bibnamefont {Clancy}}, \bibinfo {author} {\bibfnamefont {J.~P.~C.}\
  \bibnamefont {Ruff}}, \bibinfo {author} {\bibfnamefont {M.}~\bibnamefont
  {Pula}}, \bibinfo {author} {\bibfnamefont {Q.}~\bibnamefont {Chen}}, \bibinfo
  {author} {\bibfnamefont {S.}~\bibnamefont {Yoon}}, \bibinfo {author}
  {\bibfnamefont {Y.}~\bibnamefont {Cai}},\ and\ \bibinfo {author}
  {\bibfnamefont {G.~M.}\ \bibnamefont {Luke}},\ }\bibfield  {title} {\bibinfo
  {title} {Synthesis and physical and magnetic properties of \ch{CuAlCr4S8}: A
  \ch{Cr}-based breathing pyrochlore},\ }\href
  {https://doi.org/10.1103/PhysRevB.106.024407} {\bibfield  {journal} {\bibinfo
   {journal} {Phys. Rev. B}\ }\textbf {\bibinfo {volume} {106}},\ \bibinfo
  {pages} {024407} (\bibinfo {year} {2022})}\BibitemShut {NoStop}%
\bibitem [{\citenamefont {Savary}\ \emph {et~al.}(2016)\citenamefont {Savary},
  \citenamefont {Wang}, \citenamefont {Kee}, \citenamefont {Kim}, \citenamefont
  {Yu},\ and\ \citenamefont {Chen}}]{Savary2016QuantumLattice}%
  \BibitemOpen
  \bibfield  {author} {\bibinfo {author} {\bibfnamefont {L.}~\bibnamefont
  {Savary}}, \bibinfo {author} {\bibfnamefont {X.}~\bibnamefont {Wang}},
  \bibinfo {author} {\bibfnamefont {H.~Y.}\ \bibnamefont {Kee}}, \bibinfo
  {author} {\bibfnamefont {Y.~B.}\ \bibnamefont {Kim}}, \bibinfo {author}
  {\bibfnamefont {Y.}~\bibnamefont {Yu}},\ and\ \bibinfo {author}
  {\bibfnamefont {G.}~\bibnamefont {Chen}},\ }\bibfield  {title} {\bibinfo
  {title} {{Quantum spin ice on the breathing pyrochlore lattice}},\ }\href
  {https://doi.org/10.1103/PhysRevB.94.075146} {\bibfield  {journal} {\bibinfo
  {journal} {Physical Review B}\ }\textbf {\bibinfo {volume} {94}},\ \bibinfo
  {pages} {1} (\bibinfo {year} {2016})}\BibitemShut {NoStop}%
\bibitem [{\citenamefont {Yan}\ \emph {et~al.}(2020)\citenamefont {Yan},
  \citenamefont {Benton}, \citenamefont {Jaubert},\ and\ \citenamefont
  {Shannon}}]{Yan2020Rank-2Lattice}%
  \BibitemOpen
  \bibfield  {author} {\bibinfo {author} {\bibfnamefont {H.}~\bibnamefont
  {Yan}}, \bibinfo {author} {\bibfnamefont {O.}~\bibnamefont {Benton}},
  \bibinfo {author} {\bibfnamefont {L.~D.~C.}\ \bibnamefont {Jaubert}},\ and\
  \bibinfo {author} {\bibfnamefont {N.}~\bibnamefont {Shannon}},\ }\bibfield
  {title} {\bibinfo {title} {{Rank--2 {$U(1)$} Spin Liquid on the Breathing
  Pyrochlore Lattice}},\ }\href
  {https://doi.org/10.1103/PhysRevLett.124.127203} {\bibfield  {journal}
  {\bibinfo  {journal} {Phys. Rev. Lett.}\ }\textbf {\bibinfo {volume} {124}},\
  \bibinfo {pages} {127203} (\bibinfo {year} {2020})}\BibitemShut {NoStop}%
\bibitem [{\citenamefont {Aoyama}\ and\ \citenamefont
  {Kawamura}(2023)}]{aoyama2023zero}%
  \BibitemOpen
  \bibfield  {author} {\bibinfo {author} {\bibfnamefont {K.}~\bibnamefont
  {Aoyama}}\ and\ \bibinfo {author} {\bibfnamefont {H.}~\bibnamefont
  {Kawamura}},\ }\bibfield  {title} {\bibinfo {title} {{Zero-Field Miniature
  Skyrmion Crystal and Chiral Domain State in Breathing-Kagome
  Antiferromagnets}},\ }\href {https://doi.org/10.7566/JPSJ.92.033701}
  {\bibfield  {journal} {\bibinfo  {journal} {Journal of the Physical Society
  of Japan}\ }\textbf {\bibinfo {volume} {92}},\ \bibinfo {pages} {033701}
  (\bibinfo {year} {2023})}\BibitemShut {NoStop}%
\bibitem [{\citenamefont {Li}\ \emph {et~al.}(2016)\citenamefont {Li},
  \citenamefont {Li}, \citenamefont {Kim}, \citenamefont {Balents},
  \citenamefont {Yu},\ and\ \citenamefont {Chen}}]{Li2016WeylAntiferromagnets}%
  \BibitemOpen
  \bibfield  {author} {\bibinfo {author} {\bibfnamefont {F.-Y.}\ \bibnamefont
  {Li}}, \bibinfo {author} {\bibfnamefont {Y.-D.}\ \bibnamefont {Li}}, \bibinfo
  {author} {\bibfnamefont {Y.~B.}\ \bibnamefont {Kim}}, \bibinfo {author}
  {\bibfnamefont {L.}~\bibnamefont {Balents}}, \bibinfo {author} {\bibfnamefont
  {Y.}~\bibnamefont {Yu}},\ and\ \bibinfo {author} {\bibfnamefont
  {G.}~\bibnamefont {Chen}},\ }\bibfield  {title} {\bibinfo {title} {Weyl
  magnons in breathing pyrochlore antiferromagnets},\ }\href
  {https://doi.org/10.1038/ncomms12691} {\bibfield  {journal} {\bibinfo
  {journal} {Nature communications}\ }\textbf {\bibinfo {volume} {7}},\
  \bibinfo {pages} {12691} (\bibinfo {year} {2016})}\BibitemShut {NoStop}%
\bibitem [{\citenamefont {Pokharel}\ \emph {et~al.}(2020)\citenamefont
  {Pokharel}, \citenamefont {Arachchige}, \citenamefont {Williams},
  \citenamefont {May}, \citenamefont {Fishman}, \citenamefont {Sala},
  \citenamefont {Calder}, \citenamefont {Ehlers}, \citenamefont {Parker},
  \citenamefont {Hong}, \citenamefont {Wildes}, \citenamefont {Mandrus},
  \citenamefont {Paddison},\ and\ \citenamefont
  {Christianson}}]{Pokharel2020ClusterS8}%
  \BibitemOpen
  \bibfield  {author} {\bibinfo {author} {\bibfnamefont {G.}~\bibnamefont
  {Pokharel}}, \bibinfo {author} {\bibfnamefont {H.~S.}\ \bibnamefont
  {Arachchige}}, \bibinfo {author} {\bibfnamefont {T.~J.}\ \bibnamefont
  {Williams}}, \bibinfo {author} {\bibfnamefont {A.~F.}\ \bibnamefont {May}},
  \bibinfo {author} {\bibfnamefont {R.~S.}\ \bibnamefont {Fishman}}, \bibinfo
  {author} {\bibfnamefont {G.}~\bibnamefont {Sala}}, \bibinfo {author}
  {\bibfnamefont {S.}~\bibnamefont {Calder}}, \bibinfo {author} {\bibfnamefont
  {G.}~\bibnamefont {Ehlers}}, \bibinfo {author} {\bibfnamefont {D.~S.}\
  \bibnamefont {Parker}}, \bibinfo {author} {\bibfnamefont {T.}~\bibnamefont
  {Hong}}, \bibinfo {author} {\bibfnamefont {A.}~\bibnamefont {Wildes}},
  \bibinfo {author} {\bibfnamefont {D.}~\bibnamefont {Mandrus}}, \bibinfo
  {author} {\bibfnamefont {J.~A.}\ \bibnamefont {Paddison}},\ and\ \bibinfo
  {author} {\bibfnamefont {A.~D.}\ \bibnamefont {Christianson}},\ }\bibfield
  {title} {\bibinfo {title} {{Cluster Frustration in the Breathing Pyrochlore
  Magnet \ch{LiGaCr4S8}}},\ }\href
  {https://doi.org/10.1103/PhysRevLett.125.167201} {\bibfield  {journal}
  {\bibinfo  {journal} {Physical Review Letters}\ }\textbf {\bibinfo {volume}
  {125}},\ \bibinfo {pages} {1} (\bibinfo {year} {2020})}\BibitemShut {NoStop}%
\bibitem [{\citenamefont {Gen}\ \emph {et~al.}(2020)\citenamefont {Gen},
  \citenamefont {Okamoto}, \citenamefont {Mori}, \citenamefont {Takenaka},\
  and\ \citenamefont
  {Kohama}}]{GenMasakiandOkamotoYoshihikoandMoriMasakiandTakenakaKoshiandKohama2020MagnetizationT}%
  \BibitemOpen
  \bibfield  {author} {\bibinfo {author} {\bibfnamefont {M.}~\bibnamefont
  {Gen}}, \bibinfo {author} {\bibfnamefont {Y.}~\bibnamefont {Okamoto}},
  \bibinfo {author} {\bibfnamefont {M.}~\bibnamefont {Mori}}, \bibinfo {author}
  {\bibfnamefont {K.}~\bibnamefont {Takenaka}},\ and\ \bibinfo {author}
  {\bibfnamefont {Y.}~\bibnamefont {Kohama}},\ }\bibfield  {title} {\bibinfo
  {title} {{Magnetization process of the breathing pyrochlore magnet
  \ch{CuInCr4S8} in ultra-high magnetic fields up to 150 T}},\ }\href
  {https://doi.org/10.1103/PhysRevB.101.054434} {\bibfield  {journal} {\bibinfo
   {journal} {Phys. Rev. B}\ }\textbf {\bibinfo {volume} {101}},\ \bibinfo
  {pages} {054434} (\bibinfo {year} {2020})}\BibitemShut {NoStop}%
\bibitem [{\citenamefont {Haku}\ \emph
  {et~al.}(2016{\natexlab{a}})\citenamefont {Haku}, \citenamefont {Soda},
  \citenamefont {Sera}, \citenamefont {Kimura}, \citenamefont {Itoh},
  \citenamefont {Yokoo},\ and\ \citenamefont
  {Masuda}}]{Haku2016CrystalBa3Yb2Zn5O11}%
  \BibitemOpen
  \bibfield  {author} {\bibinfo {author} {\bibfnamefont {T.}~\bibnamefont
  {Haku}}, \bibinfo {author} {\bibfnamefont {M.}~\bibnamefont {Soda}}, \bibinfo
  {author} {\bibfnamefont {M.}~\bibnamefont {Sera}}, \bibinfo {author}
  {\bibfnamefont {K.}~\bibnamefont {Kimura}}, \bibinfo {author} {\bibfnamefont
  {S.}~\bibnamefont {Itoh}}, \bibinfo {author} {\bibfnamefont {T.}~\bibnamefont
  {Yokoo}},\ and\ \bibinfo {author} {\bibfnamefont {T.}~\bibnamefont
  {Masuda}},\ }\bibfield  {title} {\bibinfo {title} {{Crystal field excitations
  in the breathing pyrochlore antiferromagnet \ch{Ba3Yb2Zn5O11}}},\ }\href
  {https://doi.org/10.7566/JPSJ.85.034721} {\bibfield  {journal} {\bibinfo
  {journal} {Journal of the Physical Society of Japan}\ }\textbf {\bibinfo
  {volume} {85}},\ \bibinfo {pages} {1} (\bibinfo {year}
  {2016}{\natexlab{a}})}\BibitemShut {NoStop}%
\bibitem [{\citenamefont {Rau}\ and\ \citenamefont
  {Gingras}(2018)}]{Rau2018FrustrationOctahedra}%
  \BibitemOpen
  \bibfield  {author} {\bibinfo {author} {\bibfnamefont {J.~G.}\ \bibnamefont
  {Rau}}\ and\ \bibinfo {author} {\bibfnamefont {M.~J.~P.}\ \bibnamefont
  {Gingras}},\ }\bibfield  {title} {\bibinfo {title} {Frustration and
  anisotropic exchange in ytterbium magnets with edge-shared octahedra},\
  }\href {https://doi.org/10.1103/PhysRevB.98.054408} {\bibfield  {journal}
  {\bibinfo  {journal} {Phys. Rev. B}\ }\textbf {\bibinfo {volume} {98}},\
  \bibinfo {pages} {054408} (\bibinfo {year} {2018})}\BibitemShut {NoStop}%
\bibitem [{\citenamefont {Iwahara}\ \emph {et~al.}(2022)\citenamefont
  {Iwahara}, \citenamefont {Huang}, \citenamefont {Neefjes},\ and\
  \citenamefont {Chibotaru}}]{Chibotaru}%
  \BibitemOpen
  \bibfield  {author} {\bibinfo {author} {\bibfnamefont {N.}~\bibnamefont
  {Iwahara}}, \bibinfo {author} {\bibfnamefont {Z.}~\bibnamefont {Huang}},
  \bibinfo {author} {\bibfnamefont {I.}~\bibnamefont {Neefjes}},\ and\ \bibinfo
  {author} {\bibfnamefont {L.~F.}\ \bibnamefont {Chibotaru}},\ }\bibfield
  {title} {\bibinfo {title} {Multipolar exchange interaction and complex order
  in insulating lanthanides},\ }\href
  {https://doi.org/10.1103/PhysRevB.105.144401} {\bibfield  {journal} {\bibinfo
   {journal} {Phys. Rev. B}\ }\textbf {\bibinfo {volume} {105}},\ \bibinfo
  {pages} {144401} (\bibinfo {year} {2022})}\BibitemShut {NoStop}%
\bibitem [{\citenamefont {Rau}\ \emph {et~al.}(2018)\citenamefont {Rau},
  \citenamefont {Wu}, \citenamefont {May}, \citenamefont {Taylor},
  \citenamefont {Liu}, \citenamefont {Higgins}, \citenamefont {Butch},
  \citenamefont {Ross}, \citenamefont {Nair}, \citenamefont {Lumsden},
  \citenamefont {Gingras},\ and\ \citenamefont
  {Christianson}}]{Rau2018BehaviorField}%
  \BibitemOpen
  \bibfield  {author} {\bibinfo {author} {\bibfnamefont {J.~G.}\ \bibnamefont
  {Rau}}, \bibinfo {author} {\bibfnamefont {L.~S.}\ \bibnamefont {Wu}},
  \bibinfo {author} {\bibfnamefont {A.~F.}\ \bibnamefont {May}}, \bibinfo
  {author} {\bibfnamefont {A.~E.}\ \bibnamefont {Taylor}}, \bibinfo {author}
  {\bibfnamefont {I.~L.}\ \bibnamefont {Liu}}, \bibinfo {author} {\bibfnamefont
  {J.}~\bibnamefont {Higgins}}, \bibinfo {author} {\bibfnamefont {N.~P.}\
  \bibnamefont {Butch}}, \bibinfo {author} {\bibfnamefont {K.~A.}\ \bibnamefont
  {Ross}}, \bibinfo {author} {\bibfnamefont {H.~S.}\ \bibnamefont {Nair}},
  \bibinfo {author} {\bibfnamefont {M.~D.}\ \bibnamefont {Lumsden}}, \bibinfo
  {author} {\bibfnamefont {M.~J.}\ \bibnamefont {Gingras}},\ and\ \bibinfo
  {author} {\bibfnamefont {A.~D.}\ \bibnamefont {Christianson}},\ }\bibfield
  {title} {\bibinfo {title} {{Behavior of the breathing pyrochlore lattice
  \ch{Ba3Yb2Zn5O11} in applied magnetic field}},\ }\href
  {https://doi.org/10.1088/1361-648X/aae45a} {\bibfield  {journal} {\bibinfo
  {journal} {Journal of Physics -- Condensed Matter}\ }\textbf {\bibinfo
  {volume} {30}},\ \bibinfo {pages} {455801} (\bibinfo {year}
  {2018})}\BibitemShut {NoStop}%
\bibitem [{\citenamefont {Kimura}\ \emph {et~al.}(2014)\citenamefont {Kimura},
  \citenamefont {Nakatsuji},\ and\ \citenamefont
  {Kimura}}]{Kimura2014ExperimentalAntiferromagnet}%
  \BibitemOpen
  \bibfield  {author} {\bibinfo {author} {\bibfnamefont {K.}~\bibnamefont
  {Kimura}}, \bibinfo {author} {\bibfnamefont {S.}~\bibnamefont {Nakatsuji}},\
  and\ \bibinfo {author} {\bibfnamefont {T.}~\bibnamefont {Kimura}},\
  }\bibfield  {title} {\bibinfo {title} {Experimental realization of a quantum
  breathing pyrochlore antiferromagnet},\ }\href
  {https://doi.org/10.1103/PhysRevB.90.060414} {\bibfield  {journal} {\bibinfo
  {journal} {Phys. Rev. B}\ }\textbf {\bibinfo {volume} {90}},\ \bibinfo
  {pages} {060414} (\bibinfo {year} {2014})}\BibitemShut {NoStop}%
\bibitem [{\citenamefont {Rau}\ \emph {et~al.}(2016)\citenamefont {Rau},
  \citenamefont {Wu}, \citenamefont {May}, \citenamefont {Poudel},
  \citenamefont {Winn}, \citenamefont {Garlea}, \citenamefont {Huq},
  \citenamefont {Whitfield}, \citenamefont {Taylor}, \citenamefont {Lumsden},
  \citenamefont {Gingras},\ and\ \citenamefont
  {Christianson}}]{rau2016anisotropic}%
  \BibitemOpen
  \bibfield  {author} {\bibinfo {author} {\bibfnamefont {J.~G.}\ \bibnamefont
  {Rau}}, \bibinfo {author} {\bibfnamefont {L.~S.}\ \bibnamefont {Wu}},
  \bibinfo {author} {\bibfnamefont {A.~F.}\ \bibnamefont {May}}, \bibinfo
  {author} {\bibfnamefont {L.}~\bibnamefont {Poudel}}, \bibinfo {author}
  {\bibfnamefont {B.}~\bibnamefont {Winn}}, \bibinfo {author} {\bibfnamefont
  {V.~O.}\ \bibnamefont {Garlea}}, \bibinfo {author} {\bibfnamefont
  {A.}~\bibnamefont {Huq}}, \bibinfo {author} {\bibfnamefont {P.}~\bibnamefont
  {Whitfield}}, \bibinfo {author} {\bibfnamefont {A.~E.}\ \bibnamefont
  {Taylor}}, \bibinfo {author} {\bibfnamefont {M.~D.}\ \bibnamefont {Lumsden}},
  \bibinfo {author} {\bibfnamefont {M.~J.~P.}\ \bibnamefont {Gingras}},\ and\
  \bibinfo {author} {\bibfnamefont {A.~D.}\ \bibnamefont {Christianson}},\
  }\bibfield  {title} {\bibinfo {title} {Anisotropic exchange within decoupled
  tetrahedra in the quantum breathing pyrochlore \ch{Ba3Yb2Zn5O11}},\ }\href
  {https://doi.org/10.1103/PhysRevLett.116.257204} {\bibfield  {journal}
  {\bibinfo  {journal} {Phys. Rev. Lett.}\ }\textbf {\bibinfo {volume} {116}},\
  \bibinfo {pages} {257204} (\bibinfo {year} {2016})}\BibitemShut {NoStop}%
\bibitem [{\citenamefont {Haku}\ \emph
  {et~al.}(2016{\natexlab{b}})\citenamefont {Haku}, \citenamefont {Kimura},
  \citenamefont {Matsumoto}, \citenamefont {Soda}, \citenamefont {Sera},
  \citenamefont {Yu}, \citenamefont {Mole}, \citenamefont {Takeuchi},
  \citenamefont {Nakatsuji}, \citenamefont {Kono} \emph
  {et~al.}}]{haku2016low}%
  \BibitemOpen
  \bibfield  {author} {\bibinfo {author} {\bibfnamefont {T.}~\bibnamefont
  {Haku}}, \bibinfo {author} {\bibfnamefont {K.}~\bibnamefont {Kimura}},
  \bibinfo {author} {\bibfnamefont {Y.}~\bibnamefont {Matsumoto}}, \bibinfo
  {author} {\bibfnamefont {M.}~\bibnamefont {Soda}}, \bibinfo {author}
  {\bibfnamefont {M.}~\bibnamefont {Sera}}, \bibinfo {author} {\bibfnamefont
  {D.}~\bibnamefont {Yu}}, \bibinfo {author} {\bibfnamefont {R.}~\bibnamefont
  {Mole}}, \bibinfo {author} {\bibfnamefont {T.}~\bibnamefont {Takeuchi}},
  \bibinfo {author} {\bibfnamefont {S.}~\bibnamefont {Nakatsuji}}, \bibinfo
  {author} {\bibfnamefont {Y.}~\bibnamefont {Kono}}, \emph {et~al.},\
  }\bibfield  {title} {\bibinfo {title} {Low-energy excitations and
  ground-state selection in the quantum breathing pyrochlore antiferromagnet
  \ch{Ba3Yb2Zn5O11}},\ }\href {https://doi.org/10.1103/PhysRevB.93.220407}
  {\bibfield  {journal} {\bibinfo  {journal} {Physical Review B}\ }\textbf
  {\bibinfo {volume} {93}},\ \bibinfo {pages} {220407} (\bibinfo {year}
  {2016}{\natexlab{b}})}\BibitemShut {NoStop}%
\bibitem [{\citenamefont {Haku}\ \emph {et~al.}(2017)\citenamefont {Haku},
  \citenamefont {Soda}, \citenamefont {Sera}, \citenamefont {Kimura},
  \citenamefont {Taylor}, \citenamefont {Itoh}, \citenamefont {Yokoo},
  \citenamefont {Matsumoto}, \citenamefont {Yu}, \citenamefont {Mole},
  \citenamefont {Takeuchi}, \citenamefont {Nakatsuji}, \citenamefont {Kono},
  \citenamefont {Sakakibara}, \citenamefont {Chang},\ and\ \citenamefont
  {Masuda}}]{haku2017neutron}%
  \BibitemOpen
  \bibfield  {author} {\bibinfo {author} {\bibfnamefont {T.}~\bibnamefont
  {Haku}}, \bibinfo {author} {\bibfnamefont {M.}~\bibnamefont {Soda}}, \bibinfo
  {author} {\bibfnamefont {M.}~\bibnamefont {Sera}}, \bibinfo {author}
  {\bibfnamefont {K.}~\bibnamefont {Kimura}}, \bibinfo {author} {\bibfnamefont
  {J.}~\bibnamefont {Taylor}}, \bibinfo {author} {\bibfnamefont
  {S.}~\bibnamefont {Itoh}}, \bibinfo {author} {\bibfnamefont {T.}~\bibnamefont
  {Yokoo}}, \bibinfo {author} {\bibfnamefont {Y.}~\bibnamefont {Matsumoto}},
  \bibinfo {author} {\bibfnamefont {D.}~\bibnamefont {Yu}}, \bibinfo {author}
  {\bibfnamefont {R.~A.}\ \bibnamefont {Mole}}, \bibinfo {author}
  {\bibfnamefont {T.}~\bibnamefont {Takeuchi}}, \bibinfo {author}
  {\bibfnamefont {S.}~\bibnamefont {Nakatsuji}}, \bibinfo {author}
  {\bibfnamefont {Y.}~\bibnamefont {Kono}}, \bibinfo {author} {\bibfnamefont
  {T.}~\bibnamefont {Sakakibara}}, \bibinfo {author} {\bibfnamefont {L.-J.}\
  \bibnamefont {Chang}},\ and\ \bibinfo {author} {\bibfnamefont
  {T.}~\bibnamefont {Masuda}},\ }\bibfield  {title} {\bibinfo {title} {Neutron
  scattering study in breathing pyrochlore antiferromagnet \ch{Ba3Yb2Zn5O11}},\
  }\href {https://doi.org/10.1088/1742-6596/828/1/012018} {\bibfield  {journal}
  {\bibinfo  {journal} {Journal of Physics: Conference Series}\ }\textbf
  {\bibinfo {volume} {828}},\ \bibinfo {pages} {012018} (\bibinfo {year}
  {2017})}\BibitemShut {NoStop}%
\bibitem [{\citenamefont {Dissanayake}\ \emph {et~al.}(2022)\citenamefont
  {Dissanayake}, \citenamefont {Shi}, \citenamefont {Rau}, \citenamefont {Bag},
  \citenamefont {Steinhardt}, \citenamefont {Butch}, \citenamefont {Frontzek},
  \citenamefont {Podlesnyak}, \citenamefont {Graf}, \citenamefont {Marjerrison}
  \emph {et~al.}}]{Dissanayake2021TowardsStudy}%
  \BibitemOpen
  \bibfield  {author} {\bibinfo {author} {\bibfnamefont {S.}~\bibnamefont
  {Dissanayake}}, \bibinfo {author} {\bibfnamefont {Z.}~\bibnamefont {Shi}},
  \bibinfo {author} {\bibfnamefont {J.~G.}\ \bibnamefont {Rau}}, \bibinfo
  {author} {\bibfnamefont {R.}~\bibnamefont {Bag}}, \bibinfo {author}
  {\bibfnamefont {W.}~\bibnamefont {Steinhardt}}, \bibinfo {author}
  {\bibfnamefont {N.~P.}\ \bibnamefont {Butch}}, \bibinfo {author}
  {\bibfnamefont {M.}~\bibnamefont {Frontzek}}, \bibinfo {author}
  {\bibfnamefont {A.}~\bibnamefont {Podlesnyak}}, \bibinfo {author}
  {\bibfnamefont {D.}~\bibnamefont {Graf}}, \bibinfo {author} {\bibfnamefont
  {C.}~\bibnamefont {Marjerrison}}, \emph {et~al.},\ }\bibfield  {title}
  {\bibinfo {title} {Towards understanding the magnetic properties of the
  breathing pyrochlore compound \ch{Ba3Yb2Zn5O11} through single-crystal
  studies},\ }\href
  {https://doi.org/https://doi.org/10.1038/s41535-022-00488-w} {\bibfield
  {journal} {\bibinfo  {journal} {npj Quantum Materials}\ }\textbf {\bibinfo
  {volume} {7}},\ \bibinfo {pages} {77} (\bibinfo {year} {2022})}\BibitemShut
  {NoStop}%
\bibitem [{\citenamefont {Bag}\ \emph {et~al.}(2023)\citenamefont {Bag},
  \citenamefont {Dissanayake}, \citenamefont {Yan}, \citenamefont {Shi},
  \citenamefont {Graf}, \citenamefont {Choi}, \citenamefont {Marjerrison},
  \citenamefont {Lang}, \citenamefont {Lancaster}, \citenamefont {Qiu} \emph
  {et~al.}}]{bag2023beyond}%
  \BibitemOpen
  \bibfield  {author} {\bibinfo {author} {\bibfnamefont {R.}~\bibnamefont
  {Bag}}, \bibinfo {author} {\bibfnamefont {S.~E.}\ \bibnamefont
  {Dissanayake}}, \bibinfo {author} {\bibfnamefont {H.}~\bibnamefont {Yan}},
  \bibinfo {author} {\bibfnamefont {Z.}~\bibnamefont {Shi}}, \bibinfo {author}
  {\bibfnamefont {D.}~\bibnamefont {Graf}}, \bibinfo {author} {\bibfnamefont
  {E.~S.}\ \bibnamefont {Choi}}, \bibinfo {author} {\bibfnamefont
  {C.}~\bibnamefont {Marjerrison}}, \bibinfo {author} {\bibfnamefont
  {F.}~\bibnamefont {Lang}}, \bibinfo {author} {\bibfnamefont {T.}~\bibnamefont
  {Lancaster}}, \bibinfo {author} {\bibfnamefont {Y.}~\bibnamefont {Qiu}},
  \emph {et~al.},\ }\bibfield  {title} {\bibinfo {title} {Beyond single
  tetrahedron physics of the breathing pyrochlore compound \ch{Ba3Yb2Zn5O11}},\
  }\href {https://doi.org/10.1103/PhysRevB.107.L140408} {\bibfield  {journal}
  {\bibinfo  {journal} {Physical Review B}\ }\textbf {\bibinfo {volume}
  {107}},\ \bibinfo {pages} {L140408} (\bibinfo {year} {2023})}\BibitemShut
  {NoStop}%
\bibitem [{\citenamefont {Talanov}\ and\ \citenamefont
  {Talanov}(2020)}]{Talanov2020FormationAspects}%
  \BibitemOpen
  \bibfield  {author} {\bibinfo {author} {\bibfnamefont {M.~V.}\ \bibnamefont
  {Talanov}}\ and\ \bibinfo {author} {\bibfnamefont {V.~M.}\ \bibnamefont
  {Talanov}},\ }\bibfield  {title} {\bibinfo {title} {{Formation of breathing
  pyrochlore lattices: Structural, thermodynamic and crystal chemical
  aspects}},\ }\href {https://doi.org/10.1039/c9ce01635j} {\bibfield  {journal}
  {\bibinfo  {journal} {CrystEngComm}\ }\textbf {\bibinfo {volume} {22}},\
  \bibinfo {pages} {1176} (\bibinfo {year} {2020})}\BibitemShut {NoStop}%
\bibitem [{\citenamefont {Rodríguez-Carvajal}(1993)}]{Fulprof}%
  \BibitemOpen
  \bibfield  {author} {\bibinfo {author} {\bibfnamefont {J.}~\bibnamefont
  {Rodríguez-Carvajal}},\ }\bibfield  {title} {\bibinfo {title} {Recent
  advances in magnetic structure determination by neutron powder diffraction},\
  }\href {https://doi.org/https://doi.org/10.1016/0921-4526(93)90108-I}
  {\bibfield  {journal} {\bibinfo  {journal} {Physica B: Condensed Matter}\
  }\textbf {\bibinfo {volume} {192}},\ \bibinfo {pages} {55} (\bibinfo {year}
  {1993})}\BibitemShut {NoStop}%
\bibitem [{\citenamefont {Palatinus}\ and\ \citenamefont
  {Chapuis}(2007)}]{superflip}%
  \BibitemOpen
  \bibfield  {author} {\bibinfo {author} {\bibfnamefont {L.}~\bibnamefont
  {Palatinus}}\ and\ \bibinfo {author} {\bibfnamefont {G.}~\bibnamefont
  {Chapuis}},\ }\bibfield  {title} {\bibinfo {title} {{{\it SUPERFLIP} {--} a
  computer program for the solution of crystal structures by charge flipping in
  arbitrary dimensions}},\ }\href {https://doi.org/10.1107/S0021889807029238}
  {\bibfield  {journal} {\bibinfo  {journal} {Journal of Applied
  Crystallography}\ }\textbf {\bibinfo {volume} {40}},\ \bibinfo {pages} {786}
  (\bibinfo {year} {2007})}\BibitemShut {NoStop}%
\bibitem [{\citenamefont {Betteridge}\ \emph {et~al.}(2003)\citenamefont
  {Betteridge}, \citenamefont {Carruthers}, \citenamefont {Cooper},
  \citenamefont {Prout},\ and\ \citenamefont {Watkin}}]{crystals}%
  \BibitemOpen
  \bibfield  {author} {\bibinfo {author} {\bibfnamefont {P.~W.}\ \bibnamefont
  {Betteridge}}, \bibinfo {author} {\bibfnamefont {J.~R.}\ \bibnamefont
  {Carruthers}}, \bibinfo {author} {\bibfnamefont {R.~I.}\ \bibnamefont
  {Cooper}}, \bibinfo {author} {\bibfnamefont {K.}~\bibnamefont {Prout}},\ and\
  \bibinfo {author} {\bibfnamefont {D.~J.}\ \bibnamefont {Watkin}},\ }\bibfield
   {title} {\bibinfo {title} {{{\it CRYSTALS} version 12: software for guided
  crystal structure analysis}},\ }\href
  {https://doi.org/10.1107/S0021889803021800} {\bibfield  {journal} {\bibinfo
  {journal} {Journal of Applied Crystallography}\ }\textbf {\bibinfo {volume}
  {36}},\ \bibinfo {pages} {1487} (\bibinfo {year} {2003})}\BibitemShut
  {NoStop}%
\bibitem [{\citenamefont {Clover}\ and\ \citenamefont
  {Wolf}(1970)}]{Clover1970}%
  \BibitemOpen
  \bibfield  {author} {\bibinfo {author} {\bibfnamefont {R.~B.}\ \bibnamefont
  {Clover}}\ and\ \bibinfo {author} {\bibfnamefont {W.~P.}\ \bibnamefont
  {Wolf}},\ }\bibfield  {title} {\bibinfo {title} {{Magnetic Susceptibility
  Measurements with a Tunnel Diode Oscillator}},\ }\href
  {https://doi.org/10.1063/1.1684598} {\bibfield  {journal} {\bibinfo
  {journal} {Review of Scientific Instruments}\ }\textbf {\bibinfo {volume}
  {41}},\ \bibinfo {pages} {617} (\bibinfo {year} {1970})}\BibitemShut
  {NoStop}%
\bibitem [{\citenamefont {Shi}\ \emph {et~al.}(2019)\citenamefont {Shi},
  \citenamefont {Steinhardt}, \citenamefont {Graf}, \citenamefont {Corboz},
  \citenamefont {Weickert}, \citenamefont {Harrison}, \citenamefont {Jaime},
  \citenamefont {Marjerrison}, \citenamefont {Dabkowska}, \citenamefont
  {Mila},\ and\ \citenamefont {Haravifard}}]{Shi2019}%
  \BibitemOpen
  \bibfield  {author} {\bibinfo {author} {\bibfnamefont {Z.}~\bibnamefont
  {Shi}}, \bibinfo {author} {\bibfnamefont {W.}~\bibnamefont {Steinhardt}},
  \bibinfo {author} {\bibfnamefont {D.}~\bibnamefont {Graf}}, \bibinfo {author}
  {\bibfnamefont {P.}~\bibnamefont {Corboz}}, \bibinfo {author} {\bibfnamefont
  {F.}~\bibnamefont {Weickert}}, \bibinfo {author} {\bibfnamefont
  {N.}~\bibnamefont {Harrison}}, \bibinfo {author} {\bibfnamefont
  {M.}~\bibnamefont {Jaime}}, \bibinfo {author} {\bibfnamefont
  {C.}~\bibnamefont {Marjerrison}}, \bibinfo {author} {\bibfnamefont {H.~A.}\
  \bibnamefont {Dabkowska}}, \bibinfo {author} {\bibfnamefont {F.}~\bibnamefont
  {Mila}},\ and\ \bibinfo {author} {\bibfnamefont {S.}~\bibnamefont
  {Haravifard}},\ }\bibfield  {title} {\bibinfo {title} {{Emergent bound states
  and impurity pairs in chemically doped Shastry-Sutherland system}},\ }\href
  {https://doi.org/10.1038/s41467-019-10410-x} {\bibfield  {journal} {\bibinfo
  {journal} {Nature Communications}\ }\textbf {\bibinfo {volume} {10}},\
  \bibinfo {pages} {2439} (\bibinfo {year} {2019})}\BibitemShut {NoStop}%
\bibitem [{\citenamefont {Granroth}\ \emph {et~al.}(2010)\citenamefont
  {Granroth}, \citenamefont {Kolesnikov}, \citenamefont {Sherline},
  \citenamefont {Clancy}, \citenamefont {Ross}, \citenamefont {Ruff},
  \citenamefont {Gaulin},\ and\ \citenamefont {Nagler}}]{granroth2010sequoia}%
  \BibitemOpen
  \bibfield  {author} {\bibinfo {author} {\bibfnamefont {G.~E.}\ \bibnamefont
  {Granroth}}, \bibinfo {author} {\bibfnamefont {A.~I.}\ \bibnamefont
  {Kolesnikov}}, \bibinfo {author} {\bibfnamefont {T.~E.}\ \bibnamefont
  {Sherline}}, \bibinfo {author} {\bibfnamefont {J.~P.}\ \bibnamefont
  {Clancy}}, \bibinfo {author} {\bibfnamefont {K.~A.}\ \bibnamefont {Ross}},
  \bibinfo {author} {\bibfnamefont {J.~P.~C.}\ \bibnamefont {Ruff}}, \bibinfo
  {author} {\bibfnamefont {B.~D.}\ \bibnamefont {Gaulin}},\ and\ \bibinfo
  {author} {\bibfnamefont {S.~E.}\ \bibnamefont {Nagler}},\ }\bibfield  {title}
  {\bibinfo {title} {{SEQUOIA: A Newly Operating Chopper Spectrometer at the
  {SNS}}},\ }\href {https://doi.org/10.1088/1742-6596/251/1/012058} {\bibfield
  {journal} {\bibinfo  {journal} {Journal of Physics: Conference Series}\
  }\textbf {\bibinfo {volume} {251}},\ \bibinfo {pages} {012058} (\bibinfo
  {year} {2010})}\BibitemShut {NoStop}%
\bibitem [{\citenamefont {Azuah}\ \emph {et~al.}(2009)\citenamefont {Azuah},
  \citenamefont {Kneller}, \citenamefont {Qiu}, \citenamefont
  {Tregenna-Piggott}, \citenamefont {Brown}, \citenamefont {Copley},\ and\
  \citenamefont {Dimeo}}]{azuah2009dave}%
  \BibitemOpen
  \bibfield  {author} {\bibinfo {author} {\bibfnamefont {R.~T.}\ \bibnamefont
  {Azuah}}, \bibinfo {author} {\bibfnamefont {L.~R.}\ \bibnamefont {Kneller}},
  \bibinfo {author} {\bibfnamefont {Y.}~\bibnamefont {Qiu}}, \bibinfo {author}
  {\bibfnamefont {P.~L.}\ \bibnamefont {Tregenna-Piggott}}, \bibinfo {author}
  {\bibfnamefont {C.~M.}\ \bibnamefont {Brown}}, \bibinfo {author}
  {\bibfnamefont {J.~R.}\ \bibnamefont {Copley}},\ and\ \bibinfo {author}
  {\bibfnamefont {R.~M.}\ \bibnamefont {Dimeo}},\ }\bibfield  {title} {\bibinfo
  {title} {Dave: a comprehensive software suite for the reduction,
  visualization, and analysis of low energy neutron spectroscopic data},\
  }\href {https://doi.org/10.6028/jres.114.025} {\bibfield  {journal} {\bibinfo
   {journal} {Journal of research of the National Institute of Standards and
  Technology}\ }\textbf {\bibinfo {volume} {114}},\ \bibinfo {pages} {341}
  (\bibinfo {year} {2009})}\BibitemShut {NoStop}%
\bibitem [{\citenamefont {Scheie}(2021)}]{scheie2021pycrystalfield}%
  \BibitemOpen
  \bibfield  {author} {\bibinfo {author} {\bibfnamefont {A.}~\bibnamefont
  {Scheie}},\ }\bibfield  {title} {\bibinfo {title} {{PyCrystalField: software
  for calculation, analysis and fitting of crystal electric field
  Hamiltonians}},\ }\href {https://doi.org/10.1107/S160057672001554X}
  {\bibfield  {journal} {\bibinfo  {journal} {Journal of Applied
  Crystallography}\ }\textbf {\bibinfo {volume} {54}},\ \bibinfo {pages} {356}
  (\bibinfo {year} {2021})}\BibitemShut {NoStop}%
\bibitem [{\citenamefont {Rotter}(2004)}]{rotter2004using}%
  \BibitemOpen
  \bibfield  {author} {\bibinfo {author} {\bibfnamefont {M.}~\bibnamefont
  {Rotter}},\ }\bibfield  {title} {\bibinfo {title} {{Using McPhase to
  calculate magnetic phase diagrams of rare earth compounds}},\ }\href
  {https://doi.org/https://doi.org/10.1016/j.jmmm.2003.12.1394} {\bibfield
  {journal} {\bibinfo  {journal} {Journal of Magnetism and Magnetic Materials}\
  }\textbf {\bibinfo {volume} {272-276}},\ \bibinfo {pages} {E481} (\bibinfo
  {year} {2004})}\BibitemShut {NoStop}%
\bibitem [{\citenamefont {Ehlers}\ \emph {et~al.}(2011)\citenamefont {Ehlers},
  \citenamefont {Podlesnyak}, \citenamefont {Niedziela}, \citenamefont
  {Iverson},\ and\ \citenamefont {Sokol}}]{ehlers2011new}%
  \BibitemOpen
  \bibfield  {author} {\bibinfo {author} {\bibfnamefont {G.}~\bibnamefont
  {Ehlers}}, \bibinfo {author} {\bibfnamefont {A.~A.}\ \bibnamefont
  {Podlesnyak}}, \bibinfo {author} {\bibfnamefont {J.~L.}\ \bibnamefont
  {Niedziela}}, \bibinfo {author} {\bibfnamefont {E.~B.}\ \bibnamefont
  {Iverson}},\ and\ \bibinfo {author} {\bibfnamefont {P.~E.}\ \bibnamefont
  {Sokol}},\ }\bibfield  {title} {\bibinfo {title} {{The new cold neutron
  chopper spectrometer at the Spallation Neutron Source: Design and
  performance}},\ }\href {https://doi.org/10.1063/1.3626935} {\bibfield
  {journal} {\bibinfo  {journal} {Review of Scientific Instruments}\ }\textbf
  {\bibinfo {volume} {82}},\ \bibinfo {pages} {085108} (\bibinfo {year}
  {2011})}\BibitemShut {NoStop}%
\bibitem [{\citenamefont {Neuefeind}\ \emph {et~al.}(2012)\citenamefont
  {Neuefeind}, \citenamefont {Feygenson}, \citenamefont {Carruth},
  \citenamefont {Hoffmann},\ and\ \citenamefont {Chipley}}]{NEUEFEIND201268}%
  \BibitemOpen
  \bibfield  {author} {\bibinfo {author} {\bibfnamefont {J.}~\bibnamefont
  {Neuefeind}}, \bibinfo {author} {\bibfnamefont {M.}~\bibnamefont
  {Feygenson}}, \bibinfo {author} {\bibfnamefont {J.}~\bibnamefont {Carruth}},
  \bibinfo {author} {\bibfnamefont {R.}~\bibnamefont {Hoffmann}},\ and\
  \bibinfo {author} {\bibfnamefont {K.~K.}\ \bibnamefont {Chipley}},\
  }\bibfield  {title} {\bibinfo {title} {{The Nanoscale Ordered MAterials
  Diffractometer NOMAD at the Spallation Neutron Source SNS}},\ }\href
  {https://doi.org/https://doi.org/10.1016/j.nimb.2012.05.037} {\bibfield
  {journal} {\bibinfo  {journal} {Nuclear Instruments and Methods in Physics
  Research Section B: Beam Interactions with Materials and Atoms}\ }\textbf
  {\bibinfo {volume} {287}},\ \bibinfo {pages} {68} (\bibinfo {year}
  {2012})}\BibitemShut {NoStop}%
\bibitem [{\citenamefont {Farrow}\ \emph {et~al.}(2007)\citenamefont {Farrow},
  \citenamefont {Juhas}, \citenamefont {Liu}, \citenamefont {Bryndin},
  \citenamefont {Božin}, \citenamefont {Bloch}, \citenamefont {Proffen},\ and\
  \citenamefont {Billinge}}]{Farrow_2007}%
  \BibitemOpen
  \bibfield  {author} {\bibinfo {author} {\bibfnamefont {C.~L.}\ \bibnamefont
  {Farrow}}, \bibinfo {author} {\bibfnamefont {P.}~\bibnamefont {Juhas}},
  \bibinfo {author} {\bibfnamefont {J.~W.}\ \bibnamefont {Liu}}, \bibinfo
  {author} {\bibfnamefont {D.}~\bibnamefont {Bryndin}}, \bibinfo {author}
  {\bibfnamefont {E.~S.}\ \bibnamefont {Božin}}, \bibinfo {author}
  {\bibfnamefont {J.}~\bibnamefont {Bloch}}, \bibinfo {author} {\bibfnamefont
  {T.}~\bibnamefont {Proffen}},\ and\ \bibinfo {author} {\bibfnamefont
  {S.~J.~L.}\ \bibnamefont {Billinge}},\ }\bibfield  {title} {\bibinfo {title}
  {{PDFfit2 and PDFgui: computer programs for studying nanostructure in
  crystals}},\ }\href {https://doi.org/10.1088/0953-8984/19/33/335219}
  {\bibfield  {journal} {\bibinfo  {journal} {Journal of Physics: Condensed
  Matter}\ }\textbf {\bibinfo {volume} {19}},\ \bibinfo {pages} {335219}
  (\bibinfo {year} {2007})}\BibitemShut {NoStop}%
\bibitem [{\citenamefont {Gingras}\ \emph {et~al.}(2000)\citenamefont
  {Gingras}, \citenamefont {den Hertog}, \citenamefont {Faucher}, \citenamefont
  {Gardner}, \citenamefont {Dunsiger}, \citenamefont {Chang}, \citenamefont
  {Gaulin}, \citenamefont {Raju},\ and\ \citenamefont
  {Greedan}}]{Gingras-Tb2Ti2O7}%
  \BibitemOpen
  \bibfield  {author} {\bibinfo {author} {\bibfnamefont {M.~J.~P.}\
  \bibnamefont {Gingras}}, \bibinfo {author} {\bibfnamefont {B.~C.}\
  \bibnamefont {den Hertog}}, \bibinfo {author} {\bibfnamefont
  {M.}~\bibnamefont {Faucher}}, \bibinfo {author} {\bibfnamefont {J.~S.}\
  \bibnamefont {Gardner}}, \bibinfo {author} {\bibfnamefont {S.~R.}\
  \bibnamefont {Dunsiger}}, \bibinfo {author} {\bibfnamefont {L.~J.}\
  \bibnamefont {Chang}}, \bibinfo {author} {\bibfnamefont {B.~D.}\ \bibnamefont
  {Gaulin}}, \bibinfo {author} {\bibfnamefont {N.~P.}\ \bibnamefont {Raju}},\
  and\ \bibinfo {author} {\bibfnamefont {J.~E.}\ \bibnamefont {Greedan}},\
  }\bibfield  {title} {\bibinfo {title} {Thermodynamic and single-ion
  properties of {${\mathrm{Tb}}^{3+}$} within the collective paramagnetic-spin
  liquid state of the frustrated pyrochlore antiferromagnet
  {${\mathrm{Tb}}_{2}{\mathrm{Ti}}_{2}{\mathrm{O}}_{7}$}},\ }\href
  {https://doi.org/10.1103/PhysRevB.62.6496} {\bibfield  {journal} {\bibinfo
  {journal} {Phys. Rev. B}\ }\textbf {\bibinfo {volume} {62}},\ \bibinfo
  {pages} {6496} (\bibinfo {year} {2000})}\BibitemShut {NoStop}%
\bibitem [{\citenamefont {Hutchings}(1964)}]{hutchings1964point}%
  \BibitemOpen
  \bibfield  {author} {\bibinfo {author} {\bibfnamefont {M.~T.}\ \bibnamefont
  {Hutchings}},\ }\bibfield  {title} {\bibinfo {title} {Point-charge
  calculations of energy levels of magnetic ions in crystalline electric
  fields},\ }\href
  {https://doi.org/https://doi.org/10.1016/S0081-1947(08)60517-2} {\bibfield
  {journal} {\bibinfo  {journal} {Solid State Physics}\ }\textbf {\bibinfo
  {volume} {16}},\ \bibinfo {pages} {227} (\bibinfo {year} {1964})}\BibitemShut
  {NoStop}%
\bibitem [{\citenamefont {Stevens}(1952)}]{stevens1952matrix}%
  \BibitemOpen
  \bibfield  {author} {\bibinfo {author} {\bibfnamefont {K.}~\bibnamefont
  {Stevens}},\ }\bibfield  {title} {\bibinfo {title} {Matrix elements and
  operator equivalents connected with the magnetic properties of rare earth
  ions},\ }\href {https://doi.org/10.1088/0370-1298/65/3/308} {\bibfield
  {journal} {\bibinfo  {journal} {Proceedings of the Physical Society. Section
  A}\ }\textbf {\bibinfo {volume} {65}},\ \bibinfo {pages} {209} (\bibinfo
  {year} {1952})}\BibitemShut {NoStop}%
\bibitem [{\citenamefont {Zinkin}\ \emph {et~al.}(1996)\citenamefont {Zinkin},
  \citenamefont {Harris}, \citenamefont {Tun}, \citenamefont {Cowley},\ and\
  \citenamefont {Wanklyn}}]{Zinkin-Tm2Ti2O7}%
  \BibitemOpen
  \bibfield  {author} {\bibinfo {author} {\bibfnamefont {M.~P.}\ \bibnamefont
  {Zinkin}}, \bibinfo {author} {\bibfnamefont {M.~J.}\ \bibnamefont {Harris}},
  \bibinfo {author} {\bibfnamefont {Z.}~\bibnamefont {Tun}}, \bibinfo {author}
  {\bibfnamefont {R.~A.}\ \bibnamefont {Cowley}},\ and\ \bibinfo {author}
  {\bibfnamefont {B.~M.}\ \bibnamefont {Wanklyn}},\ }\bibfield  {title}
  {\bibinfo {title} {Lifting of the ground-state degeneracy by crystal-field
  interactions in the pyrochlore {Tm$_2$Ti$_2$O$_7$}},\ }\href
  {https://doi.org/10.1088/0953-8984/8/2/007} {\bibfield  {journal} {\bibinfo
  {journal} {Journal of Physics -- Condensed Matter}\ }\textbf {\bibinfo
  {volume} {8}},\ \bibinfo {pages} {193} (\bibinfo {year} {1996})}\BibitemShut
  {NoStop}%
\bibitem [{\citenamefont {Narayanamurti}\ and\ \citenamefont
  {Pohl}(1970)}]{RevModPhys.42.201}%
  \BibitemOpen
  \bibfield  {author} {\bibinfo {author} {\bibfnamefont {V.}~\bibnamefont
  {Narayanamurti}}\ and\ \bibinfo {author} {\bibfnamefont {R.~O.}\ \bibnamefont
  {Pohl}},\ }\bibfield  {title} {\bibinfo {title} {Tunneling states of defects
  in solids},\ }\href {https://doi.org/10.1103/RevModPhys.42.201} {\bibfield
  {journal} {\bibinfo  {journal} {Rev. Mod. Phys.}\ }\textbf {\bibinfo {volume}
  {42}},\ \bibinfo {pages} {201} (\bibinfo {year} {1970})}\BibitemShut
  {NoStop}%
\bibitem [{\citenamefont {Goto}\ \emph {et~al.}(2004)\citenamefont {Goto},
  \citenamefont {Nemoto}, \citenamefont {Yamaguchi}, \citenamefont {Akatsu},
  \citenamefont {Yanagisawa}, \citenamefont {Suzuki},\ and\ \citenamefont
  {Kitazawa}}]{tunnel1}%
  \BibitemOpen
  \bibfield  {author} {\bibinfo {author} {\bibfnamefont {T.}~\bibnamefont
  {Goto}}, \bibinfo {author} {\bibfnamefont {Y.}~\bibnamefont {Nemoto}},
  \bibinfo {author} {\bibfnamefont {T.}~\bibnamefont {Yamaguchi}}, \bibinfo
  {author} {\bibfnamefont {M.}~\bibnamefont {Akatsu}}, \bibinfo {author}
  {\bibfnamefont {T.}~\bibnamefont {Yanagisawa}}, \bibinfo {author}
  {\bibfnamefont {O.}~\bibnamefont {Suzuki}},\ and\ \bibinfo {author}
  {\bibfnamefont {H.}~\bibnamefont {Kitazawa}},\ }\bibfield  {title} {\bibinfo
  {title} {Tunneling and rattling in clathrate crystal},\ }\href
  {https://doi.org/10.1103/PhysRevB.70.184126} {\bibfield  {journal} {\bibinfo
  {journal} {Phys. Rev. B}\ }\textbf {\bibinfo {volume} {70}},\ \bibinfo
  {pages} {184126} (\bibinfo {year} {2004})}\BibitemShut {NoStop}%
\bibitem [{\citenamefont {Gou}\ \emph {et~al.}(2005)\citenamefont {Gou},
  \citenamefont {Li}, \citenamefont {Chi}, \citenamefont {Ross}, \citenamefont
  {Beekman},\ and\ \citenamefont {Nolas}}]{tunnel2}%
  \BibitemOpen
  \bibfield  {author} {\bibinfo {author} {\bibfnamefont {W.}~\bibnamefont
  {Gou}}, \bibinfo {author} {\bibfnamefont {Y.}~\bibnamefont {Li}}, \bibinfo
  {author} {\bibfnamefont {J.}~\bibnamefont {Chi}}, \bibinfo {author}
  {\bibfnamefont {J.~H.}\ \bibnamefont {Ross}}, \bibinfo {author}
  {\bibfnamefont {M.}~\bibnamefont {Beekman}},\ and\ \bibinfo {author}
  {\bibfnamefont {G.~S.}\ \bibnamefont {Nolas}},\ }\bibfield  {title} {\bibinfo
  {title} {{NMR study of slow atomic motion in
  ${\mathrm{Sr}}_{8}{\mathrm{Ga}}_{16}{\mathrm{Ge}}_{30}$ clathrate}},\ }\href
  {https://doi.org/10.1103/PhysRevB.71.174307} {\bibfield  {journal} {\bibinfo
  {journal} {Phys. Rev. B}\ }\textbf {\bibinfo {volume} {71}},\ \bibinfo
  {pages} {174307} (\bibinfo {year} {2005})}\BibitemShut {NoStop}%
\bibitem [{\citenamefont {Hermann}\ \emph {et~al.}(2006)\citenamefont
  {Hermann}, \citenamefont {Keppens}, \citenamefont {Bonville}, \citenamefont
  {Nolas}, \citenamefont {Grandjean}, \citenamefont {Long}, \citenamefont
  {Christen}, \citenamefont {Chakoumakos}, \citenamefont {Sales},\ and\
  \citenamefont {Mandrus}}]{tunnel3}%
  \BibitemOpen
  \bibfield  {author} {\bibinfo {author} {\bibfnamefont {R.~P.}\ \bibnamefont
  {Hermann}}, \bibinfo {author} {\bibfnamefont {V.}~\bibnamefont {Keppens}},
  \bibinfo {author} {\bibfnamefont {P.}~\bibnamefont {Bonville}}, \bibinfo
  {author} {\bibfnamefont {G.~S.}\ \bibnamefont {Nolas}}, \bibinfo {author}
  {\bibfnamefont {F.}~\bibnamefont {Grandjean}}, \bibinfo {author}
  {\bibfnamefont {G.~J.}\ \bibnamefont {Long}}, \bibinfo {author}
  {\bibfnamefont {H.~M.}\ \bibnamefont {Christen}}, \bibinfo {author}
  {\bibfnamefont {B.~C.}\ \bibnamefont {Chakoumakos}}, \bibinfo {author}
  {\bibfnamefont {B.~C.}\ \bibnamefont {Sales}},\ and\ \bibinfo {author}
  {\bibfnamefont {D.}~\bibnamefont {Mandrus}},\ }\bibfield  {title} {\bibinfo
  {title} {{Direct Experimental Evidence for Atomic Tunneling of Europium in
  Crystalline ${\mathrm{Eu}}_{8}{\mathrm{Ga}}_{16}{\mathrm{Ge}}_{30}$}},\
  }\href {https://doi.org/10.1103/PhysRevLett.97.017401} {\bibfield  {journal}
  {\bibinfo  {journal} {Phys. Rev. Lett.}\ }\textbf {\bibinfo {volume} {97}},\
  \bibinfo {pages} {017401} (\bibinfo {year} {2006})}\BibitemShut {NoStop}%
\bibitem [{\citenamefont {Sun}\ \emph {et~al.}(2020)\citenamefont {Sun},
  \citenamefont {Niu}, \citenamefont {Hermann}, \citenamefont {Moon},
  \citenamefont {Shulumba}, \citenamefont {Page}, \citenamefont {Zhao},
  \citenamefont {Thind}, \citenamefont {Mahalingam}, \citenamefont
  {Milam-Guerrero} \emph {et~al.}}]{sun2020high}%
  \BibitemOpen
  \bibfield  {author} {\bibinfo {author} {\bibfnamefont {B.}~\bibnamefont
  {Sun}}, \bibinfo {author} {\bibfnamefont {S.}~\bibnamefont {Niu}}, \bibinfo
  {author} {\bibfnamefont {R.~P.}\ \bibnamefont {Hermann}}, \bibinfo {author}
  {\bibfnamefont {J.}~\bibnamefont {Moon}}, \bibinfo {author} {\bibfnamefont
  {N.}~\bibnamefont {Shulumba}}, \bibinfo {author} {\bibfnamefont
  {K.}~\bibnamefont {Page}}, \bibinfo {author} {\bibfnamefont {B.}~\bibnamefont
  {Zhao}}, \bibinfo {author} {\bibfnamefont {A.~S.}\ \bibnamefont {Thind}},
  \bibinfo {author} {\bibfnamefont {K.}~\bibnamefont {Mahalingam}}, \bibinfo
  {author} {\bibfnamefont {J.}~\bibnamefont {Milam-Guerrero}}, \emph {et~al.},\
  }\bibfield  {title} {\bibinfo {title} {High frequency atomic tunneling yields
  ultralow and glass-like thermal conductivity in chalcogenide single
  crystals},\ }\href {https://doi.org/10.1038/s41467-020-19872-w} {\bibfield
  {journal} {\bibinfo  {journal} {Nature Communications}\ }\textbf {\bibinfo
  {volume} {11}},\ \bibinfo {pages} {6039} (\bibinfo {year}
  {2020})}\BibitemShut {NoStop}%
\bibitem [{\citenamefont {Adroja}\ \emph {et~al.}(2012)\citenamefont {Adroja},
  \citenamefont {del Moral}, \citenamefont {de~la Fuente}, \citenamefont
  {Fraile}, \citenamefont {Goremychkin}, \citenamefont {Taylor}, \citenamefont
  {Hillier},\ and\ \citenamefont {Fernandez-Alonso}}]{Adroja_2012}%
  \BibitemOpen
  \bibfield  {author} {\bibinfo {author} {\bibfnamefont {D.~T.}\ \bibnamefont
  {Adroja}}, \bibinfo {author} {\bibfnamefont {A.}~\bibnamefont {del Moral}},
  \bibinfo {author} {\bibfnamefont {C.}~\bibnamefont {de~la Fuente}}, \bibinfo
  {author} {\bibfnamefont {A.}~\bibnamefont {Fraile}}, \bibinfo {author}
  {\bibfnamefont {E.~A.}\ \bibnamefont {Goremychkin}}, \bibinfo {author}
  {\bibfnamefont {J.~W.}\ \bibnamefont {Taylor}}, \bibinfo {author}
  {\bibfnamefont {A.~D.}\ \bibnamefont {Hillier}},\ and\ \bibinfo {author}
  {\bibfnamefont {F.}~\bibnamefont {Fernandez-Alonso}},\ }\bibfield  {title}
  {\bibinfo {title} {{Vibron Quasibound State in the Noncentrosymmetric
  Tetragonal Heavy-Fermion Compound {Ce}{Cu}{Al}$_3$}},\ }\href
  {https://doi.org/10.1103/PhysRevLett.108.216402} {\bibfield  {journal}
  {\bibinfo  {journal} {Phys. Rev. Lett.}\ }\textbf {\bibinfo {volume} {108}},\
  \bibinfo {pages} {216402} (\bibinfo {year} {2012})}\BibitemShut {NoStop}%
\bibitem [{\citenamefont {Loewenhaupt}\ \emph {et~al.}(1979)\citenamefont
  {Loewenhaupt}, \citenamefont {Rainford},\ and\ \citenamefont
  {Steglich}}]{Loewenhaupt_1979}%
  \BibitemOpen
  \bibfield  {author} {\bibinfo {author} {\bibfnamefont {M.}~\bibnamefont
  {Loewenhaupt}}, \bibinfo {author} {\bibfnamefont {B.~D.}\ \bibnamefont
  {Rainford}},\ and\ \bibinfo {author} {\bibfnamefont {F.}~\bibnamefont
  {Steglich}},\ }\bibfield  {title} {\bibinfo {title} {{Dynamic Jahn-Teller
  Effect in a Rare-Earth Compound: {{Ce}{Al}$_2$} }},\ }\href
  {https://doi.org/10.1103/PhysRevLett.42.1709} {\bibfield  {journal} {\bibinfo
   {journal} {Phys. Rev. Lett.}\ }\textbf {\bibinfo {volume} {42}},\ \bibinfo
  {pages} {1709} (\bibinfo {year} {1979})}\BibitemShut {NoStop}%
\bibitem [{\citenamefont {Thalmeier}\ and\ \citenamefont
  {Fulde}(1982)}]{Thalmeier_1982}%
  \BibitemOpen
  \bibfield  {author} {\bibinfo {author} {\bibfnamefont {P.}~\bibnamefont
  {Thalmeier}}\ and\ \bibinfo {author} {\bibfnamefont {P.}~\bibnamefont
  {Fulde}},\ }\bibfield  {title} {\bibinfo {title} {{Bound State between a
  Crystal-Field Excitation and a Phonon in {{Ce}{Al}$_2$} }},\ }\href
  {https://doi.org/10.1103/PhysRevLett.49.1588} {\bibfield  {journal} {\bibinfo
   {journal} {Phys. Rev. Lett.}\ }\textbf {\bibinfo {volume} {49}},\ \bibinfo
  {pages} {1588} (\bibinfo {year} {1982})}\BibitemShut {NoStop}%
\bibitem [{\citenamefont {Thalmeier}(1984)}]{Thalmeier_1984}%
  \BibitemOpen
  \bibfield  {author} {\bibinfo {author} {\bibfnamefont {P.}~\bibnamefont
  {Thalmeier}},\ }\bibfield  {title} {\bibinfo {title} {{Theory of the bound
  state between phonons and a {CEF} excitation in {{Ce}{Al}$_2$} }},\ }\href
  {https://doi.org/10.1088/0022-3719/17/23/015} {\bibfield  {journal} {\bibinfo
   {journal} {Journal of Physics C: Solid State Physics}\ }\textbf {\bibinfo
  {volume} {17}},\ \bibinfo {pages} {4153} (\bibinfo {year}
  {1984})}\BibitemShut {NoStop}%
\bibitem [{\citenamefont {Loong}\ \emph {et~al.}(1999)\citenamefont {Loong},
  \citenamefont {Loewenhaupt}, \citenamefont {Nipko}, \citenamefont {Braden},\
  and\ \citenamefont {Boatner}}]{Loong_1999}%
  \BibitemOpen
  \bibfield  {author} {\bibinfo {author} {\bibfnamefont {C.-K.}\ \bibnamefont
  {Loong}}, \bibinfo {author} {\bibfnamefont {M.}~\bibnamefont {Loewenhaupt}},
  \bibinfo {author} {\bibfnamefont {J.~C.}\ \bibnamefont {Nipko}}, \bibinfo
  {author} {\bibfnamefont {M.}~\bibnamefont {Braden}},\ and\ \bibinfo {author}
  {\bibfnamefont {L.~A.}\ \bibnamefont {Boatner}},\ }\bibfield  {title}
  {\bibinfo {title} {{Dynamic coupling of crystal-field and phonon states in
  ${\mathrm{YbPO}}_{4}$}},\ }\href {https://doi.org/10.1103/PhysRevB.60.R12549}
  {\bibfield  {journal} {\bibinfo  {journal} {Phys. Rev. B}\ }\textbf {\bibinfo
  {volume} {60}},\ \bibinfo {pages} {R12549} (\bibinfo {year}
  {1999})}\BibitemShut {NoStop}%
\bibitem [{\citenamefont {Hinuma}\ \emph {et~al.}(2017)\citenamefont {Hinuma},
  \citenamefont {Pizzi}, \citenamefont {Kumagai}, \citenamefont {Oba},\ and\
  \citenamefont {Tanaka}}]{hinuma2017band}%
  \BibitemOpen
  \bibfield  {author} {\bibinfo {author} {\bibfnamefont {Y.}~\bibnamefont
  {Hinuma}}, \bibinfo {author} {\bibfnamefont {G.}~\bibnamefont {Pizzi}},
  \bibinfo {author} {\bibfnamefont {Y.}~\bibnamefont {Kumagai}}, \bibinfo
  {author} {\bibfnamefont {F.}~\bibnamefont {Oba}},\ and\ \bibinfo {author}
  {\bibfnamefont {I.}~\bibnamefont {Tanaka}},\ }\bibfield  {title} {\bibinfo
  {title} {Band structure diagram paths based on crystallography},\ }\href
  {https://doi.org/https://doi.org/10.1016/j.commatsci.2016.10.015} {\bibfield
  {journal} {\bibinfo  {journal} {Computational Materials Science}\ }\textbf
  {\bibinfo {volume} {128}},\ \bibinfo {pages} {140} (\bibinfo {year}
  {2017})}\BibitemShut {NoStop}%
\bibitem [{\citenamefont {Togo}\ and\ \citenamefont
  {Tanaka}(2018)}]{togo2018texttt}%
  \BibitemOpen
  \bibfield  {author} {\bibinfo {author} {\bibfnamefont {A.}~\bibnamefont
  {Togo}}\ and\ \bibinfo {author} {\bibfnamefont {I.}~\bibnamefont {Tanaka}},\
  }\bibfield  {title} {\bibinfo {title} {{Spglib}: a software library for
  crystal symmetry search},\ }\href {https://doi.org/10.48550/arXiv.1808.01590}
  {\bibfield  {journal} {\bibinfo  {journal} {arXiv preprint arXiv:1808.01590}\
  } (\bibinfo {year} {2018})}\BibitemShut {NoStop}%
\bibitem [{\citenamefont {Bertin}\ \emph {et~al.}(2012)\citenamefont {Bertin},
  \citenamefont {Chapuis}, \citenamefont {De~R{\'e}otier},\ and\ \citenamefont
  {Yaouanc}}]{bertin2012crystal}%
  \BibitemOpen
  \bibfield  {author} {\bibinfo {author} {\bibfnamefont {A.}~\bibnamefont
  {Bertin}}, \bibinfo {author} {\bibfnamefont {Y.}~\bibnamefont {Chapuis}},
  \bibinfo {author} {\bibfnamefont {P.~D.}\ \bibnamefont {De~R{\'e}otier}},\
  and\ \bibinfo {author} {\bibfnamefont {A.}~\bibnamefont {Yaouanc}},\
  }\bibfield  {title} {\bibinfo {title} {Crystal electric field in the
  \ch{R2Ti2O7} pyrochlore compounds},\ }\href
  {https://doi.org/10.1088/0953-8984/24/25/256003} {\bibfield  {journal}
  {\bibinfo  {journal} {Journal of Physics: Condensed Matter}\ }\textbf
  {\bibinfo {volume} {24}},\ \bibinfo {pages} {256003} (\bibinfo {year}
  {2012})}\BibitemShut {NoStop}%
\bibitem [{\citenamefont {Kresse}\ and\ \citenamefont
  {Furthm{\"u}ller}(1996{\natexlab{a}})}]{kresse1996efficient}%
  \BibitemOpen
  \bibfield  {author} {\bibinfo {author} {\bibfnamefont {G.}~\bibnamefont
  {Kresse}}\ and\ \bibinfo {author} {\bibfnamefont {J.}~\bibnamefont
  {Furthm{\"u}ller}},\ }\bibfield  {title} {\bibinfo {title} {Efficient
  iterative schemes for ab initio total-energy calculations using a plane-wave
  basis set},\ }\href {https://doi.org/10.1103/PhysRevB.54.11169} {\bibfield
  {journal} {\bibinfo  {journal} {Phys. Rev. B}\ }\textbf {\bibinfo {volume}
  {54}},\ \bibinfo {pages} {11169} (\bibinfo {year}
  {1996}{\natexlab{a}})}\BibitemShut {NoStop}%
\bibitem [{\citenamefont {Kresse}\ and\ \citenamefont
  {Furthm{\"u}ller}(1996{\natexlab{b}})}]{kresse1996efficiency}%
  \BibitemOpen
  \bibfield  {author} {\bibinfo {author} {\bibfnamefont {G.}~\bibnamefont
  {Kresse}}\ and\ \bibinfo {author} {\bibfnamefont {J.}~\bibnamefont
  {Furthm{\"u}ller}},\ }\bibfield  {title} {\bibinfo {title} {Efficiency of
  ab-initio total energy calculations for metals and semiconductors using a
  plane-wave basis set},\ }\href
  {https://doi.org/https://doi.org/10.1016/0927-0256(96)00008-0} {\bibfield
  {journal} {\bibinfo  {journal} {Comput. Mater. Sci.}\ }\textbf {\bibinfo
  {volume} {6}},\ \bibinfo {pages} {15} (\bibinfo {year}
  {1996}{\natexlab{b}})}\BibitemShut {NoStop}%
\bibitem [{\citenamefont {Kresse}\ and\ \citenamefont
  {Hafner}(1993)}]{kresse1993ab}%
  \BibitemOpen
  \bibfield  {author} {\bibinfo {author} {\bibfnamefont {G.}~\bibnamefont
  {Kresse}}\ and\ \bibinfo {author} {\bibfnamefont {J.}~\bibnamefont
  {Hafner}},\ }\bibfield  {title} {\bibinfo {title} {Ab initio molecular
  dynamics for liquid metals},\ }\href
  {https://doi.org/10.1103/PhysRevB.47.558} {\bibfield  {journal} {\bibinfo
  {journal} {Phys. Rev. B}\ }\textbf {\bibinfo {volume} {47}},\ \bibinfo
  {pages} {558} (\bibinfo {year} {1993})}\BibitemShut {NoStop}%
\bibitem [{\citenamefont {Kresse}\ and\ \citenamefont
  {Joubert}(1999)}]{kresse1999ultrasoft}%
  \BibitemOpen
  \bibfield  {author} {\bibinfo {author} {\bibfnamefont {G.}~\bibnamefont
  {Kresse}}\ and\ \bibinfo {author} {\bibfnamefont {D.}~\bibnamefont
  {Joubert}},\ }\bibfield  {title} {\bibinfo {title} {From ultrasoft
  pseudopotentials to the projector augmented-wave method},\ }\href
  {https://doi.org/10.1103/PhysRevB.59.1758} {\bibfield  {journal} {\bibinfo
  {journal} {Phys. Rev. B}\ }\textbf {\bibinfo {volume} {59}},\ \bibinfo
  {pages} {1758} (\bibinfo {year} {1999})}\BibitemShut {NoStop}%
\bibitem [{\citenamefont {Bl{\"o}chl}(1994)}]{blochl1994projector}%
  \BibitemOpen
  \bibfield  {author} {\bibinfo {author} {\bibfnamefont {P.~E.}\ \bibnamefont
  {Bl{\"o}chl}},\ }\bibfield  {title} {\bibinfo {title} {Projector
  augmented-wave method},\ }\href {https://doi.org/10.1103/PhysRevB.50.17953}
  {\bibfield  {journal} {\bibinfo  {journal} {Phys. Rev. B}\ }\textbf {\bibinfo
  {volume} {50}},\ \bibinfo {pages} {17953} (\bibinfo {year}
  {1994})}\BibitemShut {NoStop}%
\bibitem [{\citenamefont {Perdew}\ \emph {et~al.}(2008)\citenamefont {Perdew},
  \citenamefont {Ruzsinszky}, \citenamefont {Csonka}, \citenamefont {Vydrov},
  \citenamefont {Scuseria}, \citenamefont {Constantin}, \citenamefont {Zhou},\
  and\ \citenamefont {Burke}}]{perdew2008restoring}%
  \BibitemOpen
  \bibfield  {author} {\bibinfo {author} {\bibfnamefont {J.~P.}\ \bibnamefont
  {Perdew}}, \bibinfo {author} {\bibfnamefont {A.}~\bibnamefont {Ruzsinszky}},
  \bibinfo {author} {\bibfnamefont {G.~I.}\ \bibnamefont {Csonka}}, \bibinfo
  {author} {\bibfnamefont {O.~A.}\ \bibnamefont {Vydrov}}, \bibinfo {author}
  {\bibfnamefont {G.~E.}\ \bibnamefont {Scuseria}}, \bibinfo {author}
  {\bibfnamefont {L.~A.}\ \bibnamefont {Constantin}}, \bibinfo {author}
  {\bibfnamefont {X.}~\bibnamefont {Zhou}},\ and\ \bibinfo {author}
  {\bibfnamefont {K.}~\bibnamefont {Burke}},\ }\bibfield  {title} {\bibinfo
  {title} {Restoring the density-gradient expansion for exchange in solids and
  surfaces},\ }\href {https://doi.org/10.1103/PhysRevLett.100.136406}
  {\bibfield  {journal} {\bibinfo  {journal} {Phys. Rev. Lett.}\ }\textbf
  {\bibinfo {volume} {100}},\ \bibinfo {pages} {136406} (\bibinfo {year}
  {2008})}\BibitemShut {NoStop}%
\bibitem [{\citenamefont {Kresse}\ \emph {et~al.}(1995)\citenamefont {Kresse},
  \citenamefont {Furthm{\"u}ller},\ and\ \citenamefont
  {Hafner}}]{kresse1995ab}%
  \BibitemOpen
  \bibfield  {author} {\bibinfo {author} {\bibfnamefont {G.}~\bibnamefont
  {Kresse}}, \bibinfo {author} {\bibfnamefont {J.}~\bibnamefont
  {Furthm{\"u}ller}},\ and\ \bibinfo {author} {\bibfnamefont {J.}~\bibnamefont
  {Hafner}},\ }\bibfield  {title} {\bibinfo {title} {Ab initio force constant
  approach to phonon dispersion relations of diamond and graphite},\ }\href
  {https://doi.org/10.1209/0295-5075/32/9/005} {\bibfield  {journal} {\bibinfo
  {journal} {Europhys. Lett.}\ }\textbf {\bibinfo {volume} {32}},\ \bibinfo
  {pages} {729} (\bibinfo {year} {1995})}\BibitemShut {NoStop}%
\bibitem [{\citenamefont {Parlinski}\ \emph {et~al.}(1997)\citenamefont
  {Parlinski}, \citenamefont {Li},\ and\ \citenamefont
  {Kawazoe}}]{parlinski1997first}%
  \BibitemOpen
  \bibfield  {author} {\bibinfo {author} {\bibfnamefont {K.}~\bibnamefont
  {Parlinski}}, \bibinfo {author} {\bibfnamefont {Z.~Q.}\ \bibnamefont {Li}},\
  and\ \bibinfo {author} {\bibfnamefont {Y.}~\bibnamefont {Kawazoe}},\
  }\bibfield  {title} {\bibinfo {title} {First-principles determination of the
  soft mode in cubic {${\mathrm{ZrO}}_{2}$}},\ }\href
  {https://doi.org/10.1103/PhysRevLett.78.4063} {\bibfield  {journal} {\bibinfo
   {journal} {Phys. Rev. Lett.}\ }\textbf {\bibinfo {volume} {78}},\ \bibinfo
  {pages} {4063} (\bibinfo {year} {1997})}\BibitemShut {NoStop}%
\bibitem [{\citenamefont {Togo}\ \emph {et~al.}(2023)\citenamefont {Togo},
  \citenamefont {Chaput}, \citenamefont {Tadano},\ and\ \citenamefont
  {Tanaka}}]{phonopy-phono3py-JPCM}%
  \BibitemOpen
  \bibfield  {author} {\bibinfo {author} {\bibfnamefont {A.}~\bibnamefont
  {Togo}}, \bibinfo {author} {\bibfnamefont {L.}~\bibnamefont {Chaput}},
  \bibinfo {author} {\bibfnamefont {T.}~\bibnamefont {Tadano}},\ and\ \bibinfo
  {author} {\bibfnamefont {I.}~\bibnamefont {Tanaka}},\ }\bibfield  {title}
  {\bibinfo {title} {Implementation strategies in phonopy and phono3py},\
  }\href {https://doi.org/10.1088/1361-648X/acd831} {\bibfield  {journal}
  {\bibinfo  {journal} {J. Phys. Condens. Matter}\ }\textbf {\bibinfo {volume}
  {35}},\ \bibinfo {pages} {353001} (\bibinfo {year} {2023})}\BibitemShut
  {NoStop}%
\bibitem [{\citenamefont {Togo}(2023)}]{phonopy-phono3py-JPSJ}%
  \BibitemOpen
  \bibfield  {author} {\bibinfo {author} {\bibfnamefont {A.}~\bibnamefont
  {Togo}},\ }\bibfield  {title} {\bibinfo {title} {{First-principles phonon
  calculations with Phonopy and Phono3py}},\ }\href
  {https://doi.org/10.7566/JPSJ.92.012001} {\bibfield  {journal} {\bibinfo
  {journal} {J. Phys. Soc. Jpn.}\ }\textbf {\bibinfo {volume} {92}},\ \bibinfo
  {pages} {012001} (\bibinfo {year} {2023})}\BibitemShut {NoStop}%
\end{thebibliography}%
